%% file: iclr2027_conference.tex
\documentclass{article} 
\usepackage{iclr2027_conference,times}

\input{math_commands.tex}

\usepackage{hyperref}
\usepackage{url}
\usepackage{graphicx}
\usepackage{booktabs}
\usepackage{subcaption}
\usepackage{tabularx}
\usepackage{todonotes}

\title{FactorBench: A Portfolio-Aware Benchmark for Automated Factor Mining}

\author{Zhuohan Wang\\ 
Department of Informatics\\
King's College London\\
London, UK \\
\texttt{zhuohan.wang@kcl.ac.uk}
\And 
Carmine Ventre\\
Department of Informatics\\
King's College London\\
London, UK \\
\texttt{carmine.ventre@kcl.ac.uk}
}

\iclrfinalcopy 
\begin{document}

\maketitle

\begin{abstract}
Factor mining seeks to discover signals from financial data that predict future asset returns and guide portfolio construction. Automated factor mining now spans genetic programming, reinforcement learning, generative models, and large language model agents. Yet it remains unclear whether advances across these paradigms yield more generalizable, distinct, and economically useful financial signals. We introduce FactorBench, a portfolio-aware benchmark comparing roughly five thousand mined factors from nine automated mining methods across five equity markets. A shared data and evaluation contract supports both symbolic expressions and executable Python factors, connecting heterogeneous discovery algorithms to common signal combination and portfolio construction procedures. FactorBench traces the outputs of mining systems across three levels: factor validity, temporal generalization, and predictiveness beyond measured risk and style exposures; within- and across-method pool distinctness, including similarity to the benchmark Alpha101; and composite-signal quality and after-cost long-only and long--short portfolio performance. After systematically assessing whether advances in factor mining translate into signal quality and portfolio performance, FactorBench finds that no paradigm consistently dominates.
\end{abstract}

\section{Introduction}
Quantitative investment relies on factors: functions that transform information available at a given time into stock-level scores that predict future returns. These scores are used to rank securities and serve as inputs to portfolio construction. Discovering a factor traditionally requires forming a financial hypothesis, expressing it as a formula or executable program, and evaluating the resulting signal on historical data. Automated factor mining casts this process as a machine-discovery problem in which a system searches a large space of candidate transformations and returns a pool of signals for downstream selection and combination. The task is challenging because financial signals are weak and non-stationary, while repeated search over the same historical observations can readily identify patterns that do not persist.

Automated factor mining has developed through several generations of search methods. Genetic programming (GP) explores structured spaces of symbolic expressions through evolutionary operations. Reinforcement learning (RL) and neural generative methods learn how to construct or sample expressions, often considering how multiple factors work together \citep{yu2023generating,zhu2025alphaqcm,shi2025alphaforge,chen2026alphasage}. More recently, large language model (LLM) based agents have expanded the process beyond symbolic search by formulating financial hypotheses, generating expressions or executable code, and revising candidates through empirical feedback \citep{li2026r,tang2025alphaagent,han2026quantaalpha}. However, methodological innovation has outpaced standardized evaluation. Published results are produced under method-specific data, objectives, and backtesting pipelines, making performance difficult to compare across studies. More fundamentally, the score optimized during discovery is only an intermediate proxy: a mining system ultimately delivers a pool of signals whose value depends on how they behave outside the search environment and when used together in investment decisions. It therefore remains unclear whether newer search paradigms represent progress in practically useful financial discovery.

To address this gap, we introduce FactorBench, a portfolio-aware benchmark for assessing progress in automated factor discovery. Our contributions are threefold.

\textbf{We enable broad comparison across discovery systems.}
FactorBench evaluates roughly 5,000 factors from 9 methods spanning genetic programming, reinforcement learning, generative search, and LLM agents across 5 equity markets. A shared data and evaluation contract accommodates both symbolic expressions and executable code while preserving each method's native discovery process.

\textbf{We define progress beyond the search objective.}
FactorBench treats discovery performance as an intermediate result and evaluates mining systems at three connected levels: the factors they generate, the pools they deliver, and the portfolios those signals support. This systems-level view connects predictive quality to signal distinctness and downstream investment value.

\textbf{We show that advances in search do not translate consistently downstream.}
Relative advantages change as outputs move from the discovery phase to out-of-sample evaluation, from individual factors to factor pools, and from signals to portfolios. Across markets and evaluation stages, we find no consistent progression in which newer mining paradigms deliver more generalizable, distinct, and economically useful signals than earlier approaches. As the search for consistently better factors continues, FactorBench has the potential to become the evaluation toolkit for the field. The benchmark platform is accessible at 
repository\footnote{\url{https://github.com/ZhuoHan1998/FactorBench-Release}}.


\section{Related Work}

\textbf{Automated factor discovery.}
Automated factor discovery searches for signals from market observations that predict future asset returns. A discovered factor is typically represented either as a symbolic expression composed of market-data fields and mathematical operators, or as executable code that maps a market-data panel to a stock-level signal. Classical libraries such as Alpha101 collect manually specified formulaic factors and have become widely used references for developing and evaluating automated discovery methods \citep{kakushadze2016101}. Genetic programming provides an early approach to exploring this combinatorial space through mutation, crossover, and selection. More recent methods formulate expression construction as sequential decision making. AlphaGen optimizes the marginal contribution of a newly generated factor to an existing factor collection \citep{yu2023generating}, while AlphaQCM uses distributional reinforcement learning to address the sparse and non-stationary rewards encountered during factor search \citep{zhu2025alphaqcm}. AlphaCFG instead imposes an alpha-oriented context-free grammar and combines syntax-aware representations with Monte Carlo tree search \citep{yang2026alpha}. These methods primarily differ in how they represent and explore the space of symbolic factors, as well as in whether their search objective evaluates individual factors or their contribution to a collection.

Generative approaches place greater emphasis on producing diverse collections of candidate factors. AlphaForge combines a generative--predictive model with dynamic factor selection and combination \citep{shi2025alphaforge}, whereas AlphaSAGE uses structure-aware representations and generative flow networks to sample multiple high-reward expressions \citep{chen2026alphasage}. More recently, large language models have expanded factor discovery from symbolic search toward automated research workflows. R\&D-Agent coordinates hypothesis generation, executable implementation, empirical feedback, and factor--model refinement \citep{li2026r}. AlphaAgent introduces originality, semantic-alignment, and complexity constraints to reduce factor homogenization and decay \citep{tang2025alphaagent}. QuantaAlpha further combines LLM-based generation with trajectory-level evolutionary operations that revise and recombine previous discovery processes \citep{han2026quantaalpha}. Collectively, these systems extend automated discovery from searching over expression syntax to reasoning about hypotheses and revising candidates through experimental feedback.

\textbf{Benchmarks in Finance.}
Benchmarking literatures in finance have developed along two complementary directions. Financial-language benchmarks evaluate capabilities such as document understanding, numerical reasoning, forecasting, and decision making \citep{islam2023financebench,xie2024finben}. Quantitative-investment platforms and benchmarks instead standardize data processing, predictive modeling, market environments, and strategy backtesting \citep{yang2020qlib,liu2022finrl, wang2025quantbench}. At the trading-agent level, KTD-Fin controls access to memorized market information and attributes portfolio returns to market, style, and stock-selection components \citep{zhu2026knowing}. 

Recent benchmarks focus more directly on automated factor discovery. AlphaBench evaluates the factor-mining capabilities of LLMs through tasks covering factor generation, factor evaluation, and factor search \citep{luo2026alphabench}. AlphaEval instead provides a backtest-free framework for assessing generated formula factors along predictive power, temporal stability, robustness to market perturbations, financial logic, and diversity \citep{ding2026alphaeval}. Moving toward downstream strategy design, AlphaForgeBench asks LLMs to produce executable factors and factor-based strategies, enabling their outputs to be evaluated through a deterministic execution pipeline \citep{zhang2026alphaforgebench}. 
Our FactorBench broadens the unit of evaluation from individual model capabilities to the full path from discovery to investment decisions. By placing heterogeneous mining systems under a shared protocol, it examines whether advances in discovery translate into generalizable and distinct signals and, ultimately, downstream portfolio value in diverse equity markets.

\begin{figure}[t]
    \centering
    \includegraphics[width=\linewidth]{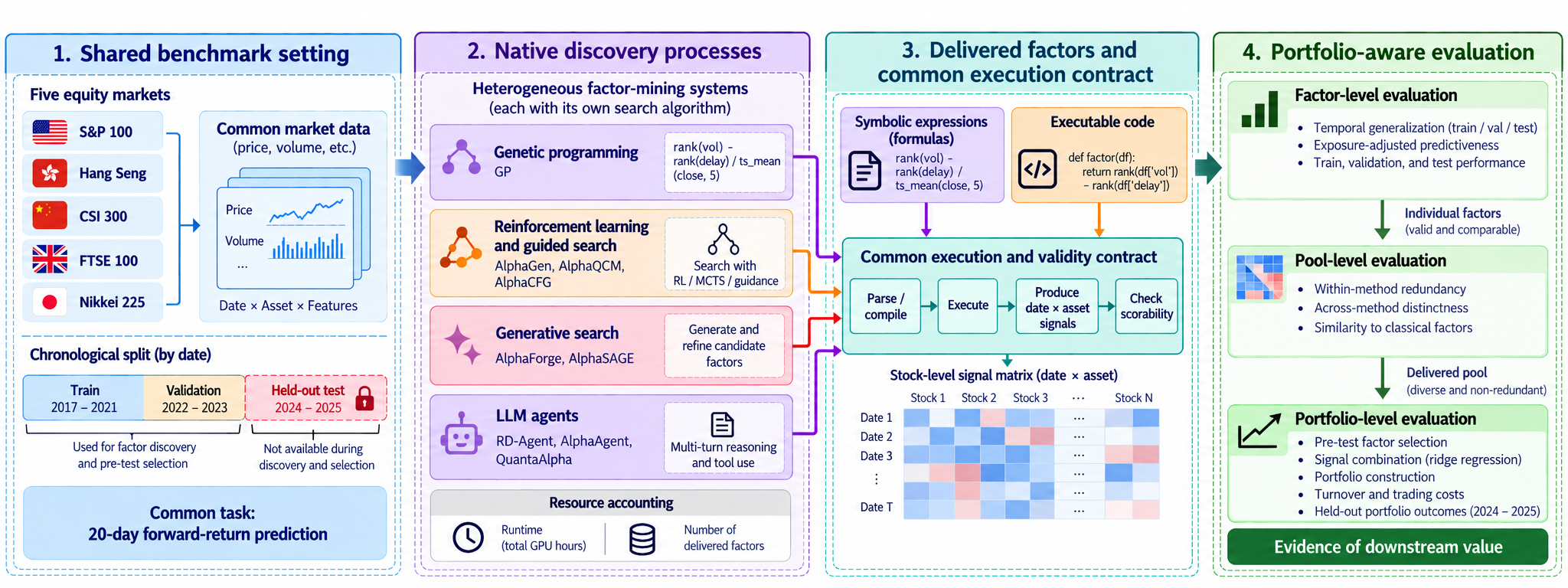}
    \caption{Overview of FactorBench.}
    \label{fig:benchmark_overview}
\end{figure}
\section{FactorBench: Benchmark Design}
This section describes how FactorBench makes heterogeneous factor-mining systems comparable while preserving their native discovery processes. We first present the benchmark workflow, then introduce the shared data setting and evaluated mining methods, and finally define the factor-, pool-, and portfolio-level evaluation protocols.
\subsection{Benchmark Overview}
\label{sec:benchmark_overview}

FactorBench is organized as a four-block pipeline that connects various factor discovery systems to common downstream evaluation, as illustrated in Figure~\ref{fig:benchmark_overview}. The first block defines the shared benchmark setting. All methods operate on the same markets, predict the same forward return, and follow a common chronological split. This shared setting ensures that methods are evaluated on the same financial prediction problem and under the same temporal information boundaries. 

The second block contains the heterogeneous mining systems. FactorBench covers genetic programming, reinforcement-learning and guided search, generative search, and LLM-based agents. Each method retains its native search algorithm, objective, and discovery procedure, allowing the benchmark to compare the systems as they are intended to operate. Their resource use and delivered output are recorded alongside the discovered factors, making the scale and cost of discovery visible. 

The third block provides the interface between discovery and evaluation. Mining systems may deliver symbolic expressions or executable programs, but a common execution contract converts both representations into standardized date-by-asset signal matrices. The downstream benchmark therefore evaluates the signals that the systems actually produce, independent of how those signals were represented or generated. 

The fourth block performs the portfolio-aware evaluation. FactorBench first examines the validity, temporal behavior, and overlap with measured style exposures. It then evaluates whether the delivered factor pool contains distinct information before testing whether selected and combined signals support portfolio performance after trading costs.

\subsection{Benchmark Setting and Factor Production}

We construct daily equity panels from five market indices: the S\&P~100 of the United States, Hang Seng Index of Hong Kong, CSI~300 of mainland China, FTSE~100 of the United Kingdom, and Nikkei~225 of Japan. Each panel contains adjusted price-and-volume observations and derived variables computed only from information available at signal formation. All mining methods receive the same input fields and predict the close-to-close return over the subsequent 20 trading days. We use a chronological split with 2017--2021 for training, 2022--2023 for validation, and 2024--2025 for testing. Training data support factor discovery, validation data may inform model selection and other pre-test decisions, and the test period is reserved for final trading evaluation. Full index constituents, preprocessing rules, field definitions, and temporal buffers are provided in Appendix~\ref{app:data_protocol}.

We evaluate nine factor-mining methods spanning four broad discovery paradigms. The comparison includes a genetic-programming baseline; reinforcement-learning and guided symbolic-search methods represented by AlphaGen, AlphaQCM, and AlphaCFG \citep{yu2023generating,zhu2025alphaqcm,yang2026alpha}; generative methods represented by AlphaForge and AlphaSAGE \citep{shi2025alphaforge,chen2026alphasage}; and LLM-based agents represented by R\&D-Agent, AlphaAgent, and QuantaAlpha \citep{li2026r,tang2025alphaagent,han2026quantaalpha}. Methods are selected to cover major developments in automated factor discovery and to provide an author-released implementation that can be adapted to the shared data contract. We preserve each method's search process and objective while recording its method-specific budget and delivered output. Eight methods produce symbolic expressions, whereas R\&D-Agent produces executable Python code; both forms are passed to the shared execution contract that maps every delivered factor to a date-by-asset signal matrix. We additionally evaluate the implementable subset of Alpha101 as a classical, non-mined reference \citep{kakushadze2016101}.  Complete implementation details, upstream versions, configurations, and deviations are reported in Appendix~\ref{app:method_implementation}.


\subsection{Evaluation Protocols}
\label{sec:evaluation_protocols}


\paragraph{Factor-level evaluation.}
We first assess \textit{factor validity}. A factor--market instance is
\emph{executable} if its symbolic expression or Python implementation runs successfully under the shared execution contract. We evaluate the resulting factor values separately on each trading date, where the cross section consists of the stocks in the corresponding market. A date is \emph{scorable} if at least two stocks have finite factor values and forward-return targets, and both the factor values and the targets have non-zero cross-sectional variance. We retain an executable factor--market instance for predictive evaluation only if it has at least 100 scorable dates across each time split. This requirement excludes factors whose outputs are predominantly missing or constant across stocks.
For each scorable factor, we assess \textit{temporal generalization}. We compute the daily cross-sectional information coefficient (IC), defined as the Pearson correlation between its signal and the subsequent 20-day stock return, and summarize it separately over the training, validation, and test periods. Factor orientation is determined from its training IC and held fixed thereafter, so validation and test performance measure temporal generalization and preserve out-of-sample sign reversals. 
We further examine whether \textit{factor predictiveness} extends beyond the measured risk and style exposures. On each date, we regress the signed factor cross-sectionally on market beta, volatility, momentum, short-term reversal, and liquidity. We use adjusted \(R^2\) to quantify exposure overlap, and residual IC to measure predictiveness after removing these exposures. More details are in Appendix~\ref{app:factor_evaluation}.

\paragraph{Pool-level evaluation.}
FactorBench evaluates whether a mining system delivers a broad and novel collection of signals rather than repeated variants of the same idea. We measure similarity from the cross-sectional correlation between factor signals, using Spearman correlation as the primary measure because the signals are ultimately used to rank stocks. Pool-level evaluation considers three forms of distinctness. First, \textit{within-method redundancy} measures whether factors produced by the same run duplicate one another. Second, \textit{across-method distinctness} examines whether different mining systems converge on the same signal space. Third, \textit{similarity to Alpha101} assesses whether mined factors reproduce previously published formulaic signals. The appendix evaluates the temporal stability of these patterns across the training, validation, and test periods and examines their robustness to replacing Spearman correlation with Pearson correlation. Further details are provided in Appendix~\ref{app:pool_evaluation}.

\paragraph{Portfolio-level evaluation.}
A mining system delivers a pool of signals, whereas investment decisions require a single portfolio-level ranking. FactorBench therefore evaluates the full translation from a delivered pool to an implementable strategy. Under a unified selection rule, it first chooses valid factors using predictive evidence available before the test period, while limiting redundancy among the retained signals. From the selected pool, FactorBench constructs composites using equal weighting, validation-IC weighting, and a regularized ridge regression that learns factor weights on training data with its regularization chosen on validation data.
The composite is evaluated first before portfolio construction, allowing us to separate the information preserved through factor selection and combination from the economic performance that remains after portfolio constraints, turnover, and trading costs.
To assess whether downstream value depends on investment constraints, FactorBench applies each composite to both long-only and dollar-neutral long--short portfolios. The long-only setting measures active stock-selection value relative to an equal-weight benchmark that holds every eligible stock in the market. The long--short setting holds equal dollar exposure in the top-ranked and bottom-ranked stocks, reducing the contribution of broad market movements and testing whether the composite distinguishes relative winners from relative losers. Comparing the two reveals whether a signal improves a fully invested portfolio and whether its cross-sectional value remains when market direction is substantially reduced.
Finally, FactorBench conducts sensitivity analysis under varying combination method, rebalancing frequency, and transaction costs, which identifies what design choices preserve or erode portfolio value. More details are in Appendix~\ref{app:portfolio_evaluation}.

\section{Experimental Evaluation}
Following the benchmark hierarchy, we first evaluate individual factors, then the distinctness of delivered pools, and finally whether the resulting composite signals produce after-cost portfolio value; additional market and configuration related results are reported in the appendix.

\subsection{Factor-Level Evaluation}
\label{sec:factor_evaluation}

\paragraph{Can mining systems reliably deliver valid factors?}
Table~\ref{tab:resource_validity} summarizes experimental coverage, resource use, and the validity of the factors delivered by each system. All nine automated methods complete 3-seed runs for each market on one A100 GPU, yielding 135 mining runs and 4,914 candidate factors. All LLM-based methods use Qwen3.8-27B~\citep{qwen2026qwen38}. Of these factors, 4,838 (98.5\%) execute successfully and 4,698 (95.6\%) satisfy the common scorability requirement of at least 100 valid dates across each split. The Alpha101 reference includes 71 formulas and excludes those requiring industry neutralization information that our data do not contain; all are executable, and 70 are scorable across all five markets.
Average wall-clock time ranges from 0.14 hours per run for GP to 27.90 hours for QuantaAlpha, while the number of delivered factors ranges from 107 to 1,619. These values describe the resource footprint and output scale of the evaluated systems. For LLM-based systems, wall-clock time also reflects external API latency and should not be interpreted solely as computation time.
Execution failures are concentrated in R\&D-Agent, AlphaAgent, and QuantaAlpha, whereas several methods with perfect execution still deliver factors that cannot be scored reliably. More details can be found in Appendix~\ref{app:factor_validity}.
\begin{table}[t]
    \centering
    \small
    \setlength{\tabcolsep}{6pt}
    \caption{Resource usage and factor validity.}
    \label{tab:resource_validity}
    \begin{tabular}{lrrrr}
        \toprule
        Method
        & Hours/Run
        & Delivered
        & \shortstack{Executable count (\%)}
        & \shortstack{Scorable count (\%)} \\
        \midrule
        Alpha101
        & -- & 71
        & 71 (100.0\%)
        & 70 (98.6\%) \\
        \midrule
        GP
        & 0.14 & 300
        & 300 (100.0\%)
        & 300 (100.0\%) \\
        AlphaGen
        & 1.66 & 150
        & 150 (100.0\%)
        & 149 (99.3\%) \\
        AlphaCFG
        & 2.37 & 150
        & 150 (100.0\%)
        & 139 (92.7\%) \\
        AlphaQCM
        & 8.19 & 143
        & 143 (100.0\%)
        & 142 (99.3\%) \\
        AlphaForge
        & 1.29 & 1{,}572
        & 1{,}572 (100.0\%)
        & 1{,}535 (97.6\%) \\
        AlphaSAGE
        & 0.31 & 750
        & 750 (100.0\%)
        & 696 (92.8\%) \\
        R\&D-Agent
        & 10.77 & 107
        & 88 (82.2\%)
        & 87 (81.3\%) \\
        AlphaAgent
        & 5.08 & 123
        & 99 (80.5\%)
        & 99 (80.5\%) \\
        QuantaAlpha
        & 27.90 & 1{,}619
        & 1{,}586 (98.0\%)
        & 1{,}551 (95.8\%) \\
        \bottomrule
    \end{tabular}
\end{table}

\begin{figure*}[t]
    \centering
    \includegraphics[width=\textwidth]{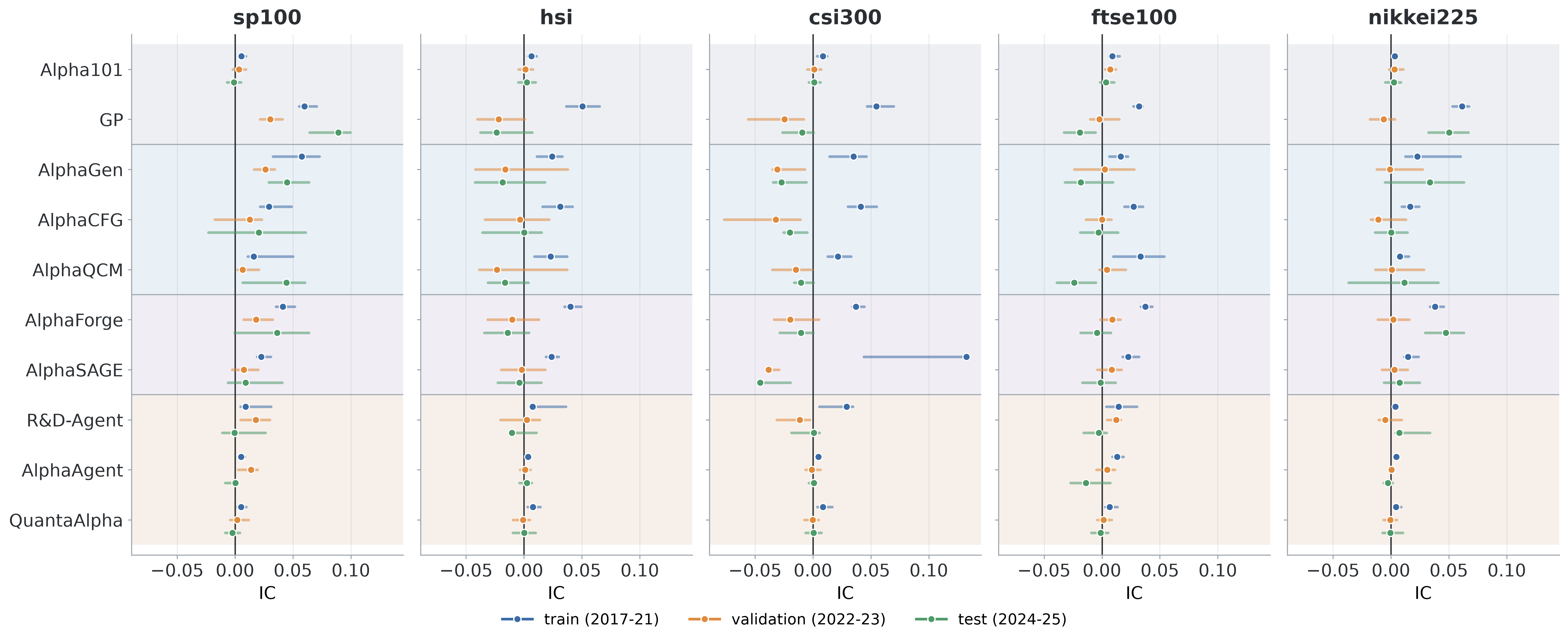}
    \caption{Temporal generalization across five equity markets. Points report the median IC across factors, and horizontal bars show the interquartile range, separately for the training, validation, and testing periods. Factor orientation is determined using training data and held fixed in subsequent periods, so negative validation or test IC indicates an out-of-sample sign reversal.}
    \label{fig:ic_generalization}
\end{figure*}

\paragraph{Do discovered factors generalize over time?}
Figure~\ref{fig:ic_generalization} compares the predictive performance of the same scorable factors across temporal splits. GP, RL, and generative methods frequently attain substantially higher IC during training, but much of this advantage disappears by validation and generally does not recover in testing. LLM-based methods start from lower training IC and consequently exhibit smaller absolute declines, but their held-out IC is not consistently higher. Their narrower train-to-test gap therefore reflects less in-sample performance to lose rather than stronger out-of-sample predictiveness.

This contrast reflects differences in how predictive feedback guides search. GP, RL, and generative methods repeatedly optimize training-period IC or related objectives, producing stronger in-sample fit but greater exposure to adaptive overfitting, while LLM agents use IC-related feedback through limited, multi-objective proposal-and-revision cycles. Temporal non-stationarity and factor decay may additionally weaken all methods after discovery. Additional RankIC, ICIR and temporal degradation results are reported in Appendix~\ref{app:factor_generalization}.
\begin{figure*}[t]
    \centering
    \includegraphics[width=\textwidth]{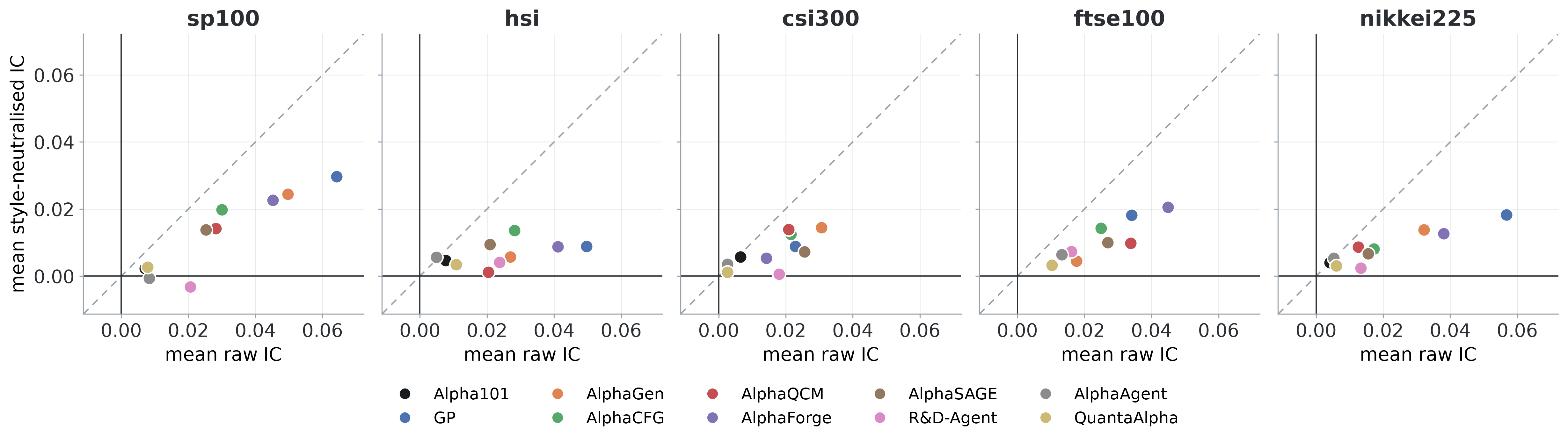}
    \caption{Predictive performance after neutralizing measured style exposures. Each point represents one mining method in one market during the training period. The horizontal axis reports mean raw IC, and the vertical axis reports mean IC after neutralization.}
    \label{fig:neutralized_ic}
\end{figure*}

\paragraph{What remains after controlling for measured exposures?}
Figure~\ref{fig:neutralized_ic} compares the method--market average training IC before and after neutralizing each factor against market beta, volatility, momentum, short-term reversal, and liquidity. Most points fall below the $45^\circ$ line, indicating that these exposures account for part of the predictive performance. The reduction is particularly consistent for GP, AlphaForge, and AlphaGen, with GP showing the largest overall decline. For R\&D-Agent, neutralized IC even becomes negative in some markets. Thus, high raw training IC does not necessarily represent predictive information beyond familiar risk and style exposures.

\begin{figure*}[t]
    \centering

    \begin{subfigure}[t]{0.325\textwidth}
        \centering
        \includegraphics[width=\linewidth]{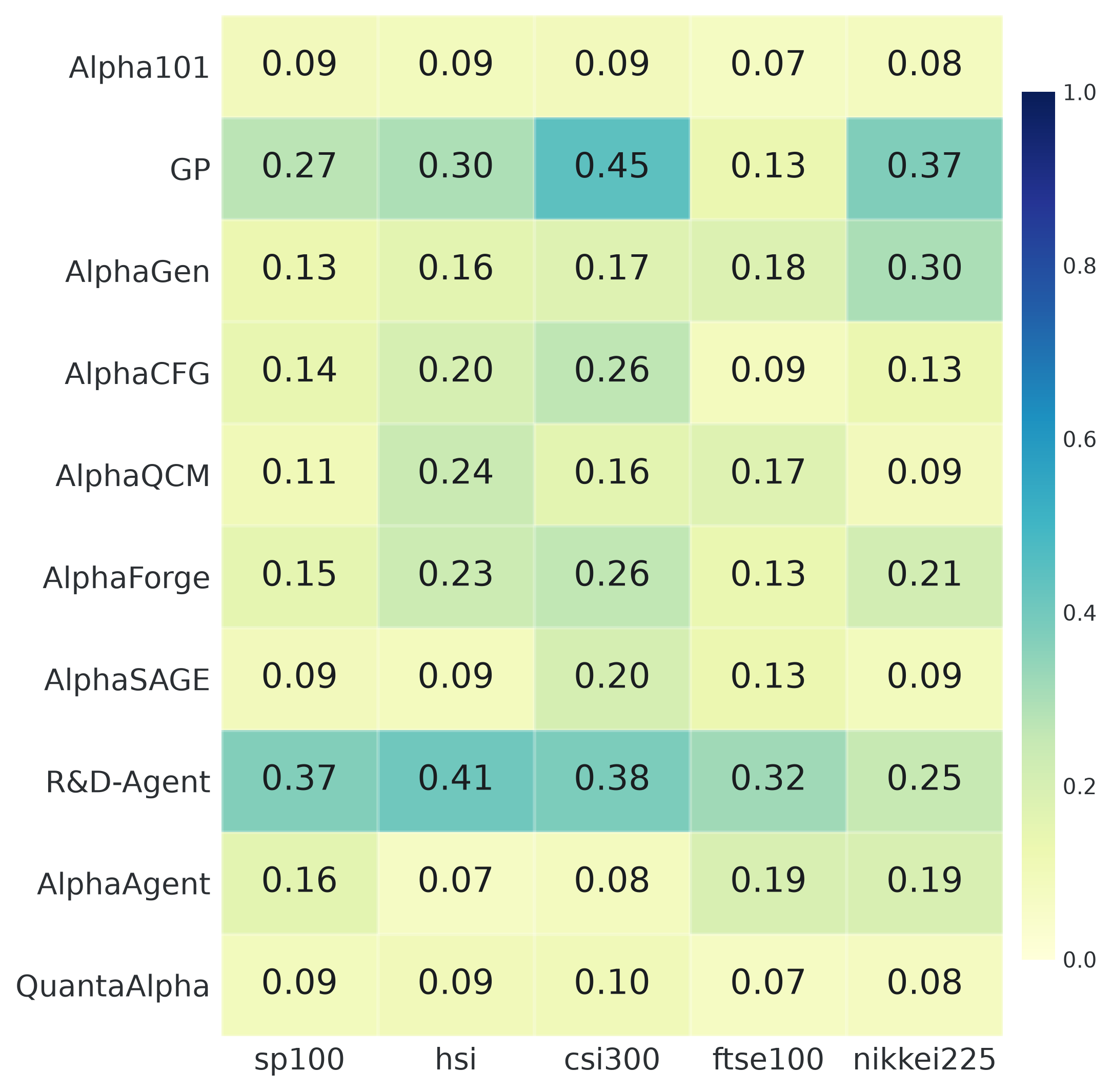}
        \caption{Training Period}
        \label{fig:style_overlap_train}
    \end{subfigure}
    \hfill
    \begin{subfigure}[t]{0.325\textwidth}
        \centering
        \includegraphics[width=\linewidth]{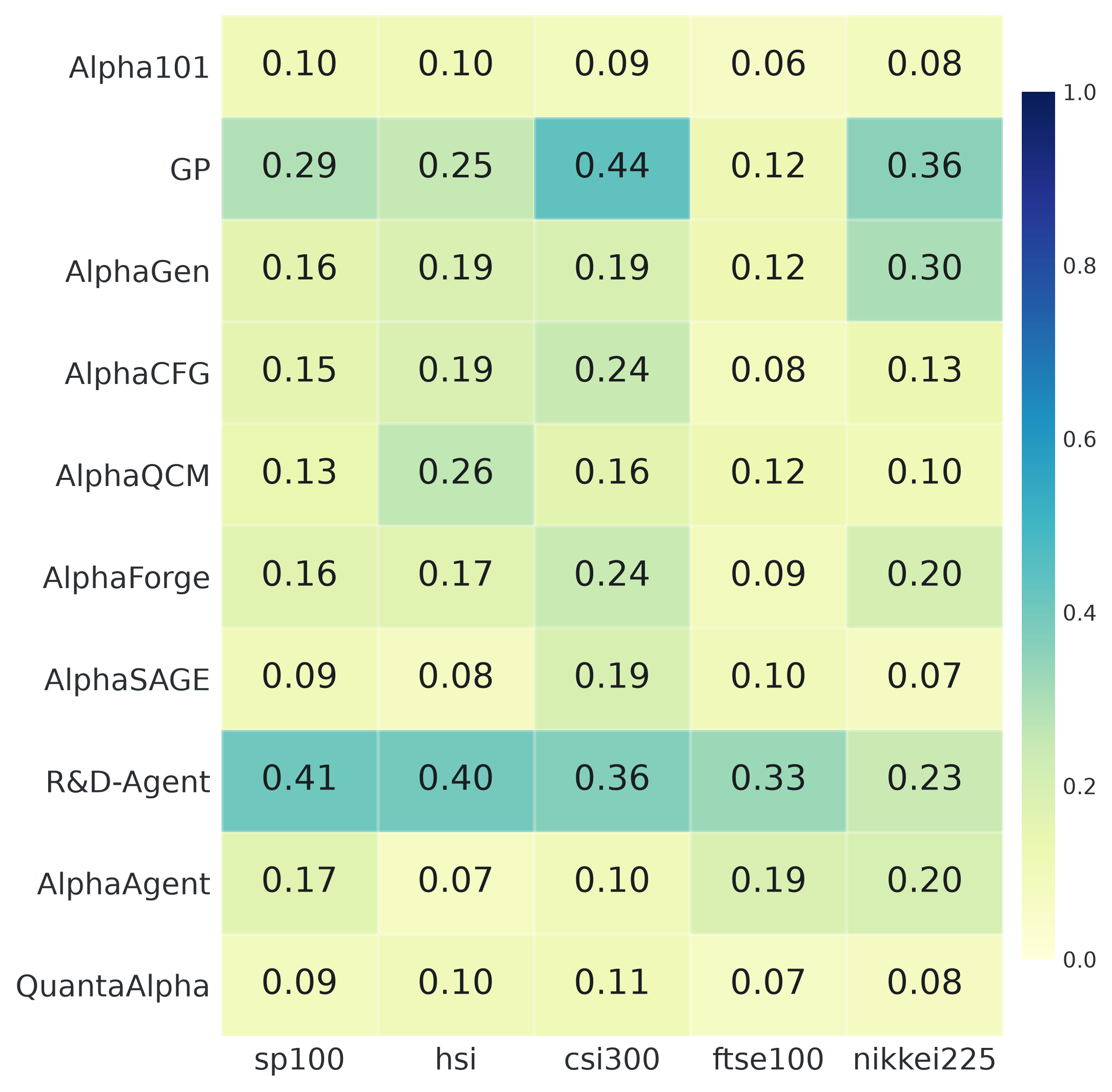}
        \caption{Validation Period}
        \label{fig:style_overlap_val}
    \end{subfigure}
    \hfill
    \begin{subfigure}[t]{0.325\textwidth}
        \centering
        \includegraphics[width=\linewidth]{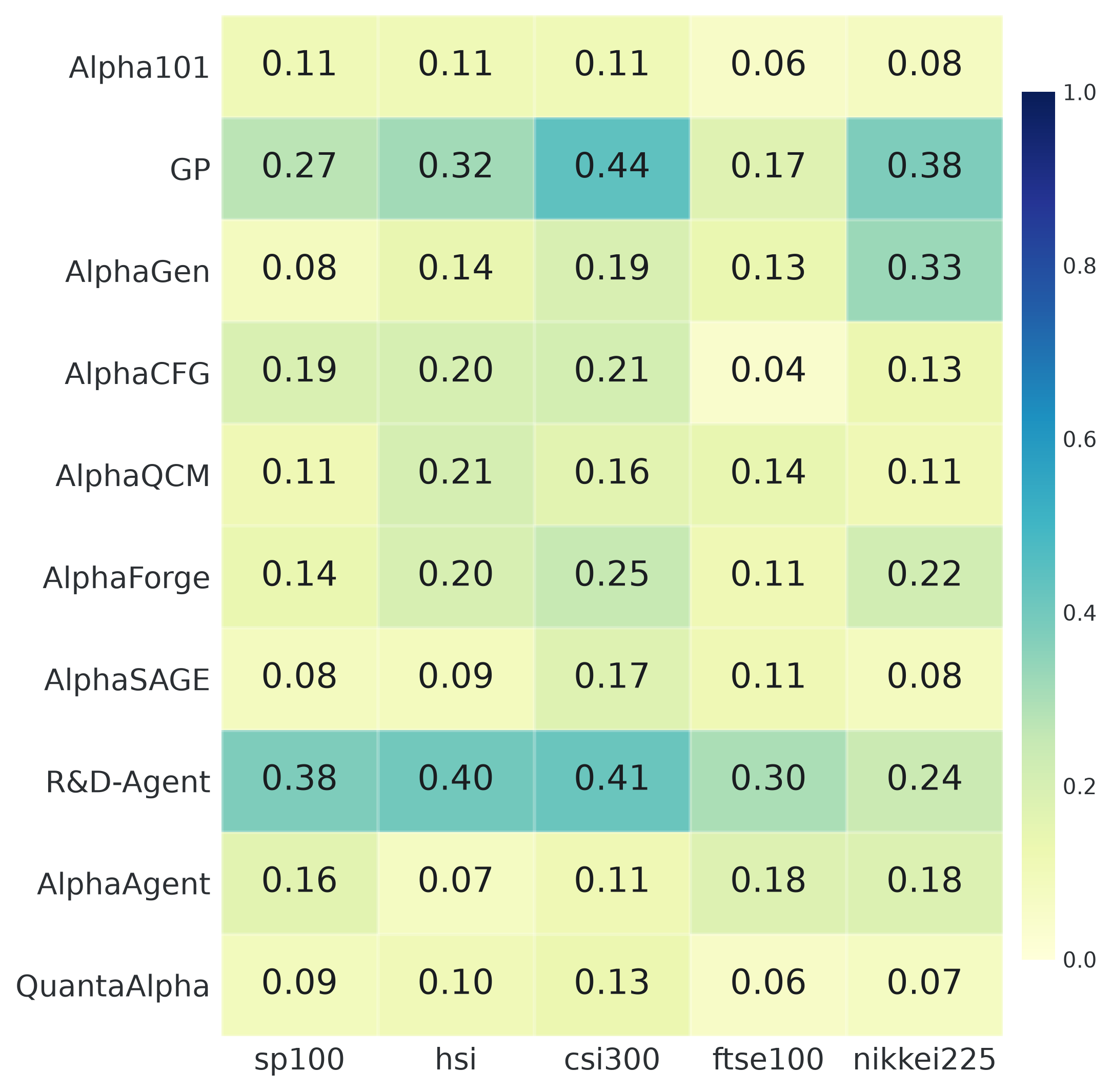}
        \caption{Test Period}
        \label{fig:style_overlap_test}
    \end{subfigure}

    \caption{Persistence of dependence on exposures. Each cell reports the adjusted \(R^2\) obtained by linearly regressing a factor on exposures. Adjusted \(R^2\) is first averaged over dates, summarized within each run, and then averaged equally across runs.}
    \label{fig:style_overlap}
\end{figure*}

Figure~\ref{fig:style_overlap} measures the extent of this overlap directly using adjusted $R^2$: higher values indicate that the measured exposures explain a larger share of a factor pool's cross-sectional variation. The method ranking is remarkably stable across training, validation, and test periods, suggesting that exposure overlap is a persistent property of the delivered signals. GP and R\&D-Agent consistently exhibit the highest overlap for different reasons. GP searches transformations of the same price-and-volume variables from which the controls are constructed, making familiar styles easy to rediscover. R\&D-Agent instead draws on pretrained financial knowledge, which may favor established factor concepts and trading heuristics. More details can be found in Appendix~\ref{app:factor_neutralization}.

\begin{figure}[t]
    \centering
    \includegraphics[width=\linewidth]{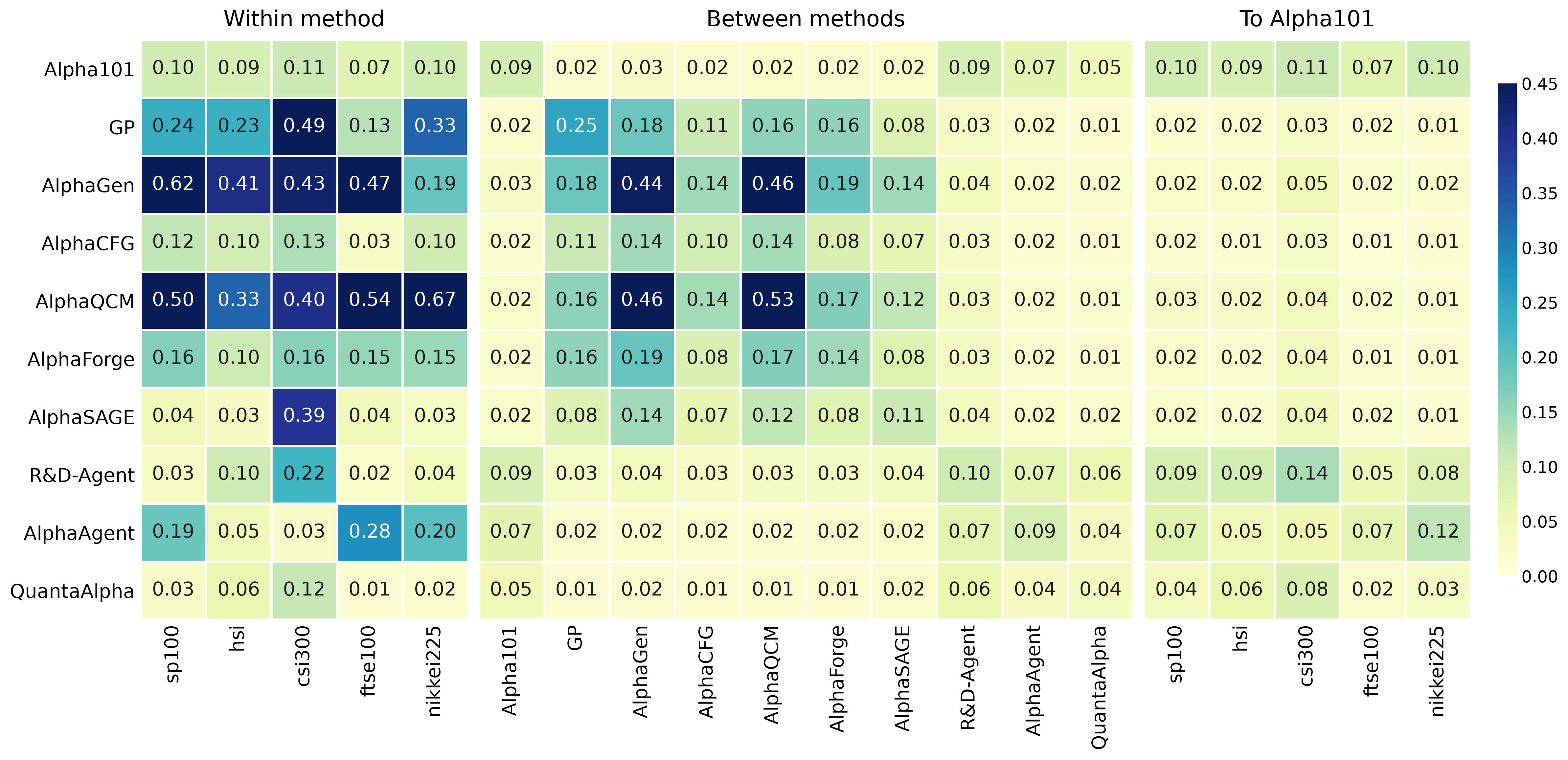}
    \caption{Pairwise similarity is the absolute value of the time-averaged daily cross-sectional Spearman correlation between factor signals, computed on matched assets and dates in training set. The left panel reports median similarity among factors within each method. The center panel reports median similarity between method pools, averaged across markets. The right panel compares each method with the Alpha101 in each market. Higher values indicate less distinct factor pools.}
    \label{fig:pool_distinctness}
\end{figure}

\subsection{Pool-Level Evaluation}
Figure~\ref{fig:pool_distinctness} first shows that nominal pool size does not necessarily represent effective signal breadth. AlphaGen and AlphaQCM exhibit the greatest within-pool redundancy, with GP also producing highly correlated pools in several markets. Other methods generally deliver more internally diverse collections, although the results remain market-dependent: AlphaSAGE, for example, is highly redundant in CSI~300 but substantially more diverse elsewhere.

The between-method panel reveals a broader family structure. Non-LLM methods tend to resemble other non-LLM methods, while the three LLM-based systems are more similar to one another than to most symbolic search methods. The strongest overlap occurs between AlphaGen and AlphaQCM, consistent with their related expression spaces, RL environments, and IC-based objectives. Similarity among the LLM methods may instead reflect their shared foundation model and common financial-language priors. These results suggest that a method's representation and search mechanism shape the region of factor space it explores.

The Alpha101 comparison further shows that LLM-generated pools are typically closer to published classical factors than those produced by non-LLM search. The pattern is strongest for R\&D-Agent and is also visible for AlphaAgent and QuantaAlpha. This is consistent with pretrained models favoring established financial concepts and familiar trading heuristics. Overall, distinctness does not improve monotonically across mining paradigms: newer methods may generate internally diverse pools while still remaining close to familiar factor families. More results can be found in Appendix~\ref{app:pool_evaluation}.

\begin{table*}[t]
    \centering
    \small
    \setlength{\tabcolsep}{3.2pt}
    \caption{\textbf{Test-period composite predictiveness and portfolio
    performance.} IC, RankIC, and ICIR describe the raw composite signal. IC (res) is computed after neutralizing the composite against market beta, volatility, momentum, short-term reversal, and liquidity. Portfolio statistics are reported after transaction costs; long--short results additionally include short-borrow costs. CAGR (compound annual growth rate) and maximum drawdown are percentages, and annual turnover is expressed in multiples of portfolio capital. Results first average seeds within each market and then weight the five markets equally.}
    \label{tab:portfolio_performance}
    \resizebox{\textwidth}{!}{
    \begin{tabular}{lrrrrrrrrrrrr}
        \toprule
        & \multicolumn{4}{c}{Composite predictiveness}
        & \multicolumn{4}{c}{Long-only portfolio}
        & \multicolumn{4}{c}{Long--short portfolio} \\
        \cmidrule(lr){2-5}
        \cmidrule(lr){6-9}
        \cmidrule(lr){10-13}
        Method
        & IC
        & RankIC
        & ICIR
        & IC (res)
        & \shortstack{Net\\CAGR}
        & \shortstack{Net\\Sharpe}
        & \shortstack{Max.\\drawdown}
        & \shortstack{Annual\\turnover}
        & \shortstack{Net\\CAGR}
        & \shortstack{Net\\Sharpe}
        & \shortstack{Max.\\drawdown}
        & \shortstack{Annual\\turnover} \\
        \midrule
        Alpha101
        & 0.0238 &  0.0176 &  0.1814 &  0.0237
        & 26.5 & 1.241 & -19.3 & 8.8
        & 2.6  &  0.319 &  -6.8 & 8.7 \\
        \midrule
        GP
        & 0.0165 & -0.0058 &  0.0806 & -0.0067
        & 30.4 & 1.173 & -23.2 & 6.3
        & 3.7  &  0.340 & -11.5 & 6.4 \\
        AlphaGen
        & 0.0064 & -0.0128 &  0.0435 & -0.0080
        & 28.0 & 1.127 & -23.2 & 5.5
        & 2.3  &  0.297 & -12.2 & 5.6 \\
        AlphaCFG
        & 0.0111 & -0.0007 &  0.0894 &  0.0007
        & 23.5 & 1.048 & -21.9 & 7.5
        & 1.0  &  0.169 & -10.4 & 7.5 \\
        AlphaQCM
        & 0.0191 &  0.0056 &  0.1691 &  0.0175
        & 29.6 & 1.313 & -20.4 & 4.7
        & 3.7  &  0.668 &  -8.5 & 4.6 \\
        AlphaForge
        & 0.0138 & -0.0036 &  0.0734 & -0.0019
        & 28.2 & 1.208 & -20.9 & 7.7
        & 2.9  &  0.380 &  -8.4 & 7.6 \\
        AlphaSAGE
        & 0.0107 & -0.0041 &  0.0709 & -0.0025
        & 29.2 & 1.244 & -21.0 & 7.4
        & 2.9  &  0.394 &  -8.4 & 7.5 \\
        R\&D-Agent
        & 0.0253 &  0.0046 &  0.1296 &  0.0011
        & 31.2 & 1.193 & -22.8 & 7.9
        & 6.5  &  0.679 &  -9.8 & 7.8 \\
        AlphaAgent
        & 0.0030 & -0.0002 & -0.0009 & -0.0025
        & 22.4 & 1.027 & -19.7 & 9.8
        & 0.3  & -0.087 &  -9.6 & 9.8 \\
        QuantaAlpha
        & 0.0153 &  0.0044 &  0.0886 &  0.0029
        & 25.3 & 1.106 & -20.8 & 8.9
        & 2.7  &  0.324 &  -8.9 & 8.8 \\
        \bottomrule
    \end{tabular}}
\end{table*}
\subsection{Portfolio-Level Evaluation}
\label{sec:portfolio_evaluation}

The preceding evaluations examine whether mined factors remain predictive, distinct, and informative beyond measured exposures. We now ask whether this information survives a unified factor selection, signal combination, portfolio construction, and backtesting with trading costs. Using training and validation information, FactorBench removes highly correlated candidates and combines the selected factors into a composite signal. The results reported in main text use ridge regression, a 20-day rebalancing interval, and a cost of 10 basis points per dollar traded. 
We report the composite factor's IC, RankIC, ICIR and residual IC neutralized against same styles introduced in Section~\ref{sec:factor_evaluation}.
Table~\ref{tab:portfolio_performance} follows each composite from test-period predictiveness to long-only and long--short portfolio performance. The long-only portfolios all produce positive average net returns, with annualized returns between 22.4\% and 31.2\% and Sharpe ratios between 1.03 and 1.31. R\&D-Agent obtains the highest net return, while AlphaQCM obtains the highest Sharpe ratio and lowest turnover. Alpha101 remains competitive with the mined factor pools. 
The long--short results provide a more discriminating test by reducing the contribution of broad market movements. Returns are substantially smaller and differences across methods become more visible. R\&D-Agent and AlphaQCM obtain the highest Sharpe ratios, while AlphaAgent produces a slightly negative Sharpe ratio. 

The composite diagnostics help explain why portfolio returns alone provide an incomplete account of progress. Although most composites have positive raw IC, their RankIC values are generally small and are negative for several methods, indicating limited monotonic ordering of the stock cross-section. 
Predictive performance also declines markedly after controlling for measured exposures. For example, R\&D-Agent has the highest raw composite IC, but its residual IC falls from 0.0253 to 0.0011. GP, AlphaGen, AlphaForge, AlphaSAGE and AlphaAgent have negative residual IC despite positive raw IC. In contrast, AlphaQCM retains most of its raw IC after neutralization, while Alpha101 is nearly unchanged. While R\&D-Agent and AlphaQCM have similar raw ICs and long--short Sharpe ratios, their residual ICs suggest different sources of performance: R\&D-Agent's raw predictiveness is largely associated with the measured exposures, whereas AlphaQCM retains predictive information beyond them and keeps lowest turnover.  
More generally, the transition from raw IC to residual IC and then to after-cost portfolio performance does not produce a consistent ordering across methods or mining paradigms. The experiments therefore provide no evidence of monotonic downstream improvement from genetic programming and reinforcement learning to generative methods and LLM agents. Results by market and seed, along with sensitivity to transaction costs, rebalancing frequency, and combination rule are reported in Appendix~\ref{app:portfolio_evaluation}.

\section{Discussion and Limitations}

\paragraph{What do our results imply about progress in factor mining?}
Across the three evaluation levels, FactorBench does not identify a clear winner among factor-mining paradigms. At the factor level, methods differ in their ability to deliver valid signals, and stronger discovery-period performance does not consistently survive temporal evaluation or adjustment for measured exposures. At the pool level, larger collections do not necessarily provide greater effective signal breadth, and distinctness does not improve monotonically across paradigms. At the portfolio level, broadly positive long-only performance contrasts with weaker long--short outcomes, and method rankings do not align consistently across predictive quality, exposure-neutralized predictiveness, and portfolio performance. Newer search paradigms therefore exhibit strengths on particular dimensions, but none consistently dominates across the full evaluation chain.

\paragraph{How far do these conclusions generalize?}
Our conclusions apply to the evaluated implementations under FactorBench's shared task rather than to the fundamental capabilities of their broader paradigms. The evidence covers five large-cap equity markets, one daily 20-day prediction target, one principal chronological split, and three random seeds per automated method. Because methods retain their native discovery processes, differences in runtime and output scale do not constitute a controlled comparison of computational efficiency. Residual IC controls only the five measured exposures and should not be interpreted as pure alpha, while the absence of industry classifications limits both neutralization and Alpha101 coverage. Finally, the portfolio tests use static universes and simplified cost and borrowing assumptions without fully modeling survivorship, capacity, market impact, or borrow availability. The results therefore support the conclusion that no paradigm consistently dominates within this benchmark, rather than a universal ranking. 


\section{Conclusions}

We introduced FactorBench, a portfolio-aware benchmark that evaluates approximately 5,000 factors produced by nine automated mining methods across five equity markets. FactorBench preserves each method's native discovery process while placing its outputs under a shared data, target, temporal-split, and execution contract. Its three-level evaluation follows the delivered factors from validity and temporal generalization, through exposure-neutralized predictiveness and pool distinctness, to long-only and long--short portfolio performance after trading costs.

The results do not identify a clear winner among genetic programming, reinforcement-learning, generative, and LLM-based approaches. Stronger discovery-period performance frequently deteriorates out of sample; larger factor pools do not necessarily contain broader information; and much of the composite predictiveness diminishes after measured exposures are removed. Broadly positive long-only outcomes also do not translate into a consistent advantage in the long--short setting. These findings show that progress in factor mining is multidimensional and cannot be established by a search objective or isolated predictive score alone. Future work can extend FactorBench through rolling temporal evaluations, point-in-time investment universes, richer exposure and market-impact models, and controlled studies of how factor quality changes with model and search compute.

\bibliography{iclr2027_conference}
\bibliographystyle{iclr2027_conference}

\clearpage

\appendix

\section{Data and Processing Protocol}
\label{app:data_protocol}

FactorBench uses daily equity data from Yahoo Finance for 5 fixed large-cap universes: the S\&P~100, Hang Seng Index, CSI~300, FTSE~100, and Nikkei~225. The constituent lists contain 100, 85, 272, 100, and 225 tickers, respectively, for a total of 782 securities. Because CSI 300 membership changes over time, we use a fixed subset of 272 constituents to maintain a consistent evaluation universe throughout the sample. The same constituent lists are used throughout the experiments. The exact identifiers are listed below.

\begingroup
\footnotesize
\sloppy

\noindent\textbf{S\&P 100 (100 tickers).}
\texttt{[AAPL, ABBV, ABT, ACN, ADBE, AMAT, AMD, AMGN, AMT, AMZN, AVGO,
AXP, BA, BAC, BK, BKNG, BLK, BMY, BRK-B, C, CAT, CL, CMCSA, COF, COP,
COST, CRM, CSCO, CVS, CVX, DE, DHR, DIS, DUK, EMR, FDX, GD, GE, GEV,
GILD, GM, GOOG, GOOGL, GS, HD, HON, IBM, INTC, INTU, ISRG, JNJ, JPM,
KO, LIN, LLY, LMT, LOW, LRCX, MA, MCD, MDLZ, MDT, META, MMM, MO, MRK,
MS, MSFT, MU, NEE, NFLX, NKE, NOW, NVDA, ORCL, PEP, PFE, PG, PLTR, PM,
QCOM, RTX, SBUX, SCHW, SO, SPG, T, TMO, TMUS, TSLA, TXN, UBER, UNH,
UNP, UPS, USB, V, VZ, WFC, WMT].}

\medskip

\noindent\textbf{Hang Seng Index (85 tickers).}
\texttt{[0005.HK, 0388.HK, 0939.HK, 1299.HK, 1398.HK, 2318.HK,
2388.HK, 2628.HK, 3968.HK, 3988.HK, 0002.HK, 0003.HK, 0006.HK,
0836.HK, 1038.HK, 2688.HK, 0012.HK, 0016.HK, 0101.HK, 0688.HK,
0823.HK, 0960.HK, 1109.HK, 1113.HK, 1209.HK, 1997.HK, 0001.HK,
0027.HK, 0066.HK, 0175.HK, 0241.HK, 0267.HK, 0285.HK, 0288.HK,
0291.HK, 0300.HK, 0316.HK, 0322.HK, 0386.HK, 0669.HK, 0700.HK,
0762.HK, 0857.HK, 0868.HK, 0881.HK, 0883.HK, 0941.HK, 0968.HK,
0981.HK, 0992.HK, 1024.HK, 1044.HK, 1088.HK, 1093.HK, 1099.HK,
1177.HK, 1211.HK, 1378.HK, 1810.HK, 1876.HK, 1928.HK, 1929.HK,
2015.HK, 2020.HK, 2057.HK, 2269.HK, 2313.HK, 2319.HK, 2331.HK,
2359.HK, 2382.HK, 2618.HK, 2899.HK, 3690.HK, 3692.HK, 6618.HK,
6690.HK, 6862.HK, 9618.HK, 9633.HK, 9888.HK, 9961.HK, 9988.HK,
9992.HK, 9999.HK].}

\medskip

\noindent\textbf{CSI 300 (272 tickers).}
\texttt{[600000.SS, 600009.SS, 600010.SS, 600011.SS, 600015.SS,
600016.SS, 600018.SS, 600019.SS, 600023.SS, 600025.SS, 600028.SS,
600029.SS, 600030.SS, 600031.SS, 600036.SS, 600039.SS, 600048.SS,
600050.SS, 600061.SS, 600085.SS, 600089.SS, 600104.SS, 600111.SS,
600115.SS, 600132.SS, 600150.SS, 600183.SS, 600188.SS, 600196.SS,
600219.SS, 600233.SS, 600276.SS, 600309.SS, 600346.SS, 600362.SS,
600406.SS, 600426.SS, 600436.SS, 600438.SS, 600460.SS, 600489.SS,
600515.SS, 600519.SS, 600547.SS, 600570.SS, 600584.SS, 600585.SS,
600588.SS, 600600.SS, 600606.SS, 600660.SS, 600674.SS, 600690.SS,
600732.SS, 600741.SS, 600745.SS, 600754.SS, 600760.SS, 600795.SS,
600803.SS, 600809.SS, 600845.SS, 600875.SS, 600886.SS, 600887.SS,
600893.SS, 600900.SS, 600905.SS, 600918.SS, 600919.SS, 600926.SS,
600938.SS, 600941.SS, 600958.SS, 600989.SS, 600999.SS, 601006.SS,
601009.SS, 601012.SS, 601021.SS, 601059.SS, 601066.SS, 601088.SS,
601100.SS, 601111.SS, 601117.SS, 601138.SS, 601155.SS, 601166.SS,
601169.SS, 601186.SS, 601211.SS, 601225.SS, 601229.SS, 601236.SS,
601238.SS, 601288.SS, 601318.SS, 601319.SS, 601336.SS, 601377.SS,
601390.SS, 601398.SS, 601600.SS, 601601.SS, 601607.SS, 601615.SS,
601628.SS, 601633.SS, 601658.SS, 601668.SS, 601669.SS, 601688.SS,
601689.SS, 601698.SS, 601699.SS, 601728.SS, 601766.SS, 601788.SS,
601799.SS, 601800.SS, 601808.SS, 601816.SS, 601818.SS, 601838.SS,
601857.SS, 601865.SS, 601868.SS, 601877.SS, 601878.SS, 601881.SS,
601888.SS, 601916.SS, 601919.SS, 601939.SS, 601985.SS, 601988.SS,
601995.SS, 601998.SS, 603019.SS, 603195.SS, 603259.SS, 603260.SS,
603288.SS, 603290.SS, 603369.SS, 603392.SS, 603486.SS, 603501.SS,
603659.SS, 603799.SS, 603806.SS, 603833.SS, 603899.SS, 603986.SS,
605117.SS, 605499.SS, 688008.SS, 688012.SS, 688036.SS, 688041.SS,
688065.SS, 688111.SS, 688126.SS, 688187.SS, 688223.SS, 688256.SS,
688271.SS, 688303.SS, 688363.SS, 688561.SS, 688599.SS, 688981.SS,
000001.SZ, 000002.SZ, 000063.SZ, 000069.SZ, 000100.SZ, 000157.SZ,
000301.SZ, 000333.SZ, 000408.SZ, 000538.SZ, 000568.SZ, 000596.SZ,
000617.SZ, 000625.SZ, 000651.SZ, 000708.SZ, 000725.SZ, 000733.SZ,
000768.SZ, 000786.SZ, 000792.SZ, 000800.SZ, 000858.SZ, 000877.SZ,
000895.SZ, 000938.SZ, 000963.SZ, 000977.SZ, 000983.SZ, 000999.SZ,
001289.SZ, 001979.SZ, 002001.SZ, 002007.SZ, 002027.SZ, 002049.SZ,
002074.SZ, 002129.SZ, 002142.SZ, 002179.SZ, 002180.SZ, 002202.SZ,
002230.SZ, 002241.SZ, 002252.SZ, 002304.SZ, 002311.SZ, 002352.SZ,
002371.SZ, 002410.SZ, 002415.SZ, 002460.SZ, 002466.SZ, 002475.SZ,
002493.SZ, 002555.SZ, 002594.SZ, 002603.SZ, 002648.SZ, 002709.SZ,
002714.SZ, 002736.SZ, 002812.SZ, 002821.SZ, 002841.SZ, 002920.SZ,
002938.SZ, 002916.SZ, 003816.SZ, 300015.SZ, 300033.SZ, 300059.SZ,
300124.SZ, 300142.SZ, 300223.SZ, 300274.SZ, 300308.SZ, 300316.SZ,
300347.SZ, 300408.SZ, 300413.SZ, 300433.SZ, 300450.SZ, 300454.SZ,
300496.SZ, 300498.SZ, 300628.SZ, 300661.SZ, 300750.SZ, 300751.SZ,
300759.SZ, 300760.SZ, 300763.SZ, 300782.SZ, 300896.SZ, 300919.SZ,
300957.SZ, 300979.SZ, 300999.SZ].}

\medskip

\noindent\textbf{FTSE 100 (100 tickers).}
\texttt{[III.L, ADM.L, AAF.L, ALW.L, AAL.L, ANTO.L, ABF.L, AZN.L,
AUTO.L, AV.L, BAB.L, BA.L, BARC.L, BTRW.L, BEZ.L, BKG.L, BP.L,
BATS.L, BLND.L, BT-A.L, BNZL.L, BRBY.L, CNA.L, CCEP.L, CCH.L, CPG.L,
CTEC.L, CRDA.L, DCC.L, DGE.L, DPLM.L, EDV.L, ENT.L, EXPN.L, FCIT.L,
FRES.L, GAW.L, GLEN.L, GSK.L, HLN.L, HLMA.L, HSX.L, HWDN.L, HSBA.L,
ICG.L, IGG.L, IHG.L, IMI.L, IMB.L, INF.L, IAG.L, ITRK.L, JD.L,
BGEO.L, KGF.L, LAND.L, LGEN.L, LLOY.L, LMP.L, LSEG.L, MNG.L, MKS.L,
MRO.L, MTLN.L, MNDI.L, NG.L, NWG.L, NXT.L, PSON.L, PSH.L, PSN.L,
PCT.L, PRU.L, RKT.L, REL.L, RTO.L, RMV.L, RIO.L, RR.L, SGE.L, SBRY.L,
SDR.L, SMT.L, SGRO.L, SVT.L, SHEL.L, SMIN.L, SN.L, SPX.L, SSE.L,
STAN.L, SDLF.L, STJ.L, TSCO.L, BBOX.L, ULVR.L, UU.L, VOD.L, WEIR.L,
WTB.L].}

\medskip

\noindent\textbf{Nikkei 225 (225 tickers).}
\texttt{[9202.T, 9201.T, 7267.T, 7202.T, 7261.T, 7211.T, 7201.T,
7270.T, 7269.T, 7203.T, 7272.T, 8304.T, 8331.T, 8354.T, 8306.T,
8411.T, 8308.T, 5831.T, 8316.T, 8309.T, 7186.T, 3407.T, 4061.T,
4901.T, 4452.T, 3405.T, 4188.T, 4183.T, 4021.T, 6988.T, 4004.T,
4063.T, 4911.T, 4005.T, 4043.T, 4042.T, 4208.T, 9433.T, 9432.T,
9434.T, 9984.T, 1721.T, 1925.T, 1808.T, 1963.T, 1812.T, 1802.T,
1928.T, 1803.T, 1801.T, 6857.T, 6770.T, 7751.T, 6902.T, 6954.T,
6504.T, 6702.T, 6501.T, 6861.T, 285A.T, 6971.T, 6920.T, 6479.T,
6503.T, 6981.T, 6701.T, 6594.T, 6645.T, 6752.T, 6723.T, 7752.T,
6963.T, 7735.T, 6724.T, 6753.T, 6758.T, 6526.T, 6976.T, 6762.T,
8035.T, 6506.T, 6841.T, 9502.T, 9503.T, 9501.T, 1332.T, 2802.T,
2502.T, 2914.T, 2801.T, 2503.T, 2269.T, 2282.T, 2871.T, 2002.T,
2501.T, 9532.T, 9531.T, 5201.T, 5333.T, 5214.T, 5233.T, 5301.T,
5332.T, 8750.T, 8725.T, 8630.T, 8795.T, 8766.T, 9147.T, 9064.T,
6113.T, 6367.T, 6361.T, 6305.T, 7004.T, 7013.T, 5631.T, 6473.T,
6301.T, 6326.T, 7011.T, 6471.T, 6472.T, 6103.T, 6302.T, 6273.T,
9107.T, 9104.T, 9101.T, 1605.T, 5714.T, 5803.T, 5801.T, 5711.T,
5706.T, 3436.T, 5802.T, 5713.T, 8253.T, 8697.T, 8591.T, 7832.T,
7912.T, 7911.T, 7951.T, 5020.T, 5019.T, 4503.T, 4519.T, 4568.T,
4523.T, 4151.T, 4578.T, 4506.T, 4507.T, 4502.T, 6146.T, 7741.T,
4902.T, 7731.T, 7733.T, 4543.T, 3861.T, 9022.T, 9020.T, 9008.T,
9009.T, 9007.T, 9001.T, 9005.T, 9021.T, 8802.T, 8801.T, 8830.T,
8804.T, 3289.T, 8267.T, 9983.T, 3099.T, 3086.T, 8252.T, 7453.T,
9843.T, 7532.T, 3382.T, 8233.T, 3092.T, 5108.T, 5101.T, 8601.T,
8604.T, 6532.T, 4751.T, 2432.T, 4324.T, 6178.T, 9766.T, 4689.T,
4385.T, 2413.T, 3659.T, 7974.T, 4307.T, 4661.T, 4755.T, 6098.T,
9735.T, 3697.T, 9602.T, 4704.T, 7012.T, 5411.T, 5406.T, 5401.T,
3401.T, 3402.T, 8001.T, 8002.T, 8058.T, 8031.T, 2768.T, 8053.T,
8015.T, 9301.T].}

Daily open, high, low, close, and volume observations are downloaded through \texttt{yfinance} with \texttt{auto\_adjust=True}. Observations without a valid closing price are removed. Factor-mining methods receive the following common input fields:

\begin{itemize}
    \item \textbf{Open, high, low, and close:} daily prices adjusted according
    to the Yahoo Finance adjustment procedure;
    \item \textbf{Volume:} daily traded share volume;
    \item \textbf{VWAP proxy:}
    $(\mathrm{high}+\mathrm{low}+\mathrm{close})/3$;
    \item \textbf{One-day return:}
    $\mathrm{close}_{i,t}/\mathrm{close}_{i,t-1}-1$; and
    \item \textbf{Average daily volume:} the 20-day rolling mean of volume,
    requiring at least five observations.
\end{itemize}

The common data layer does not apply additional winsorization, cross-sectional normalization, or general missing-value imputation. These transformations, when required, are performed within the corresponding mining method or evaluation procedure.

For every stock $i$ and date $t$, the prediction target is the 20-trading-day forward close-to-close return,
\begin{equation}
    y_{i,t}^{(20)}=
    \frac{C_{i,t+20}}{C_{i,t}} - 1,
\end{equation}
where $C_{i,t}$ is the adjusted closing price. The horizon is measured using each market's trading calendar. The target is stored separately from the input fields and is not exposed to factor expressions or executable factor programs.

Daily observations cover January 2016 through December 2025. The benchmark uses 2016 as warm-up history and evaluates factors from January 2017 through October 2025. Each market is processed independently on its own trading calendar, without aligning trading days or converting prices across currencies. The constituent sets are held fixed throughout the sample and therefore do not reconstruct historical index membership. This design provides a consistent cross-sectional universe but may introduce survivorship bias into the reported results.

\section{Method Implementation}
\label{app:method_implementation}
\subsection{Common Setup and Implementation Summary}
\label{app:common_implementation}

\paragraph{Shared experimental setup.}
We evaluate the nine automated mining methods under the same market panels, input fields, 20-day forward-return target, and chronological partition: 2017--2021 for training, 2022--2023 for validation, and January 2024--October 2025 for testing. Each method is run independently on each of the five markets with three repetitions, producing 15 runs per method (135 in total). Hyperparameters and search budgets are held fixed across markets. For stochastic non-LLM methods, the repetitions use random seeds 0, 1, and 2. For LLM-based methods, these identifiers denote independent runs but do not ensure exact reproducibility because model generation and API execution remain nondeterministic.

Methods retain their native search algorithms, objectives, expression spaces, and pool-construction procedures. FactorBench replaces method-specific data loading with adapters to the shared market panels and maps each method's prediction target to the benchmark target. Training data may be used for search and parameter estimation, while validation data may be used when required by a method's native feedback or model-selection procedure. Reported test outcomes are not used for factor acceptance, pool selection, hyperparameter tuning, or downstream composite construction.

\paragraph{Implementation provenance.}
We use the authors' released implementations of AlphaGen, AlphaCFG, AlphaQCM,
AlphaForge, AlphaSAGE, R\&D-Agent, AlphaAgent, and QuantaAlpha. Each
implementation is vendored at a fixed upstream commit, and the corresponding
FactorBench wrapper records the repository, commit, configuration, search
budget, runtime, and delivered factors for every run. GP follows the baseline
implementation released with AlphaGen. Alpha101 is evaluated as a fixed
reference library rather than as a mining method. Exact
operator spaces, hyperparameters, and adaptations are documented in the
method-specific descriptions below.

Table~\ref{tab:implementation_summary} summarizes the principal configuration
used for each method. The reported budgets apply to one market--seed run.
They preserve the scale specified by the released implementation or paper
where feasible; consequently, they are not intended to equalize the number of
candidate evaluations, generated factors, or units of compute across
paradigms.

\begin{table}[t]
    \centering
    \footnotesize
    \setlength{\tabcolsep}{5pt}
    \renewcommand{\arraystretch}{1.16}
    \caption{\textbf{Implementation summary.}
    Search configurations are reported per market--seed run.
    Capacities are target pool sizes rather than guaranteed output counts.
    All methods return symbolic expressions except R\&D-Agent, which returns
    executable Python code.}
    \label{tab:implementation_summary}

    \begin{tabularx}{\textwidth}{
        @{}
        >{\raggedright\arraybackslash}p{2.25cm}
        >{\raggedright\arraybackslash}p{3.10cm}
        >{\raggedright\arraybackslash}X
        @{}
    }
        \toprule
        \textbf{Method}
        & \textbf{Paradigm}
        & \textbf{Configuration per run} \\
        \midrule

        \multicolumn{3}{@{}l}{\textit{Reference library}} \\
        \addlinespace[2pt]
        Alpha101
        & Fixed library
        & 71 retained formulas; no mining budget \\
        
        \addlinespace[3pt]
        \midrule
        \multicolumn{3}{@{}l}{\textit{Non-LLM mining methods}} \\
        \addlinespace[2pt]
        GP
        & Genetic programming
        & 1,000 individuals $\times$ 40 generations; pool capacity 20 \\

        AlphaGen
        & Reinforcement learning
        & 200,000 PPO steps; pool capacity 10 \\

        AlphaCFG
        & Guided tree search
        & 200 iterations; 50 games per iteration; 64 MCTS simulations;
          pool capacity 10 \\

        AlphaQCM
        & Distributional RL
        & IQN agent; 2 million steps; pool capacity 10;
          $\lambda_{\mathrm{std}}=1$ \\

        AlphaForge
        & Generative search
        & Target zoo size 100 (not a hard cap); at most 50 generation rounds \\

        AlphaSAGE
        & GFlowNet
        & 10,000 sampling episodes; GNN encoder; pool capacity 50 \\

        \addlinespace[3pt]
        \midrule
        \multicolumn{3}{@{}l}{\textit{LLM-based mining methods}} \\
        \addlinespace[2pt]
        R\&D-Agent
        & Research--development agent
        & Five research--development--feedback cycles \\

        AlphaAgent
        & Multi-agent search
        & Three proposal--construction--evaluation--feedback rounds \\

        QuantaAlpha
        & Evolutionary search
        & 10 research directions; 11 evolutionary rounds;
          10 candidates per size-2 crossover \\
        \bottomrule
    \end{tabularx}
\end{table}

\paragraph{LLM configuration.}
R\&D-Agent, AlphaAgent, and QuantaAlpha use the same Qwen3.8-27B model
endpoint, so differences among them arise from their agent workflows,
prompts, feedback mechanisms, and search budgets rather than from the
underlying language model. The methods retain their native prompting and
reasoning structures. R\&D-Agent generates executable Python factors, whereas
AlphaAgent and QuantaAlpha generate expressions in their respective
domain-specific languages. Model and endpoint settings are fixed across
markets, although external API latency and nondeterministic generation affect
runtime and exact reproducibility. For AlphaAgent and QuantaAlpha, IC-based feedback is computed against the
common 20-trading-day forward return. Their native one-day portfolio metrics
are retained as additional feedback signals, while the original prompts refer
more generally to predicting returns over the next several days.

\paragraph{Common output contract.}
Every run is serialized with its method, market, repetition identifier, temporal split, target, search budget, runtime, provenance, and complete set of delivered factors. Symbolic factors retain both their expression and the grammar needed for execution; code-based factors retain their Python source and entry point. Method-specific adapters convert both representations into a date-by-asset signal under the shared execution interface. A delivered factor is marked executable when this procedure produces a valid output and scorable when it provides sufficient observations to compute IC on at least 100 dates in every temporal split. Failed or unscorable outputs remain part of the run record and are included in the validity analysis.

\paragraph{Compute environment.}
The non-LLM methods are executed in Python environments containing the
dependency versions required by their released implementations. Separate
environments are used where incompatible dependencies prevent a shared
installation, particularly for AlphaSAGE and the three LLM systems. Device
selection is exposed through the FactorBench wrappers, while the underlying
optimization algorithms remain unchanged. Because methods use different
software stacks, hardware resources, and external API calls, recorded
wall-clock time characterizes their practical resource footprint in our
setting rather than a controlled measure of computational efficiency.

\subsection{Method-Specific Implementations}
\label{app:method_specific_implementations}

This subsection records the implementation choices that distinguish the
evaluated methods. Unless stated otherwise, the released source code is
vendored without modification, and FactorBench supplies an external wrapper
for loading the shared market panels, recording run metadata, and exporting
the delivered factors. Changes that alter the method's target, feedback, or
stopping rule are stated explicitly.

\paragraph{Alpha101.}
Alpha101 serves as a fixed reference library rather than an automated mining
method. We implement the published formula definitions and evaluate every
applicable formula on each market under the same execution and scoring
contract as the mined factors. Because the benchmark data do not contain
industry classifications, formulas requiring industry neutralization are
excluded, leaving 71 of the original 101 formulas. The implementation raises
execution failures rather than silently removing formulas. For rolling
correlations, the required number of observations is bounded by the requested
window length, and undefined correlations are replaced by zero, following the
implementation used to construct this reference library. Alpha101 involves no
search, model fitting, or random seed; each formula identifier is dispatched
to its fixed implementation.

\paragraph{Genetic programming.}
The GP baseline is based on the implementation released with AlphaGen
and searches directly over AlphaGen expressions.
Programs are constructed from the six price-and-volume fields
\{\texttt{open}, \texttt{high}, \texttt{low}, \texttt{close},
\texttt{volume}, \texttt{vwap}\}, constants, arithmetic operations,
cross-sectional transformations, and rolling time-series operators. Fitness
is the mean training-period IC with the shared 20-day forward return. We use a
population of 1,000 programs for 40 generations, tournament size 600, initial
tree depths between 2 and 6, crossover probability 0.3, subtree-mutation
probability 0.1, hoist-mutation probability 0.01, and point-mutation
probability 0.1. Expressions are limited to 20 operator nodes. The delivered
pool contains at most 20 high-IC expressions subject to a pairwise-correlation
threshold of 0.7.

The original GP implementation can assign fitness to expressions that cannot
subsequently be parsed or that yield effectively constant signals. Before
export, we therefore require every candidate to contain an input field, pass a
round trip through the shared AlphaGen parser, and exhibit cross-sectional
dispersion on at least 95\% of training dates. These checks use training data
only and are treated as output-validity requirements rather than additional
predictive selection criteria.

\paragraph{AlphaGen.}
We use the official AlphaGen implementation from
\citet{yu2023generating}. AlphaGen formulates expression construction as a
masked sequential decision problem and trains a PPO policy to generate valid
expressions. The policy uses a two-layer LSTM feature extractor with hidden
dimension 128 and dropout 0.1. PPO is trained for 200,000 environment steps
with discount factor 1, entropy coefficient 0.01, and batch size 128. Candidate
factors are inserted into an MSE-based linear alpha pool of capacity 10, and
the reward reflects the improvement in the training-period pool objective.
The official expression grammar, masking rules, pool update, and optimization
procedure are retained.

FactorBench replaces the original qlib data loader with an adapter to the
shared panels. The method's native target,
\(\operatorname{close}_{t+20}/\operatorname{close}_{t}-1\), is identical to
the benchmark target and therefore requires no substitution. Pool updates and
factor generation depend on training data. Split-level diagnostics produced
by the upstream logging callback do not alter the policy, reward, or exported
pool.

\paragraph{AlphaCFG.}
We use the official AlphaCFG implementation from
\citet{yang2026alpha}. The evaluated configuration is the paper's
\texttt{cfg-sem-k} variant in pool mode with a TreeLSTM policy and value
network. Its context-free grammar restricts the search to syntactically and
semantically valid expressions, while Monte Carlo tree search constructs
expressions using guidance learned from self-play. We use a maximum expression
length of 10, a pool capacity of 10, 200 training iterations, 50 self-play
games per iteration, and 64 MCTS simulations per action.

FactorBench invokes the released experiment entry point with the common
temporal splits and 20-day prediction horizon. The qlib-backed stock-data
class is rebound to the shared panel adapter; the grammar, tree search,
policy/value training, and pool construction are unchanged. The final pool is
read from the last recorded pool snapshot and exported using the shared
AlphaGen-compatible expression grammar.

\paragraph{AlphaQCM.}
We use the official AlphaQCM implementation from
\citet{zhu2025alphaqcm}. Among its distributional RL variants, we evaluate the
implicit quantile network (IQN) agent with the released configuration:
2 million environment steps, batch size 128, 64 sampled quantiles, prioritized
experience replay, discount factor 1, pool capacity 10, and
\(\lambda_{\mathrm{std}}=1\). The agent operates over the method's
AlphaGen-derived expression space and estimates the distribution of the
pool-level reward rather than only its expectation.

The shared data adapter replaces qlib input while preserving the official
target and pool construction. Following the released model-selection rule, we
export the pool snapshot with the best validation IC, falling back to the
final pool if no validation-best snapshot is available. We also guard two rare
numerical failures in the released implementation: a singular quantile-moment
calculation and a boundary error in prioritized replay sampling. In both
cases, the wrapper retries the stochastic draw; computations that succeed
under the original implementation are unchanged.

\paragraph{AlphaForge.}
We use the official AlphaForge implementation from
\citet{shi2025alphaforge}. AlphaForge alternates between a predictor that
estimates candidate quality and a generator trained to produce promising and
diverse expressions. Expressions contain at most 20 tokens. The generator and
predictor are trained for at most 200 and 100 epochs per round, respectively.
Candidate admission uses an absolute-IC threshold of 0.03, an ICIR threshold
of 0.1, and a correlation threshold of 0.7. The target factor-zoo size is 100.

Two adaptations affect the research configuration. First, the original
delayed VWAP-to-VWAP target is replaced by the common 20-day close-to-close
return so that AlphaForge optimizes the quantity on which every method is
evaluated. Second, the released mining loop has no stopping condition when the
zoo cannot reach its target size. We impose a maximum of 50 rounds and export
the factors available when this limit is reached. FactorBench evaluates the
stage-one factor zoo; AlphaForge's native dynamic combiner is not used because
all methods are passed to the benchmark's common signal-combination
procedures. Because AlphaForge uses its own fork of the expression grammar,
its factors are tagged with a separate grammar and evaluated in an isolated
process.

\paragraph{AlphaSAGE.}
We use the official AlphaSAGE implementation from
\citet{chen2026alphasage}. AlphaSAGE represents a partially constructed
expression as a graph and uses a relational graph convolutional encoder within
a trajectory-balance GFlowNet. Its dense reward combines predictive quality,
a self-supervised structural signal, and novelty relative to the current pool.
We run 10,000 sampling episodes with pool capacity 50, optimization frequency
64, entropy coefficient 0.01, initial self-supervision weight 1, and initial
novelty weight 0.3. The two auxiliary weights decay linearly to zero over the
run. The search space permits expressions of length at most 20 over six input
fields and 36 operators.

The released implementation's hard-coded qlib paths and calendar windows are
replaced by the shared data adapter and split. Its native 20-day
close-to-close target already matches the benchmark target. AlphaSAGE uses a
method-specific expression grammar and is therefore executed in a separate
environment with the released \texttt{torchgfn} and
\texttt{torch-geometric} dependency versions. We export the factor pool
produced by the GFlowNet and do not run the method's optional adaptive
combination stage.

\paragraph{Shared LLM settings.}
R\&D-Agent, AlphaAgent, and QuantaAlpha use the same Qwen3.8-27B endpoint
through an OpenAI-compatible API. 
We preserve each method's released
system prompts, user prompts, structured-output requests, and feedback
templates; thus, their differences arise from the surrounding agent workflow
and search procedure. Calls use the upstream default sampling temperature of
0.5 unless a method explicitly supplies another value. No provider-level
sampling seed is imposed, so the three reported repetitions represent
independent executions rather than exactly reproducible generations. 

\paragraph{R\&D-Agent.}
We use the official R\&D-Agent implementation from \citet{li2026r}. R\&D-Agent alternates among hypothesis generation, factor development, execution, and feedback. Its Co-STEER developer produces executable Python programs rather than symbolic expressions and can revise failed implementations using execution feedback and retrieved experience. We run five research--development--feedback loops per market and repetition and retain every unique factor program produced in the loop trace.

The released qlib/LGBM experiment runner is replaced through R\&D-Agent's component-injection interface. The FactorBench runner executes each generated program on the shared panel and reports training and validation IC and RankIC to the native feedback mechanism. A candidate pool is summarized by the equal weight average of its daily standardized signals when feedback requires a pool-level result. The scenario, research proposal, Co-STEER code generation, and feedback prompts remain unchanged. Test-period metrics are never supplied to the agent. Generated programs retain the official file contract: they read the supplied daily market panel and write a date-by-asset factor signal.

\paragraph{AlphaAgent.}
We use the official implementation from \citet{tang2025alphaagent}. AlphaAgent organizes discovery into an Idea Agent, a Factor Agent, an expression constructor, an evaluation stage, and a feedback agent. Its FactorRegulator discourages structural duplication and rejects expressions that rely excessively on free constants or a single input variable. We run three complete discovery rounds. The method generates expressions in its approximately 70-operator domain-specific language using the native fields \{\texttt{open}, \texttt{high}, \texttt{low}, \texttt{close}, \texttt{volume}, \texttt{return}\}.

The original qlib backtest is replaced through the method's runner interface. Feedback is computed on the validation segment and contains composite IC, gross top-minus-bottom-quintile next-day return, information ratio, and maximum drawdown, matching the metric labels expected by the released feedback prompts. Test data are not used. We retain the Idea Agent, Factor Agent, FactorRegulator, expression parser, and feedback summarizer. A small execution guard corrects a released-code failure in which expression repair indexes an empty successful-factor memory during the first round; it changes only the empty-memory case and leaves the prompts and repair procedure unchanged.

\paragraph{QuantaAlpha.}
We use the official QuantaAlpha implementation from \citet{han2026quantaalpha}. QuantaAlpha extends the AlphaAgent workflow with LLM-generated research directions, trajectory-level reflection, mutation and crossover, and semantic-consistency checks. Each run begins from 10 research directions and proceeds for at most 11 evolutionary rounds. Crossover combines two parent trajectories and proposes 10 candidates. The official expression language extends AlphaAgent with conditional, comparison, skewness, and kurtosis operators, so QuantaAlpha expressions are evaluated with a separate grammar.

We preserve the released planning, proposal, mutation, crossover, quality-control, and feedback components. The qlib/LGBM runner is replaced through the official injection point by a FactorBench runner that evaluates the current composite on validation IC, RankIC, ICIR, RankICIR, and gross top-minus-bottom-quintile next-day performance. These validation results drive trajectory selection and feedback; the test segment remains unavailable to the search process. The complete factor library and the mutation/crossover lineage are retained with each run.

\paragraph{Method selection and exclusions.}
The reported comparison includes methods for which an official implementation can be connected to the shared data and execution contracts at a meaningful search budget. We additionally implemented Alpha Jungle \citep{shi2026navigating} and CogAlpha \citep{liu2026cognitive} from their papers, but exclude them from the reported experiments because neither provides official source code and their paper-scale configurations require orders of magnitude more LLM calls than can be reproduced under the common endpoint. Reporting substantially reduced versions would confound method quality with an arbitrary reduction in search budget. Their implementations are therefore retained as supplementary artifacts but do not contribute to the benchmark results.

\section{Factor-Level Evaluation}
\label{app:factor_evaluation}
This section supplements the factor-level results in the main paper with run-level reliability diagnostics, alternative predictive metrics, paired temporal comparisons, and complete exposure-neutralization specifications. All analyses use the same delivered factors, temporal splits, and training-period signal orientation as the main evaluation.

\subsection{Factor Validity}
\label{app:factor_validity}

The aggregate results in Table~\ref{tab:resource_validity} describe the average resource footprint and total output of each method. A \emph{delivered} factor is any output returned by a mining system. An \emph{executable} factor can be parsed or compiled and produces a date-by-asset signal under the shared execution contract. A \emph{scorable} factor additionally yields a finite IC on at least 100 dates in each of the training, validation and testing period. Requiring sufficient coverage in all three periods defines a common factor cohort for paired temporal comparisons: the same factors are evaluated in every split, preventing apparent changes in aggregate performance from being driven by factors entering or leaving the sample because of missing, constant, or numerically degenerate outputs.

\paragraph{Validity funnel and failure modes.}
Across the nine automated methods, FactorBench records 4,914 delivered factor instances. Of these, 4,838 (98.5\%) execute successfully and 4,698 (95.6\%) satisfy the common scorability requirement. Alpha101 contributes 71 unique applicable formulas; all execute successfully, and 70 are scorable across all five markets. Table~\ref{tab:factor_attrition} reports these unique formulas rather than counting their market-specific evaluations separately.

We divide attrition after delivery into three mutually exclusive categories. An \emph{execution failure} is a factor that fails during parsing, compilation, or execution and never produces a usable signal. An \emph{IC-incomplete} factor executes but fails to produce a finite IC in at least one temporal split, most commonly because its cross-section is constant or numerically degenerate. An \emph{insufficient-history} factor produces an IC in all three splits but has fewer than 100 valid dates in at least one split. The remaining factors form the scorable sample used by the paired temporal analysis. 
The failure composition differs across method families. All 76 execution failures occur in the LLM-based pipelines: 19 for R\&D-Agent, 24 for AlphaAgent, and 33 for QuantaAlpha. By contrast, the principal source of attrition for several symbolic search methods is numerical degeneracy after successful execution. AlphaCFG, AlphaForge, and AlphaSAGE produce 11, 25, and 47 IC-incomplete factors, respectively. QuantaAlpha exhibits both failure types, with 33 execution failures and 27 executable factors that do not yield an IC in every split.
\begin{table}[t]
    \centering
    \small
    \setlength{\tabcolsep}{4.5pt}
    \caption{\textbf{Delivered-factor attrition by failure mode.}
    Delivered counts factors returned across automated mining runs; for
    Alpha101, it counts the 71 unique applicable formulas. Failure categories
    are mutually exclusive. IC-incomplete denotes executable factors without a
    finite IC in every split, while \(<100\) dates denotes insufficient temporal
    coverage. An Alpha101 formula is scorable only if it satisfies the criterion
    in all five markets. Percentages are relative to Delivered.}
    \label{tab:factor_attrition}
    \begin{tabular}{lrrrrr}
        \toprule
        Method
        & Delivered
        & \shortstack{Execution\\failure}
        & \shortstack{IC-\\incomplete}
        & \shortstack{\(<100\)\\dates}
        & \shortstack{Scorable\\count (\%)} \\
        \midrule
        Alpha101
        & 71 & 0 & 0 & 1 & 70 (98.6\%) \\
        \midrule
        GP
        & 300 & 0 & 0 & 0 & 300 (100.0\%) \\
        AlphaGen
        & 150 & 0 & 0 & 1 & 149 (99.3\%) \\
        AlphaCFG
        & 150 & 0 & 11 & 0 & 139 (92.7\%) \\
        AlphaQCM
        & 143 & 0 & 0 & 1 & 142 (99.3\%) \\
        AlphaForge
        & 1{,}572 & 0 & 25 & 12 & 1{,}535 (97.6\%) \\
        AlphaSAGE
        & 750 & 0 & 47 & 7 & 696 (92.8\%) \\
        R\&D-Agent
        & 107 & 19 & 1 & 0 & 87 (81.3\%) \\
        AlphaAgent
        & 123 & 24 & 0 & 0 & 99 (80.5\%) \\
        QuantaAlpha
        & 1{,}619 & 33 & 27 & 8 & 1{,}551 (95.8\%) \\
        \midrule
        Total
        & 4{,}985 & 76 & 111 & 30 & 4{,}768 (95.6\%) \\
        \bottomrule
    \end{tabular}
\end{table}

These results show why executable and scorable outputs should be treated as properties of a mining system rather than as preprocessing details. Removing failed factors before reporting results would hide the execution risk of agentic pipelines, while assigning zero IC to constant signals would conflate an undefined prediction statistic with evidence of no predictiveness. We count all delivered factors when assessing validity, but evaluate predictive performance only for factors that satisfy the common scorability requirement in the subsequent factor-level analysis.

\subsection{Factor Generalization}
\label{app:factor_generalization}

The main text evaluates temporal generalization using Pearson IC. We examine whether its conclusions depend on this metric by considering RankIC and ICIR. For each factor and temporal split, RankIC is the mean daily cross-sectional Spearman correlation between the factor signal and future returns. ICIR is the mean daily Pearson IC divided by its time-series standard deviation and therefore describes the temporal consistency of predictiveness. The factor sign is determined from the training period and then held fixed in validation and test. All results use factors satisfying the common requirement of at least 100 scorable dates in every split.

Figure~\ref{fig:rankic_generalization} presents the RankIC counterpart to the IC results in the main text. GP, reinforcement-learning, and generative methods often achieve substantially higher training RankIC than validation or test RankIC, particularly in the HSI, CSI~300, and FTSE~100 markets. Thus, the observed degradation is not driven solely by a few extreme return observations that affect Pearson correlation. Some methods retain stronger rank-based predictiveness in individual markets, especially the S\&P~100 and Nikkei~225, but no method maintains a consistent advantage across all five markets. LLM-based methods generally begin with lower training RankIC and consequently exhibit smaller absolute changes, without attaining systematically higher held-out RankIC.
\begin{figure*}[t]
    \centering
    \includegraphics[width=\textwidth]{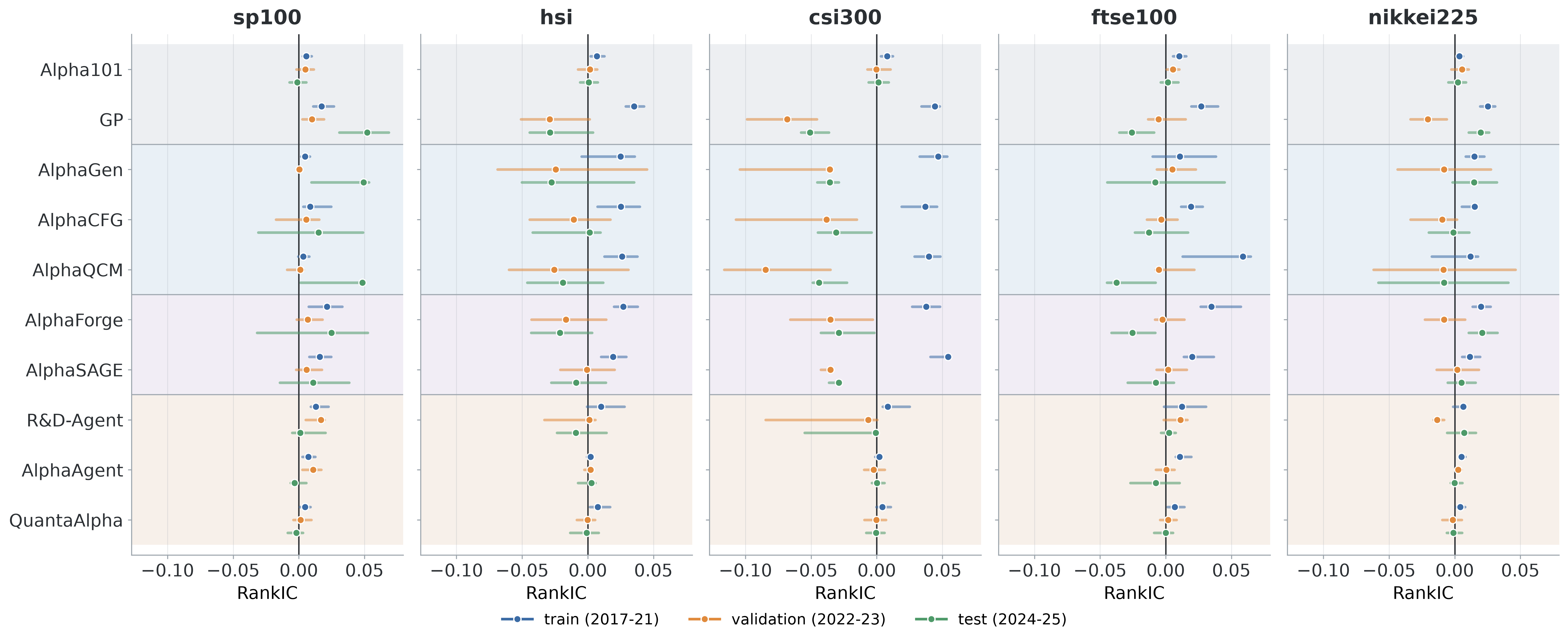}
    \caption{\textbf{RankIC across temporal splits.}
    Each point reports the median factor-level RankIC for a method and market; horizontal bars show the interquartile range across factors. Factor signs are fixed using the training period and applied unchanged to validation and test.}
    \label{fig:rankic_generalization}
\end{figure*}

Figure~\ref{fig:icir_generalization} examines the stability of daily IC rather than only its average level. The separation between training and subsequent periods remains visible for many GP, reinforcement-learning, and generative methods. Large training IC values are therefore frequently accompanied by training-specific temporal consistency that also weakens outside the discovery period. The magnitude varies considerably across markets, and several methods retain positive test ICIR in individual cases. ICIR should nevertheless be interpreted descriptively: daily IC observations are serially dependent because the 20-day return targets overlap, and the interquartile ranges describe dispersion across factors rather than statistical uncertainty.

\begin{figure*}[t]
    \centering
    \includegraphics[width=\textwidth]{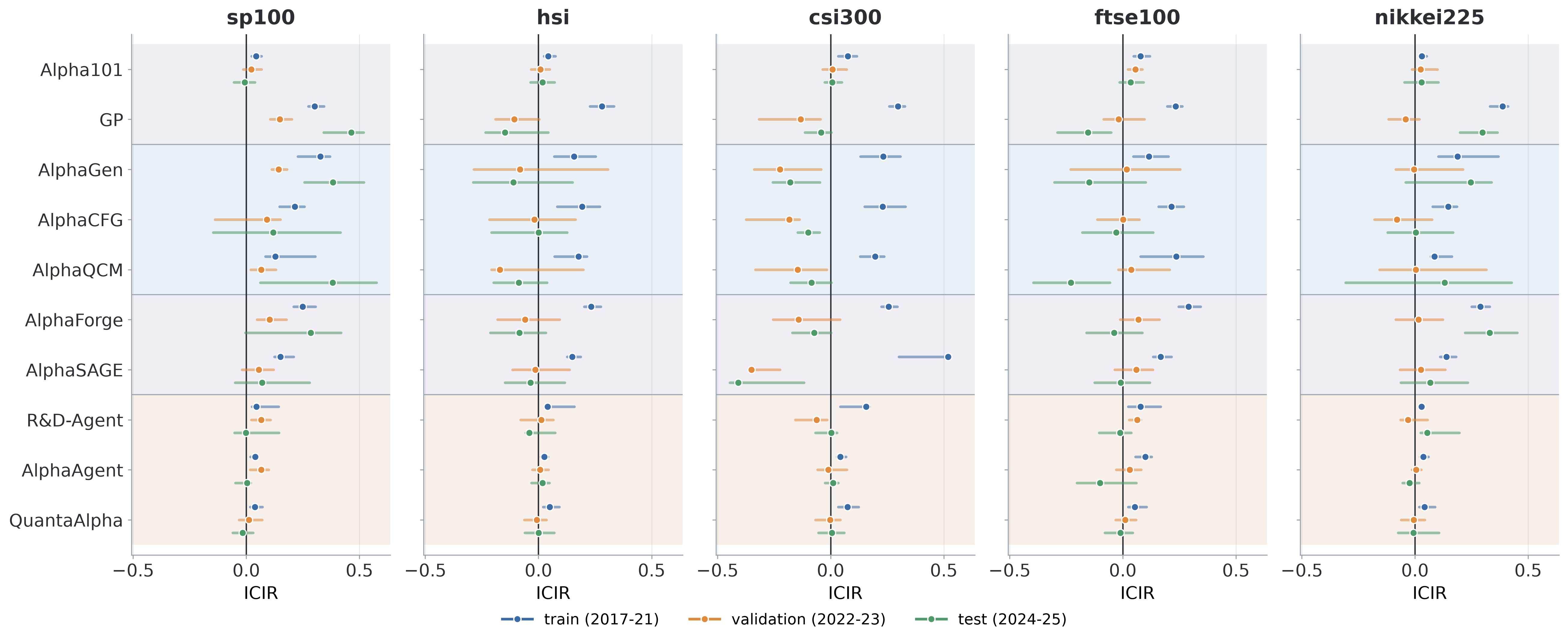}
    \caption{\textbf{ICIR across temporal splits.} Points report median factor-level ICIR and horizontal bars report the interquartile range across factors. ICIR is the mean daily Pearson IC divided by its time-series standard deviation. The same training-fixed signs, scorability requirement, method ordering, and temporal partitions used in Figure~\ref{fig:rankic_generalization} are applied.}
    \label{fig:icir_generalization}
\end{figure*}

The split-level summaries compare distributions and do not directly measure how much the same factor deteriorates. We therefore calculate paired degradation for metric \(M\) as
\[
    D_{m,k}(M)
    =
    \operatorname{median}_{f \in \mathcal{F}_{m,k}}
    \left(
        M_{f,\mathrm{train}}-M_{f,\mathrm{test}}
    \right),
\]
where \(\mathcal{F}_{m,k}\) is the common scorable factor set for method \(m\)
and market \(k\). The difference is computed within each factor before taking the median. Positive values indicate deterioration, while negative values indicate that the test-period value exceeds the training-period value. Figure~\ref{fig:paired_degradation} shows this paired statistic for IC, RankIC, and ICIR. Positive degradation is widespread across all three measures, especially among GP, reinforcement-learning, and generative methods in the HSI, CSI~300, and FTSE~100 markets. The exceptions in the S\&P~100 and Nikkei~225 show that degradation is not universal at the method--market level. LLM-based methods generally display smaller absolute degradation, but this primarily reflects their lower training-period performance rather than a consistent held-out advantage. The agreement between IC and RankIC shows that the main conclusion is robust to using either linear association or cross-sectional ordering, while the ICIR panel shows that temporal consistency often deteriorates alongside average predictiveness.
\begin{figure*}[t]
    \centering

    \begin{subfigure}[t]{0.325\textwidth}
        \centering
        \includegraphics[width=\linewidth]{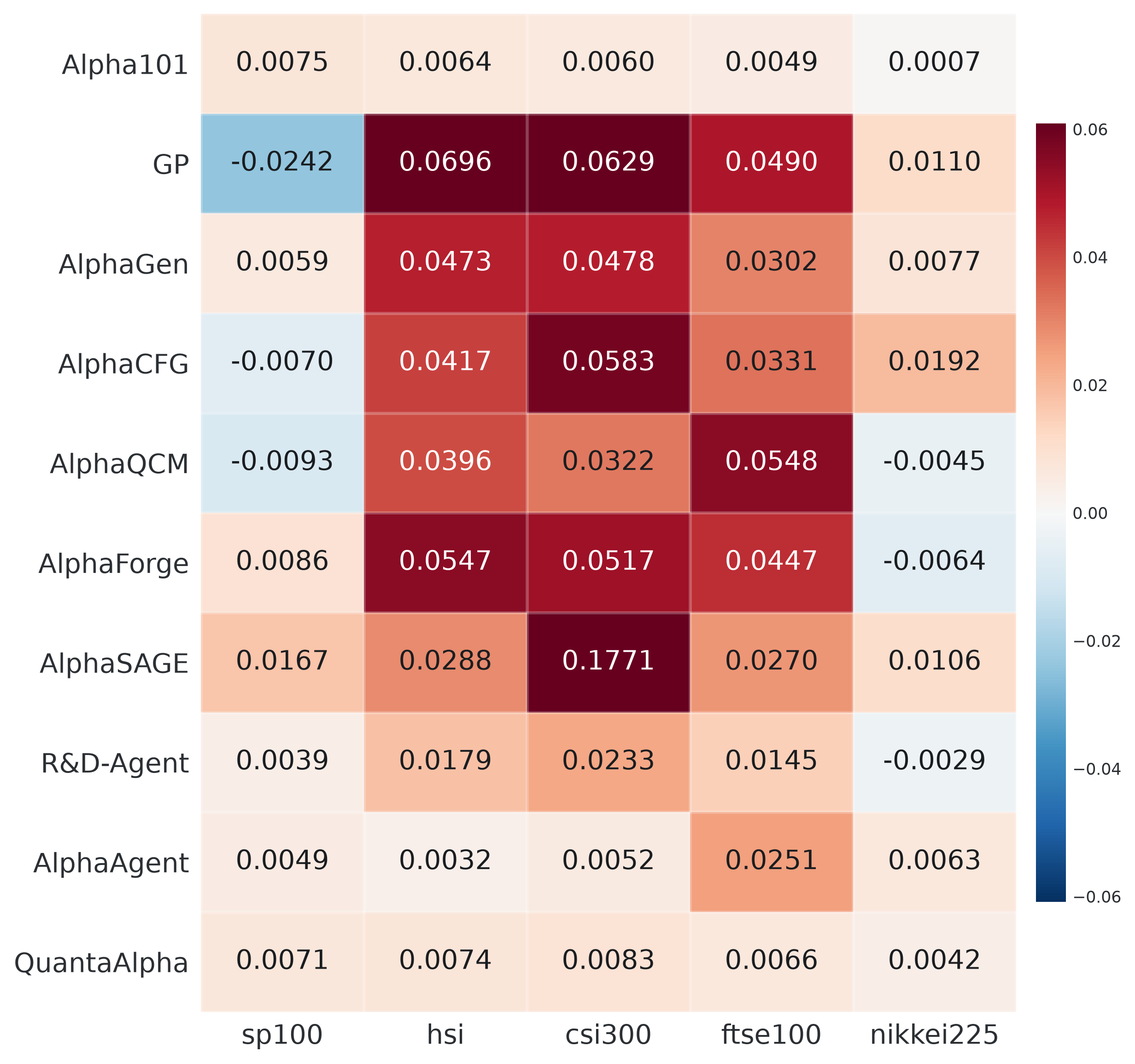}
        \caption{IC}
    \end{subfigure}
    \hfill
    \begin{subfigure}[t]{0.325\textwidth}
        \centering
        \includegraphics[width=\linewidth]{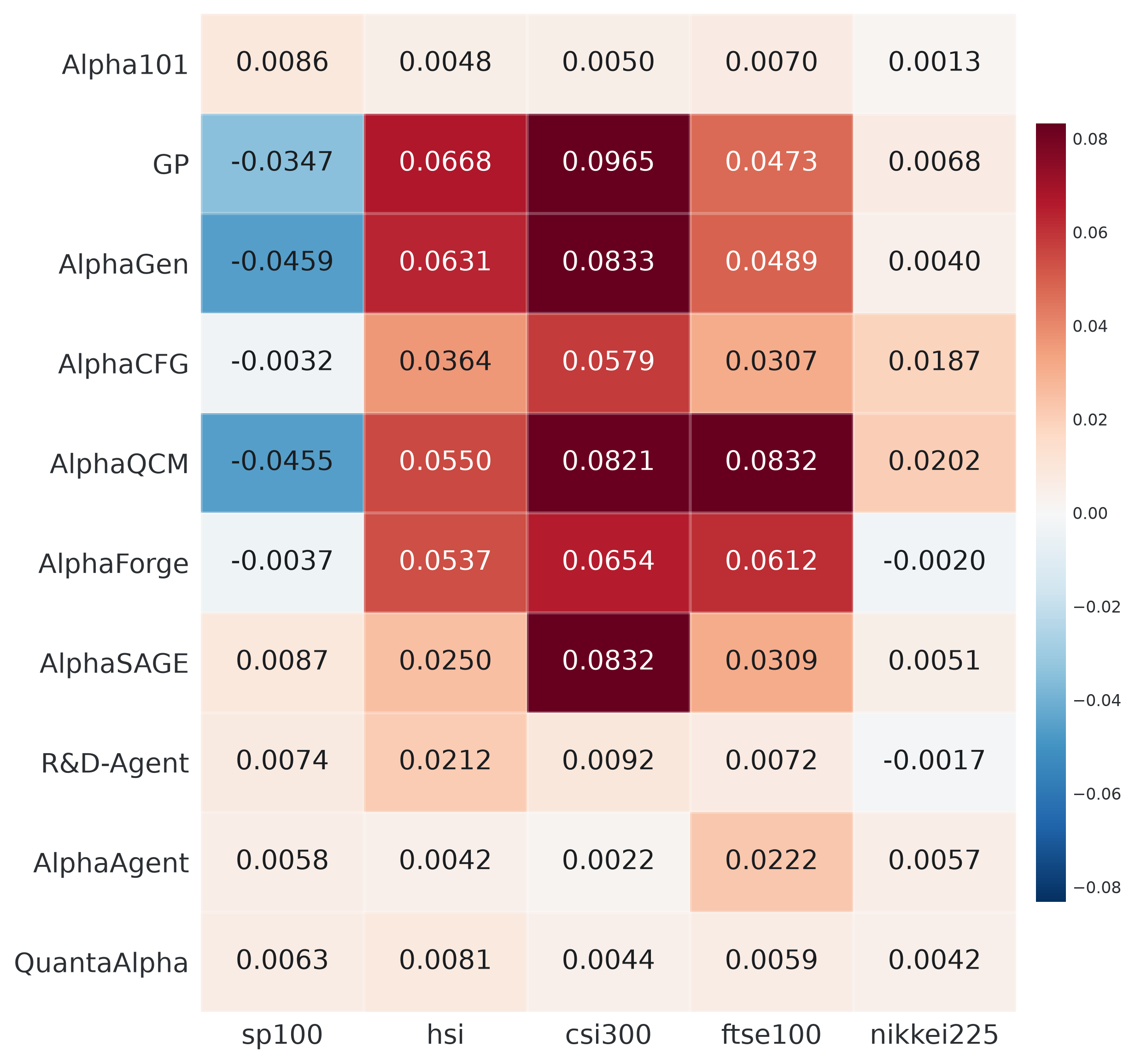}
        \caption{RankIC}
    \end{subfigure}
    \hfill
    \begin{subfigure}[t]{0.325\textwidth}
        \centering
        \includegraphics[width=\linewidth]{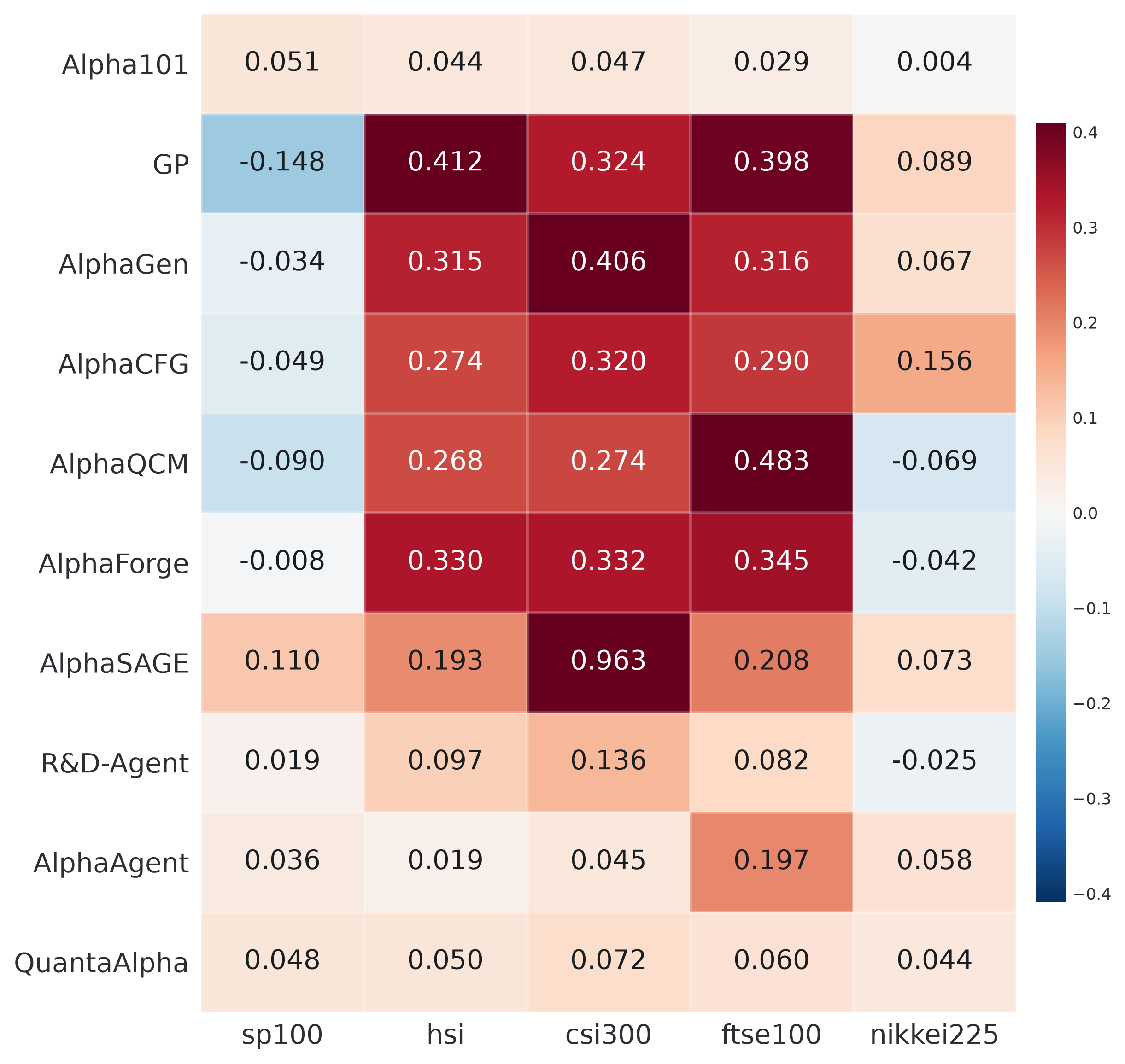}
        \caption{ICIR}
    \end{subfigure}

    \caption{\textbf{Paired train-to-test degradation.}
    Each cell reports the median within-factor difference \(M_{\mathrm{train}}-M_{\mathrm{test}}\) for a method and market. Positive values indicate deterioration and negative values indicate improvement. Differences are formed within factors before aggregation. Colors use metric-specific symmetric scales, so color intensity should be compared within each panel; the printed values provide the exact magnitudes.}
    \label{fig:paired_degradation}
\end{figure*}

\subsection{Factor Neutralization by Styles}
\label{app:factor_neutralization}

The main text asks whether factor predictiveness remains after controlling for measured risk and style exposures. We construct five exposures using only information available: market beta \citep{sharpe1964capital}, volatility \citep{ang2006cross}, momentum \citep{jegadeesh1993returns}, short-term reversal \citep{jegadeesh1990evidence}, and liquidity \citep{chordia2001trading}. Market beta is estimated over the preceding 252 trading days relative to the equal-weight market return; volatility is the standard deviation of daily returns over 60 days; momentum is the cumulative return from \(t-252\) to \(t-21\); short-term reversal is the negative cumulative return over the preceding 21 days; and liquidity is the logarithm of mean daily dollar volume over 20 days. Each exposure is winsorized at the first and ninety-ninth cross-sectional percentiles and standardized independently on each date.

For every factor and date, we estimate the cross-sectional regression
\[
    f_{i,t}
    =
    \alpha_t
    +
    \boldsymbol{\beta}_t^{\top}\mathbf{s}_{i,t}
    +
    \varepsilon_{i,t},
\]
where \(f_{i,t}\) is the training-sign-adjusted factor value and \(\mathbf{s}_{i,t}\) contains the five measured exposures. The regression includes an intercept and requires at least 30 eligible stocks. The residual \(\varepsilon_{i,t}\) defines the neutralized factor. We leave the return target unchanged because the question is whether the component of the factor that is orthogonal to the measured exposures continues to predict returns. Neutralizing both the factor and the target would instead estimate a partial correlation and answer a different question.

Raw and neutralized IC are computed on exactly the same date--asset observations. This matched-sample construction prevents differences in data availability, including the 252-day beta warm-up period, from being attributed to neutralization. As elsewhere in FactorBench, a result is retained only when at least 100 dates can be scored in every temporal split. The change in predictiveness is defined as
\[
    \Delta IC
    =
    IC_{\mathrm{raw}}-IC_{\mathrm{neutralized}},
\]
so a positive value indicates that predictive performance declines after removing the measured exposures. This change should not be interpreted as the percentage of alpha explained: when raw IC is already zero or negative, a further decline does not represent the removal of a profitable signal.

The adjusted coefficient of determination provides a complementary measure of exposure overlap. For a daily regression with \(n_t\) eligible stocks and
\(p=5\) exposures, it is
\[
    \overline{R}^{\,2}_t
    =
    1-(1-R_t^2)\frac{n_t-1}{n_t-p-1}.
\]
Adjusted \(R^2\) penalizes the explanatory power obtained mechanically from
including several regressors. The main-text heatmaps average this quantity over
dates, take the median across factors within each run, and then average runs
equally. Adjusted \(R^2\) can be negative when the controls explain less
variation than an intercept-only model; it should therefore be interpreted as
an overlap diagnostic rather than a bounded measure of explained alpha.

Figure~\ref{fig:neutralized_ic_valid} extends the training-period comparison in
the main text to validation. Most method--market points lie below the
\(45^\circ\) line or move toward zero, indicating that part of their validation
predictiveness is associated with the measured exposures. The amount of
attenuation varies across both methods and markets, and neutralization
occasionally increases IC. In several markets, raw validation IC is already
close to zero or negative, so these cases provide little evidence of profitable
predictiveness either before or after neutralization.

\begin{figure*}[t]
    \centering
    \includegraphics[width=\textwidth]{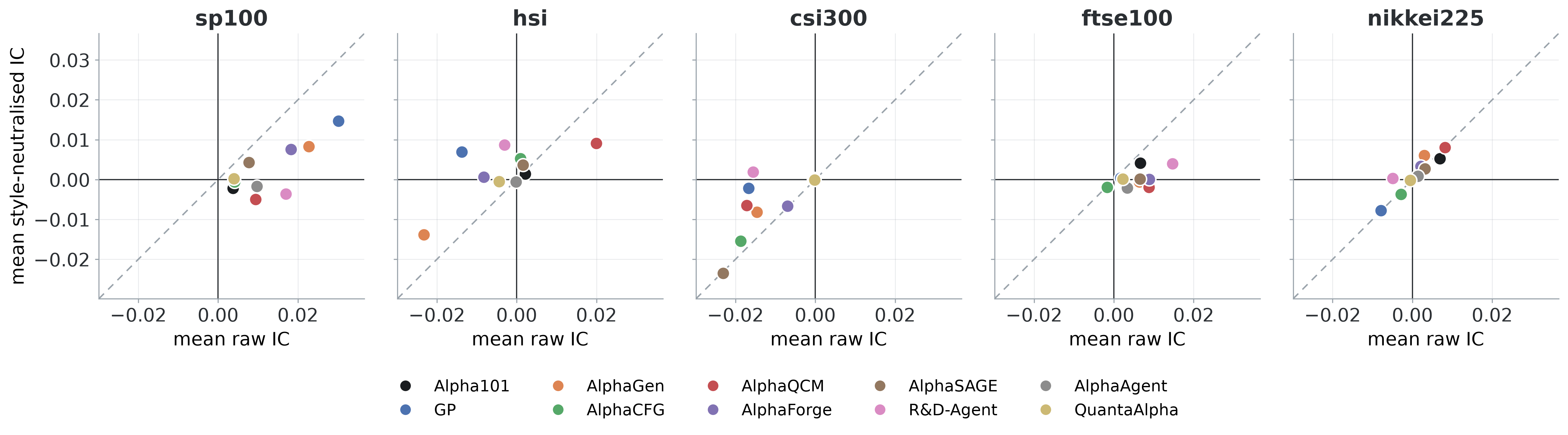}
    \caption{\textbf{Raw and style-neutralized IC in the validation period.}
    Each point represents one method in one market. The horizontal coordinate
    is mean raw IC and the vertical coordinate is mean IC after neutralizing
    factors against market beta, volatility, momentum, short-term reversal, and
    liquidity. Factor-level values are averaged within each run and then
    averaged equally across runs. Raw and neutralized IC use matched
    observations and training-fixed factor signs. Points below the
    \(45^\circ\) line lose IC after neutralization.}
    \label{fig:neutralized_ic_valid}
\end{figure*}

Figure~\ref{fig:neutralized_ic_test} shows that the effect of neutralization persists in the held-out test period, although its magnitude differs across markets. GP and R\&D-Agent exhibit comparatively large IC reductions in several markets, consistent with their higher adjusted-\(R^2\) overlap. Several other methods retain positive residual IC in individual markets, but none does so consistently across all five. Together with the stable adjusted-\(R^2\) patterns across temporal splits, these results indicate that overlap with the measured exposures often persists beyond the discovery period.
The points exhibit an approximately linear pattern because neutralization removes the exposure-related component while retaining the remainder of the same factor. Raw and neutralized signals consequently preserve similar cross-sectional rankings, so factors with higher raw IC also tend to have higher residual IC. The relevant evidence of exposure dependence is therefore the downward displacement from the \(45^\circ\) line, which measures the loss of predictiveness, together with adjusted \(R^2\), which measures how much factor variation is reproduced by the exposures.

\begin{figure*}[t]
    \centering
    \includegraphics[width=\textwidth]{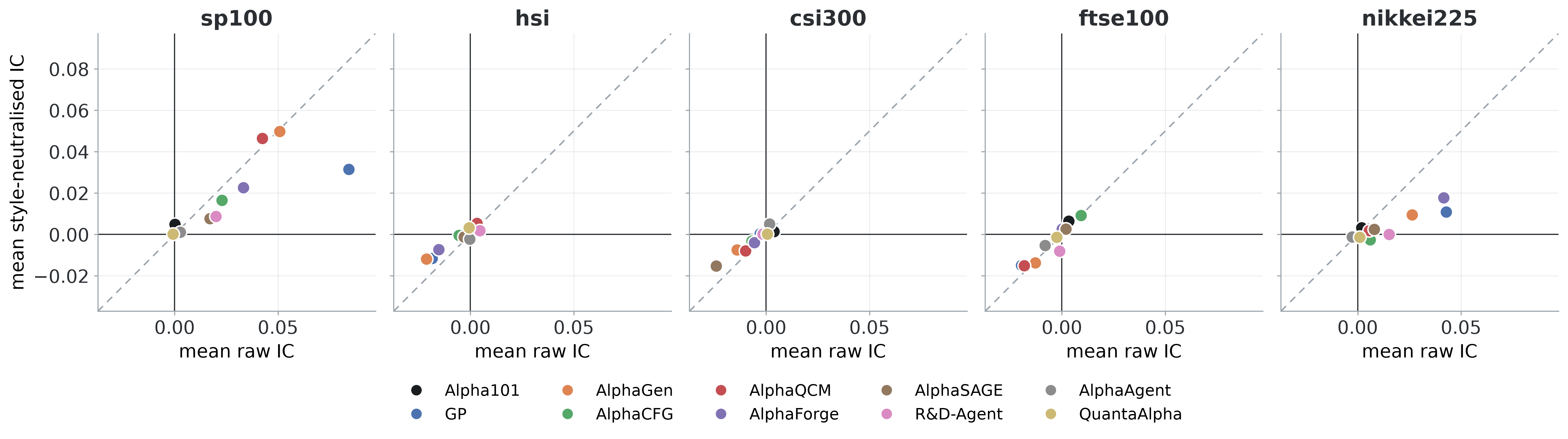}
    \caption{\textbf{Raw and style-neutralized IC in the held-out test period.} The construction follows Figure~\ref{fig:neutralized_ic_valid}. Points below the \(45^\circ\) line indicate lower predictive performance after removing the five measured exposures. Values below the horizontal zero line indicate that the residual factor is negatively associated with future returns under its training-fixed sign.}
    \label{fig:neutralized_ic_test}
\end{figure*}

These controls cover prominent price- and volume-derived exposures but do not constitute an exhaustive risk model. A positive neutralized IC indicates predictive information beyond this particular control set; it does not by itself establish exposure-free alpha.

\section{Pool-Level Evaluation}
\label{app:pool_evaluation}

The main text evaluates pool distinctness using median absolute Spearman correlation during the training period. This appendix provides the full construction of the similarity measures and examines whether the resulting patterns persist across temporal splits and correlation definitions.

\paragraph{Construction and aggregation.}
For factors \(f_i\) and \(f_j\), we first compute their cross-sectional correlation on each eligible date and then average it over the split:
\begin{equation}
    C_{ij}
    =
    \frac{1}{|\mathcal{T}_{ij}|}
    \sum_{t\in\mathcal{T}_{ij}}
    \operatorname{corr}
    \left(
        f_i(t,\cdot), f_j(t,\cdot)
    \right),
    \label{eq:factor_similarity}
\end{equation}
where \(\mathcal{T}_{ij}\) contains dates on which both factors are available for at least 20 common stocks. Constant cross-sections are treated as missing rather than assigned zero correlation, and factors available on fewer than half of the dates in a split are excluded. We report \(\lvert C_{ij}\rvert\), so that two factors carrying the same information with opposite signs are treated as similar.

Each figure summarizes this measure in three ways. \emph{Within method} reports, for each market, the median pairwise similarity among factors delivered by the same run, followed by an equal-weight average across seeds. \emph{Between methods} computes the median similarity over every factor pair in two delivered pools and then averages these block medians across eligible run pairs and markets. This construction produces a symmetric matrix and reduces sensitivity to differences in pool size. \emph{To Alpha101} applies the same block-median calculation between each delivered pool and the 71 applicable Alpha101 formulas, reported separately by market. Self-correlations are excluded throughout.

\paragraph{Temporal persistence of rank similarity.}
Figures~\ref{fig:pool_spearman_valid} and \ref{fig:pool_spearman_test} extend the training-period Spearman analysis in the main text to validation and test. The principal patterns remain stable over time. AlphaGen and AlphaQCM exhibit consistently high within-method similarity, while GP also produces relatively redundant pools in several markets. AlphaForge, AlphaCFG, and most LLM-based pools are generally more internally distinct, although the pattern remains market dependent. Thus, nominal pool size does not necessarily represent the breadth of information contained in the pool.

The between-method panels also preserve a clear family structure. Earlier formula-based methods tend to resemble one another, with particularly high similarity between AlphaGen and AlphaQCM. Similarities involving the LLM-based methods are generally smaller and are most pronounced within the LLM group. Their overlap with Alpha101 is nevertheless higher than that of most formula-based methods in several markets, especially for R\&D-Agent and AlphaAgent. These relationships change little from training to validation and testing, indicating that the observed pool structure is not confined to the discovery period.
\begin{figure*}[t]
    \centering
    \includegraphics[width=\textwidth]
        {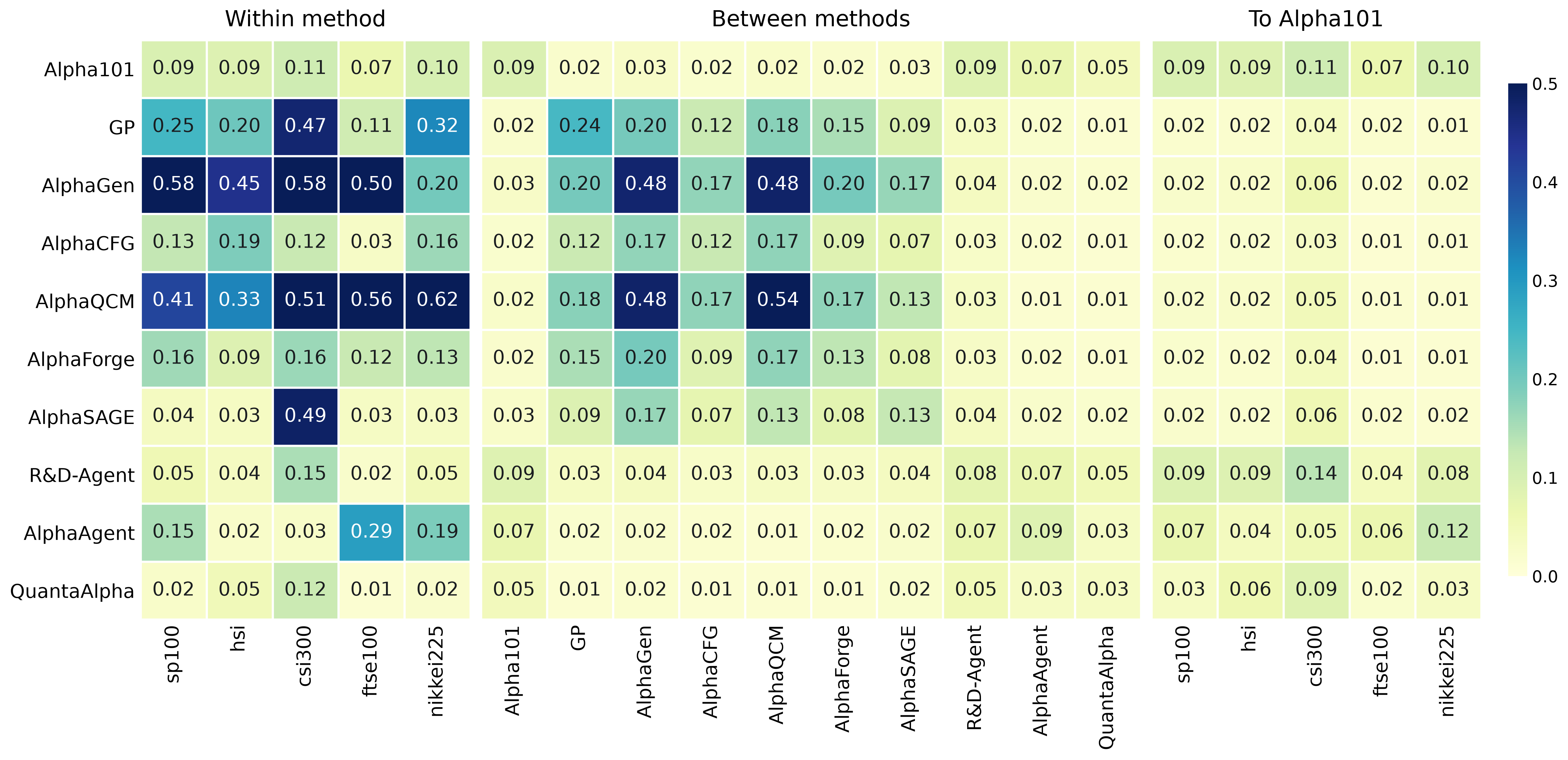}
    \caption{\textbf{Pool similarity under Spearman correlation in the
    validation period.} Each factor-pair correlation is computed cross-sectionally by date and averaged over time before taking its absolute value. The panels report median similarity within delivered pools, between methods, and to the applicable Alpha101 library. Within-method and Alpha101 comparisons are reported by market; between-method entries aggregate eligible run-pair comparisons across markets.}
    \label{fig:pool_spearman_valid}
\end{figure*}
\begin{figure*}[t]
    \centering
    \includegraphics[width=\textwidth]
        {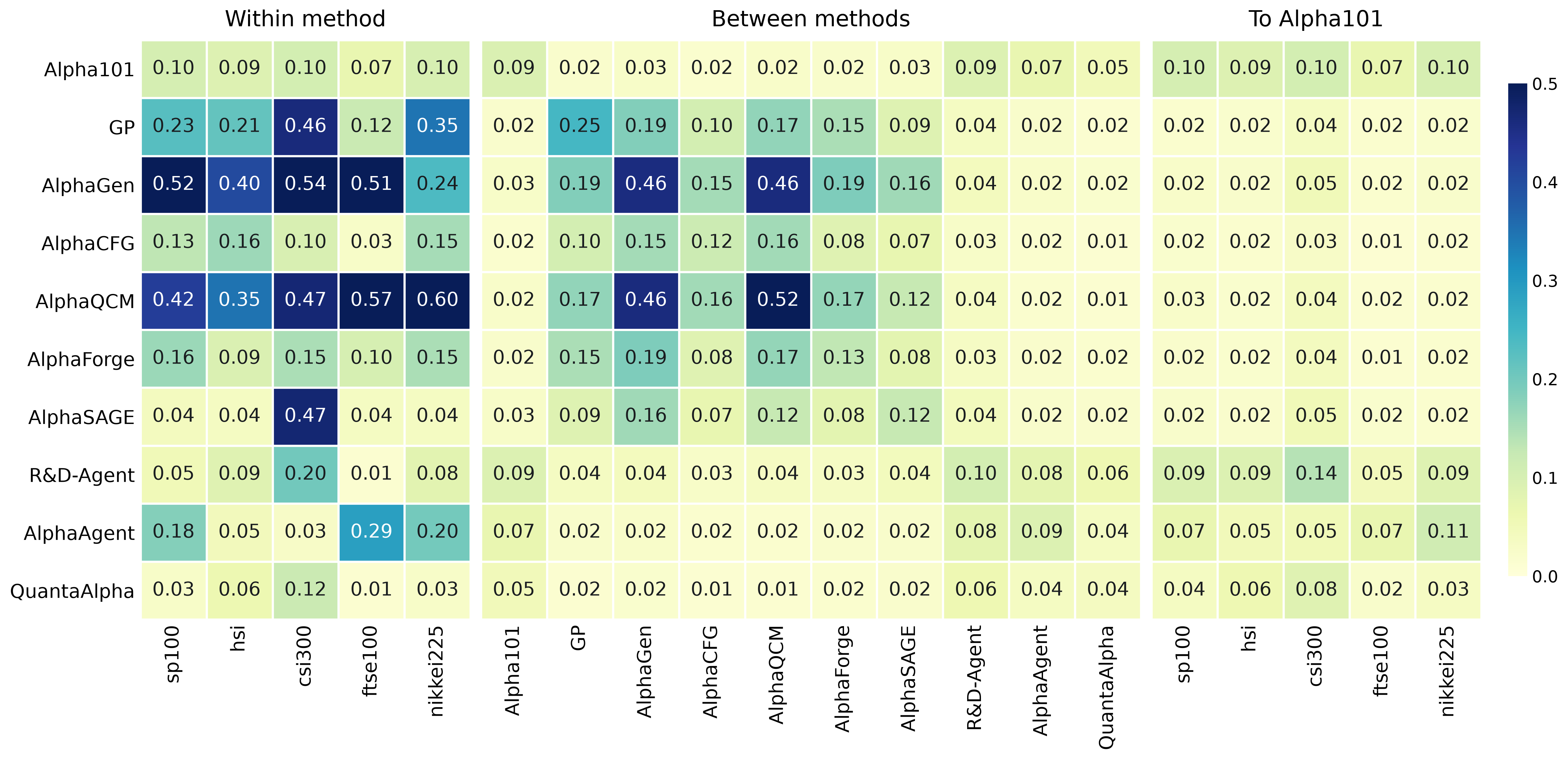}
    \caption{\textbf{Pool similarity under Spearman correlation in the
    held-out test period.}
    The construction follows Figure~\ref{fig:pool_spearman_valid}. The
    persistence of the within-method, between-method, and Alpha101 similarity
    patterns indicates that pool redundancy is a relatively stable property
    of the delivered signals.}
    \label{fig:pool_spearman_test}
\end{figure*}

\paragraph{Robustness to Pearson correlation.}
Spearman correlation measures similarity in cross-sectional rankings, which matches the use of factors for stock selection. Figures \ref{fig:pool_pearson_train}--\ref{fig:pool_pearson_test} repeat the analysis using Pearson correlation, which additionally requires similarity in the relative magnitudes of factor values.

The qualitative conclusions remain unchanged. AlphaGen, AlphaQCM, and GP continue to exhibit comparatively high internal redundancy, and the between-method structure remains concentrated among the earlier formula-based methods. R\&D-Agent and AlphaAgent also retain comparatively greater similarity to Alpha101 than most automated formula-based methods. Pearson similarities are lower than Spearman similarities for several pools, particularly AlphaGen, AlphaQCM, and AlphaSAGE. Their factors therefore agree more strongly in the ordering of stocks than in the magnitudes assigned to those stocks. This distinction is relevant for portfolio construction: signals can induce similar rankings, and hence similar trades, even when their numerical values are not linearly proportional.
\begin{figure*}[t]
    \centering
    \includegraphics[width=\textwidth]
        {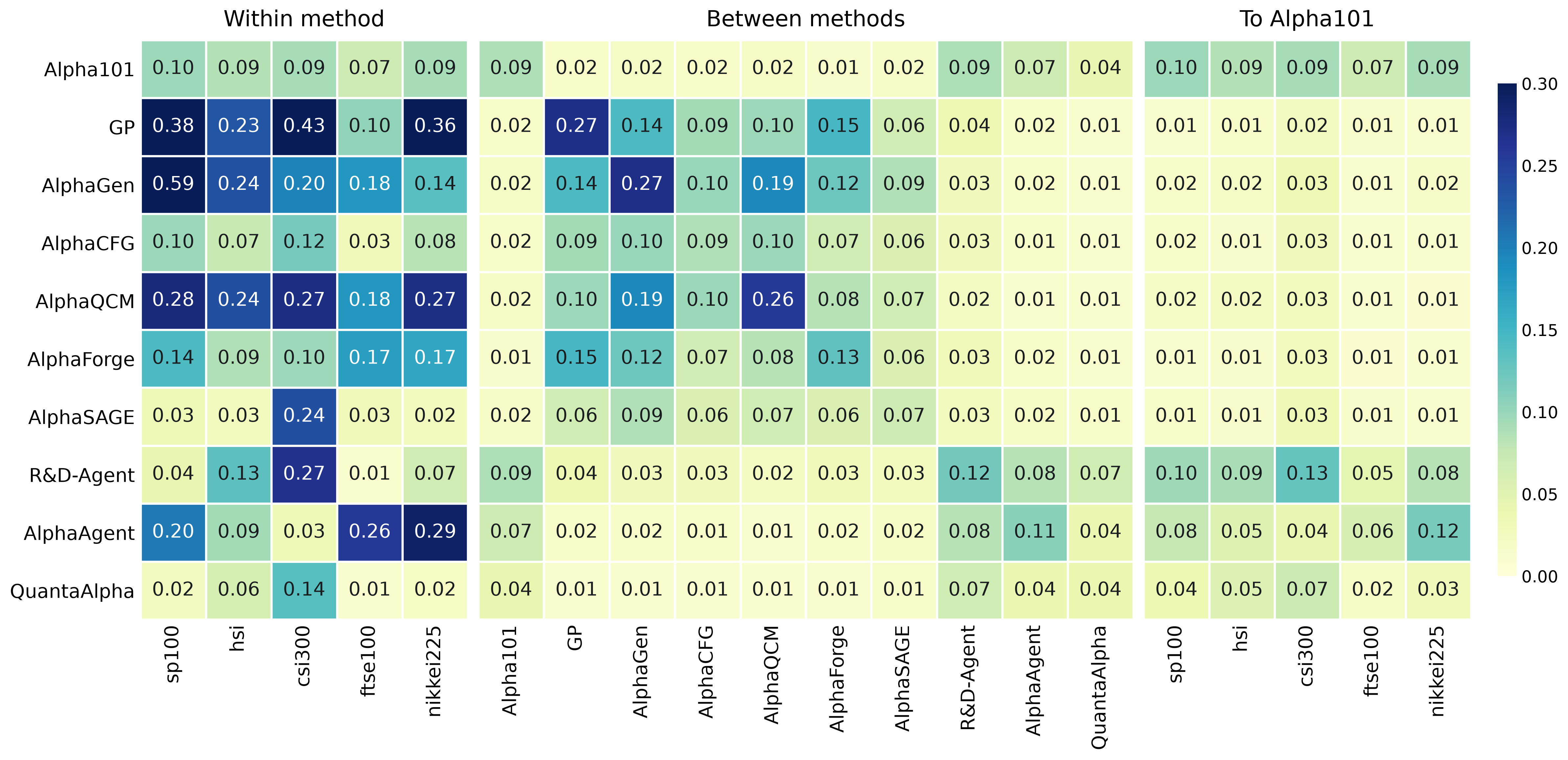}
    \caption{\textbf{Pool similarity under Pearson correlation in the
    training period.} The aggregation follows Equation~\ref{eq:factor_similarity}, replacing cross-sectional Spearman correlation with Pearson correlation. The three panels summarize within-method similarity, between-method similarity, and similarity to the applicable Alpha101 factors.}
    \label{fig:pool_pearson_train}
\end{figure*}
\begin{figure*}[t]
    \centering
    \includegraphics[width=\textwidth]
        {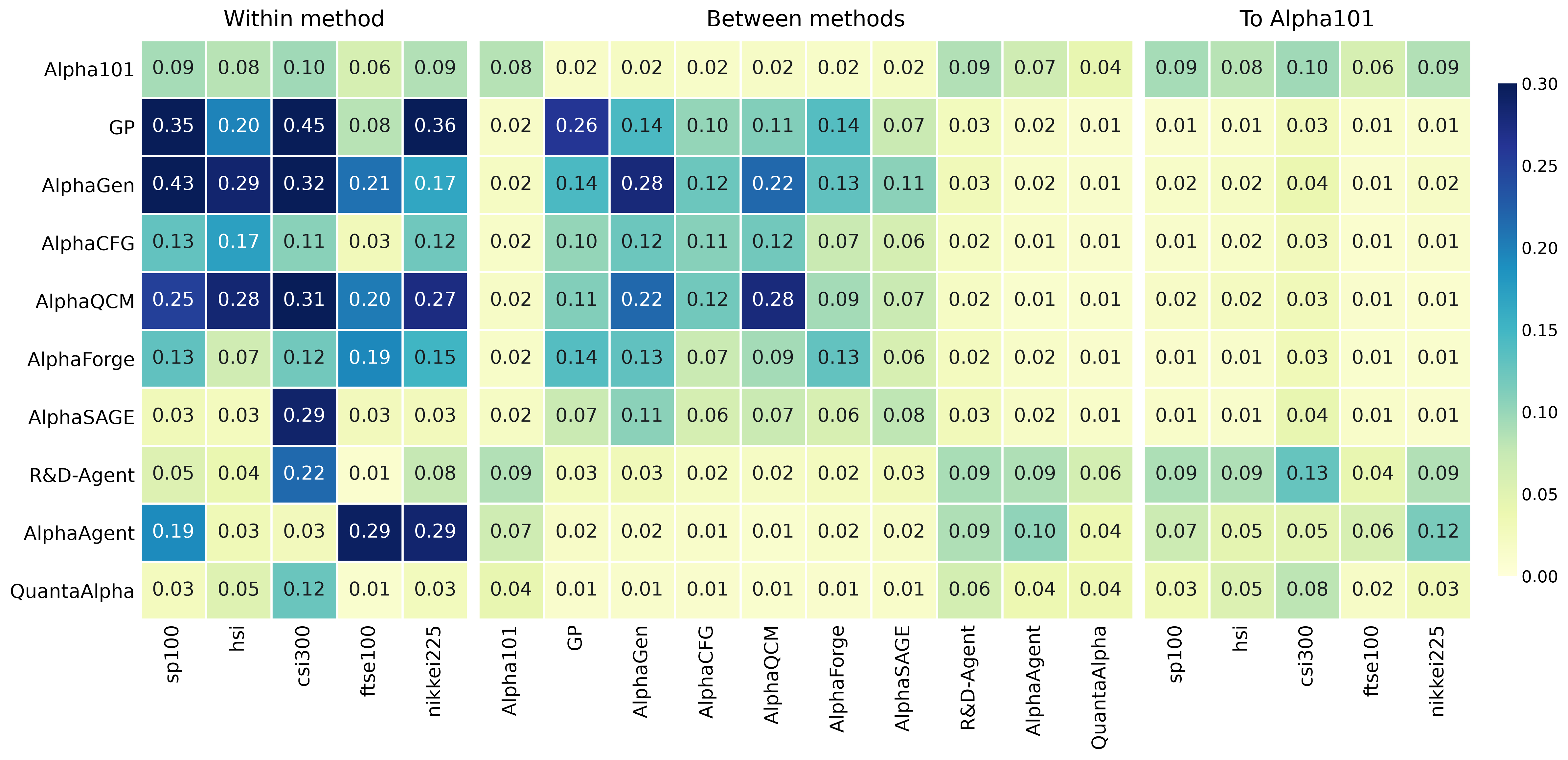}
    \caption{\textbf{Pool similarity under Pearson correlation in the validation period.} The broad method-level structure remains similar to that observed during training, although Pearson values are generally lower than their Spearman counterparts for pools whose factors preserve rankings more strongly than raw signal magnitudes.}
    \label{fig:pool_pearson_valid}
\end{figure*}
\begin{figure*}[t]
    \centering
    \includegraphics[width=\textwidth]
        {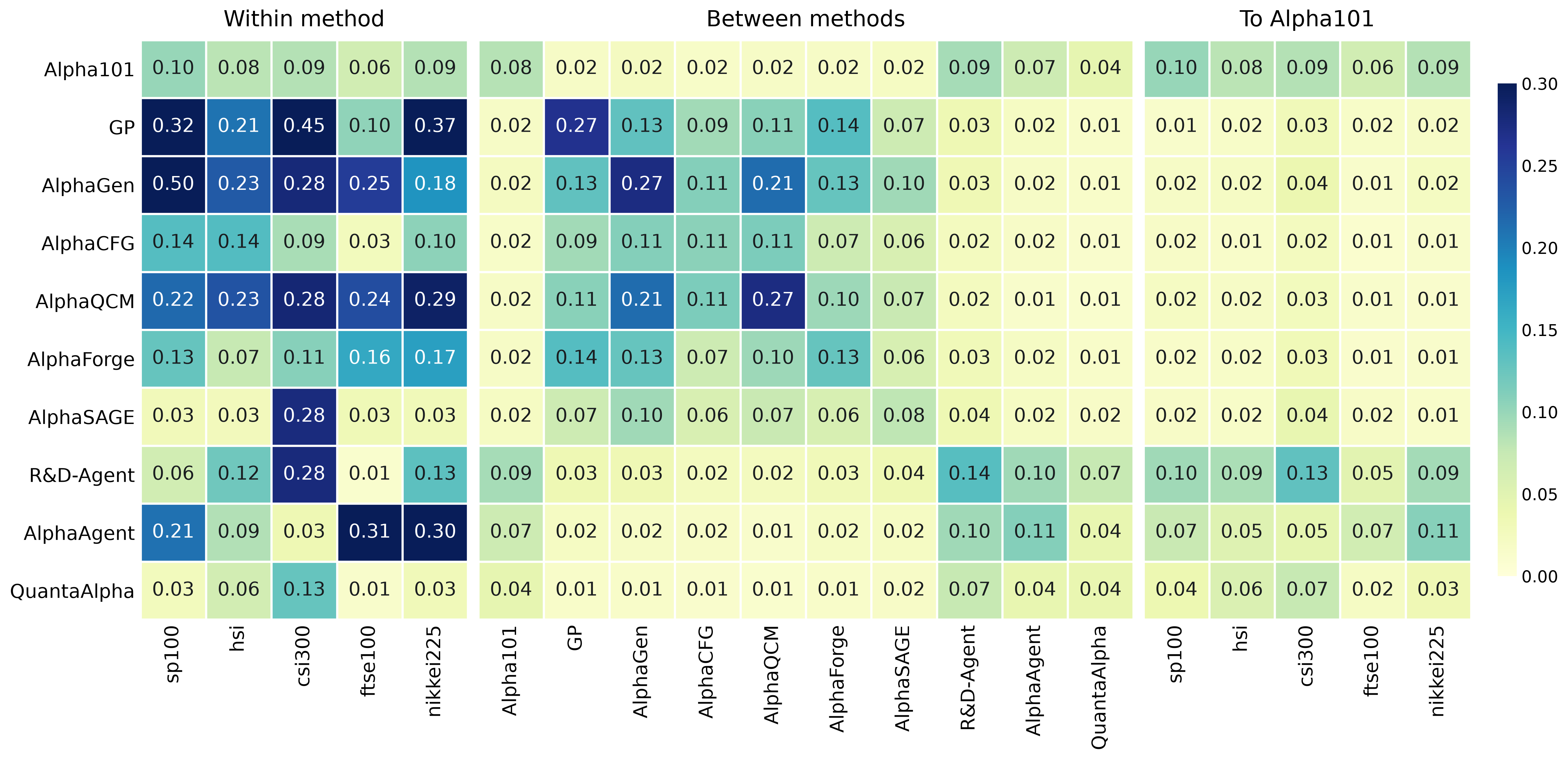}
    \caption{\textbf{Pool similarity under Pearson correlation in the held-out test period.} Within-method redundancy, the similarity among formula-based methods, and the comparatively greater Alpha101 overlap of several LLM-based methods remain visible under an alternative correlation definition.}
    \label{fig:pool_pearson_test}
\end{figure*}
Taken together, these results show that pool distinctness does not improve monotonically across mining paradigms. Some methods deliver many factors that occupy a comparatively narrow signal space, while other methods deliver smaller but more heterogeneous pools. The persistence of these patterns across temporal splits and correlation measures suggests that they primarily reflect the search spaces and inductive biases of the mining systems rather than temporary relationships in the training sample. Correlation nevertheless measures similarity in realized signals rather than common provenance: high similarity does not establish that one method reproduces another, and low similarity does not by itself imply independent predictive or economic value.

\section{Portfolio-Level Evaluation}
\label{app:portfolio_evaluation}

This section provides the disaggregated results and robustness analyses behind the portfolio-level comparison in the main text.
Before portfolio construction, we apply the same factor-selection procedure separately to each method--market--seed pool. We begin with factors that satisfy the common validity and coverage requirements. After fixing each factor's direction on the training split, we rank the eligible factors by validation IC. We then traverse this ranking and retain a candidate only if its absolute validation-period Spearman correlation with every previously selected factor is below \(0.8\). Candidates whose correlations cannot be estimated because of insufficient overlap are excluded. The default configuration imposes no limit on the number of retained factors and does not require validation IC to be positive; thus, the full selected pool contains all eligible factors that pass the redundancy constraint. No test-period predictive metric is used in this selection procedure.
Unless otherwise stated, the reported configuration uses the full selected pool, ridge combination, 20-trading-day rebalancing, and a transaction cost of 10 basis points per dollar traded.

Long-only portfolios invest equally in the top 20\% of stocks ranked by the composite signal. Long--short portfolios hold the top 20\% long and the bottom 20\% short, with \(+50\%\) and \(-50\%\) exposure at each rebalance. Long-only performance is interpreted relative to the equal-weight universe and the corresponding market index because the portfolio remains fully invested, whereas long--short performance provides a more direct assessment of cross-sectional ranking value after reducing broad market-direction exposure.

\subsection{Default Performance and Variation}
\label{app:portfolio_default}

Tables~\ref{tab:long_only_full_results} and
\ref{tab:long_short_full_results} report the complete results under the default configuration. Aggregation proceeds in two stages: seeds are first averaged within each market, after which the five market means receive equal weight. The first dispersion term is the average within-market standard deviation across random seeds, and the bracketed term is the standard deviation across market means. Alpha101 has no random seeds, so only its cross-market dispersion is reported.
\begin{table*}[t]
    \centering
    \scriptsize
    \setlength{\tabcolsep}{4pt}
    \caption{\textbf{Long-only performance under the default configuration.}
    Entries report the market-pooled mean, the typical within-market seed
    standard deviation after \(\pm\), and the cross-market standard deviation
    in brackets. CAGR and maximum drawdown are reported in percent. Turnover is
    annualized and reported as multiples of portfolio value.}
    \label{tab:long_only_full_results}
    \resizebox{\textwidth}{!}{
    \begin{tabular}{lcccc}
        \toprule
        Method
        & \shortstack{Net CAGR\\(\%)}
        & Net Sharpe
        & \shortstack{Maximum drawdown\\(\%)}
        & \shortstack{Annual turnover\\(\(\times\))} \\
        \midrule
        Alpha101
        & \(26.5\,[10.3]\)
        & \(1.241\,[0.327]\)
        & \(-19.3\,[4.5]\)
        & \(8.80\,[1.04]\) \\
        GP
        & \(30.4\pm3.0\,[14.3]\)
        & \(1.173\pm0.112\,[0.442]\)
        & \(-23.2\pm1.4\,[3.8]\)
        & \(6.32\pm0.72\,[0.80]\) \\
        AlphaGen
        & \(28.0\pm4.3\,[16.1]\)
        & \(1.127\pm0.130\,[0.511]\)
        & \(-23.2\pm1.4\,[5.3]\)
        & \(5.52\pm0.74\,[1.37]\) \\
        AlphaCFG
        & \(23.5\pm6.4\,[10.2]\)
        & \(1.048\pm0.242\,[0.388]\)
        & \(-21.9\pm2.6\,[3.8]\)
        & \(7.53\pm0.36\,[0.76]\) \\
        AlphaQCM
        & \(29.6\pm4.4\,[12.0]\)
        & \(1.313\pm0.251\,[0.494]\)
        & \(-20.4\pm2.3\,[4.2]\)
        & \(4.74\pm1.51\,[1.35]\) \\
        AlphaForge
        & \(28.2\pm4.3\,[10.7]\)
        & \(1.208\pm0.156\,[0.384]\)
        & \(-20.9\pm1.0\,[4.7]\)
        & \(7.68\pm0.36\,[0.87]\) \\
        AlphaSAGE
        & \(29.2\pm7.9\,[7.4]\)
        & \(1.244\pm0.205\,[0.319]\)
        & \(-21.0\pm2.0\,[3.0]\)
        & \(7.45\pm1.07\,[1.15]\) \\
        R\&D-Agent
        & \(31.2\pm3.9\,[10.7]\)
        & \(1.193\pm0.129\,[0.350]\)
        & \(-22.8\pm2.0\,[2.9]\)
        & \(7.89\pm0.85\,[1.09]\) \\
        AlphaAgent
        & \(22.4\pm6.6\,[8.2]\)
        & \(1.027\pm0.179\,[0.269]\)
        & \(-19.7\pm1.9\,[3.8]\)
        & \(9.82\pm0.55\,[0.51]\) \\
        QuantaAlpha
        & \(25.3\pm6.7\,[8.6]\)
        & \(1.106\pm0.215\,[0.292]\)
        & \(-20.8\pm1.7\,[4.8]\)
        & \(8.87\pm0.75\,[0.58]\) \\
        \bottomrule
    \end{tabular}}
\end{table*}
\begin{table*}[t]
    \centering
    \scriptsize
    \setlength{\tabcolsep}{4pt}
    \caption{\textbf{Long--short performance under the default configuration.}
    The aggregation and uncertainty terms follow
    Table~\ref{tab:long_only_full_results}. CAGR and maximum drawdown are
    reported in percent. Turnover is annualized and reported as multiples of
    portfolio value. The strategy includes transaction costs and the benchmark
    short-financing assumption.}
    \label{tab:long_short_full_results}
    \resizebox{\textwidth}{!}{
    \begin{tabular}{lcccc}
        \toprule
        Method
        & \shortstack{Net CAGR\\(\%)}
        & Net Sharpe
        & \shortstack{Maximum drawdown\\(\%)}
        & \shortstack{Annual turnover\\(\(\times\))} \\
        \midrule
        Alpha101
        & \(2.6\,[5.6]\)
        & \(0.319\,[0.789]\)
        & \(-6.8\,[1.3]\)
        & \(8.73\,[0.71]\) \\
        GP
        & \(3.7\pm1.4\,[8.4]\)
        & \(0.340\pm0.176\,[1.013]\)
        & \(-11.5\pm1.1\,[3.9]\)
        & \(6.37\pm0.66\,[0.83]\) \\
        AlphaGen
        & \(2.3\pm2.3\,[9.6]\)
        & \(0.297\pm0.271\,[0.989]\)
        & \(-12.2\pm2.2\,[5.0]\)
        & \(5.57\pm0.79\,[1.00]\) \\
        AlphaCFG
        & \(1.0\pm3.1\,[6.7]\)
        & \(0.169\pm0.421\,[0.901]\)
        & \(-10.4\pm2.4\,[4.3]\)
        & \(7.48\pm0.32\,[0.87]\) \\
        AlphaQCM
        & \(3.7\pm2.3\,[6.9]\)
        & \(0.668\pm0.388\,[1.032]\)
        & \(-8.5\pm3.0\,[3.6]\)
        & \(4.58\pm1.43\,[1.23]\) \\
        AlphaForge
        & \(2.9\pm2.3\,[6.5]\)
        & \(0.380\pm0.271\,[0.836]\)
        & \(-8.4\pm1.6\,[2.7]\)
        & \(7.62\pm0.43\,[0.99]\) \\
        AlphaSAGE
        & \(2.9\pm4.2\,[3.9]\)
        & \(0.394\pm0.579\,[0.486]\)
        & \(-8.4\pm1.7\,[2.8]\)
        & \(7.54\pm1.12\,[1.32]\) \\
        R\&D-Agent
        & \(6.5\pm2.2\,[7.5]\)
        & \(0.679\pm0.231\,[0.588]\)
        & \(-9.8\pm1.7\,[3.9]\)
        & \(7.84\pm0.86\,[1.12]\) \\
        AlphaAgent
        & \(0.3\pm3.5\,[4.4]\)
        & \(-0.087\pm0.432\,[0.562]\)
        & \(-9.6\pm1.9\,[1.5]\)
        & \(9.81\pm0.55\,[0.52]\) \\
        QuantaAlpha
        & \(2.7\pm4.5\,[5.1]\)
        & \(0.324\pm0.555\,[0.560]\)
        & \(-8.9\pm3.4\,[2.5]\)
        & \(8.85\pm0.79\,[0.59]\) \\
        \bottomrule
    \end{tabular}}
\end{table*}
Long-only Sharpe ratios occupy a comparatively narrow range, while their cross-market variation is often larger than the differences between method means. Positive long-only CAGR should not be read as evidence of alpha by itself: every strategy remains fully invested and therefore participates in the market return. Comparisons with the equal-weight universe are needed to separate stock-selection value from this common component.
The long--short results provide a stricter contrast. AlphaQCM and R\&D-Agent have the largest pooled Sharpe ratios, but neither establishes a stable lead relative to its seed and market dispersion. AlphaQCM operates at substantially lower turnover, while R\&D-Agent earns a larger average CAGR but exhibits greater exposure dependence below. More broadly, cross-market dispersion is large relative to the differences between methods, preventing a reliable global ranking.
Table~\ref{tab:portfolio_market_sharpe} makes this heterogeneity explicit. The S\&P~100 produces positive long--short Sharpe ratios for every method, whereas several methods become negative in the HSI, CSI~300, or FTSE~100. The Nikkei~225 favors several formula-based methods. Consequently, the pooled average summarizes materially different market-specific outcomes.

\begin{table*}[t]
    \centering
    \scriptsize
    \setlength{\tabcolsep}{3.5pt}
    \caption{\textbf{Net Sharpe ratio by market under the default configuration.} Automated-method cells report the mean and standard deviation across three random seeds. Alpha101 has one fixed evaluation per market.}
    \label{tab:portfolio_market_sharpe}

    \textbf{(a) Long-only}
    \vspace{2pt}

    \resizebox{0.8\textwidth}{!}{
    \begin{tabular}{lccccc}
        \toprule
        Method & S\&P 100 & HSI & CSI 300 & FTSE 100 & Nikkei 225 \\
        \midrule
        Alpha101     & \(1.75\)           & \(1.35\)           & \(0.98\)           & \(1.16\)           & \(0.96\) \\
        GP           & \(1.92\pm0.23\)    & \(0.87\pm0.09\)    & \(0.92\pm0.05\)    & \(0.91\pm0.11\)    & \(1.24\pm0.08\) \\
        AlphaGen     & \(1.90\pm0.12\)    & \(0.65\pm0.10\)    & \(0.66\pm0.15\)    & \(1.18\pm0.22\)    & \(1.25\pm0.05\) \\
        AlphaCFG     & \(1.56\pm0.19\)    & \(0.62\pm0.10\)    & \(0.70\pm0.21\)    & \(1.16\pm0.59\)    & \(1.20\pm0.12\) \\
        AlphaQCM     & \(2.02\pm0.17\)    & \(1.02\pm0.27\)    & \(0.71\pm0.15\)    & \(1.35\pm0.51\)    & \(1.47\pm0.15\) \\
        AlphaForge   & \(1.76\pm0.15\)    & \(0.96\pm0.08\)    & \(0.75\pm0.15\)    & \(1.28\pm0.33\)    & \(1.29\pm0.08\) \\
        AlphaSAGE    & \(1.73\pm0.18\)    & \(1.04\pm0.35\)    & \(0.91\pm0.13\)    & \(1.34\pm0.16\)    & \(1.19\pm0.20\) \\
        R\&D-Agent   & \(1.67\pm0.04\)    & \(1.00\pm0.18\)    & \(0.77\pm0.07\)    & \(1.39\pm0.21\)    & \(1.13\pm0.15\) \\
        AlphaAgent   & \(1.46\pm0.34\)    & \(1.01\pm0.14\)    & \(0.74\pm0.12\)    & \(1.03\pm0.19\)    & \(0.90\pm0.12\) \\
        QuantaAlpha  & \(1.55\pm0.38\)    & \(0.75\pm0.19\)    & \(1.17\pm0.28\)    & \(1.02\pm0.17\)    & \(1.03\pm0.06\) \\
        \bottomrule
    \end{tabular}}

    \vspace{6pt}
    \textbf{(b) Long--short}
    \vspace{2pt}

    \resizebox{0.8\textwidth}{!}{
    \begin{tabular}{lccccc}
        \toprule
        Method & S\&P 100 & HSI & CSI 300 & FTSE 100 & Nikkei 225 \\
        \midrule
        Alpha101     & \(0.97\)           & \(1.32\)           & \(0.02\)            & \(-0.18\)          & \(-0.53\) \\
        GP           & \(1.59\pm0.24\)    & \(-0.23\pm0.13\)   & \(0.21\pm0.11\)     & \(-0.95\pm0.21\)  & \(1.08\pm0.19\) \\
        AlphaGen     & \(1.46\pm0.34\)    & \(-0.66\pm0.19\)   & \(-0.44\pm0.19\)    & \(-0.14\pm0.59\)  & \(1.26\pm0.04\) \\
        AlphaCFG     & \(1.13\pm0.38\)    & \(-0.77\pm0.04\)   & \(-0.22\pm0.34\)    & \(-0.43\pm1.04\)  & \(1.14\pm0.30\) \\
        AlphaQCM     & \(1.85\pm0.30\)    & \(-0.11\pm0.32\)   & \(-0.16\pm0.27\)    & \(0.01\pm0.64\)   & \(1.74\pm0.41\) \\
        AlphaForge   & \(1.32\pm0.36\)    & \(-0.30\pm0.17\)   & \(-0.28\pm0.52\)    & \(-0.10\pm0.08\)  & \(1.26\pm0.22\) \\
        AlphaSAGE    & \(1.24\pm0.39\)    & \(0.10\pm0.94\)    & \(0.08\pm0.19\)     & \(0.20\pm0.28\)   & \(0.35\pm1.09\) \\
        R\&D-Agent   & \(1.59\pm0.08\)    & \(0.42\pm0.14\)    & \(-0.02\pm0.12\)    & \(0.67\pm0.46\)   & \(0.74\pm0.34\) \\
        AlphaAgent   & \(0.79\pm0.62\)    & \(0.02\pm0.33\)    & \(-0.27\pm0.74\)    & \(-0.23\pm0.32\)  & \(-0.74\pm0.15\) \\
        QuantaAlpha  & \(1.18\pm0.53\)    & \(-0.19\pm0.80\)   & \(0.38\pm0.74\)     & \(-0.18\pm0.46\)  & \(0.42\pm0.24\) \\
        \bottomrule
    \end{tabular}}
\end{table*}

\subsection{Wealth Dynamics and Exposure Dependence}
\label{app:portfolio_wealth}

\begin{figure*}[t]
    \centering
    \includegraphics[width=\textwidth]
        {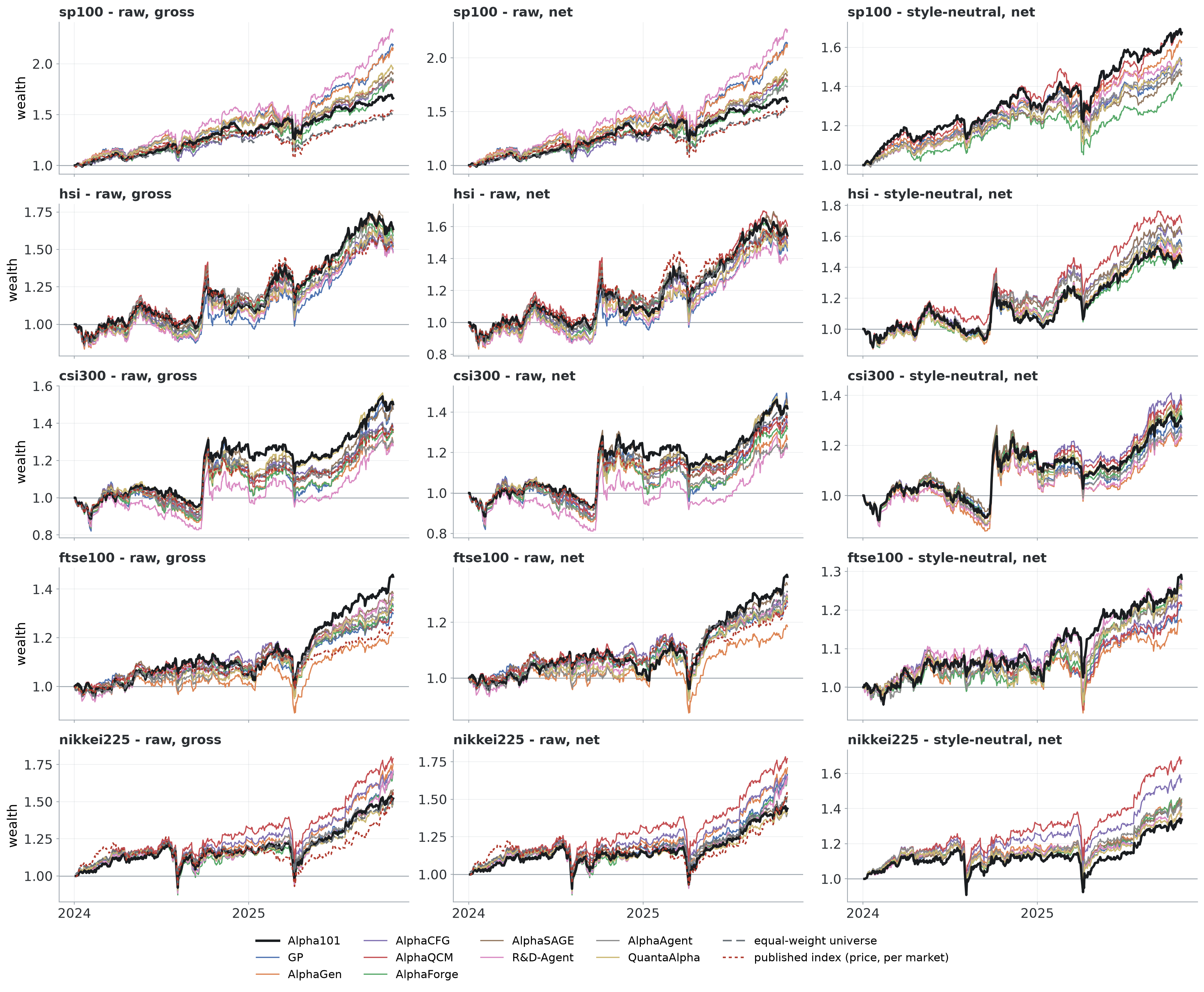}
    \caption{\textbf{Long-only test-period wealth under ridge combination.} Rows correspond to markets. Columns show raw gross wealth, raw net wealth, and net wealth obtained from style-neutralized composites. Method curves average available random seeds. The equal-weight universe and published market index provide reference paths where available.}
    \label{fig:long_only_wealth}
\end{figure*}

\begin{figure*}[t]
    \centering
    \includegraphics[width=\textwidth]
        {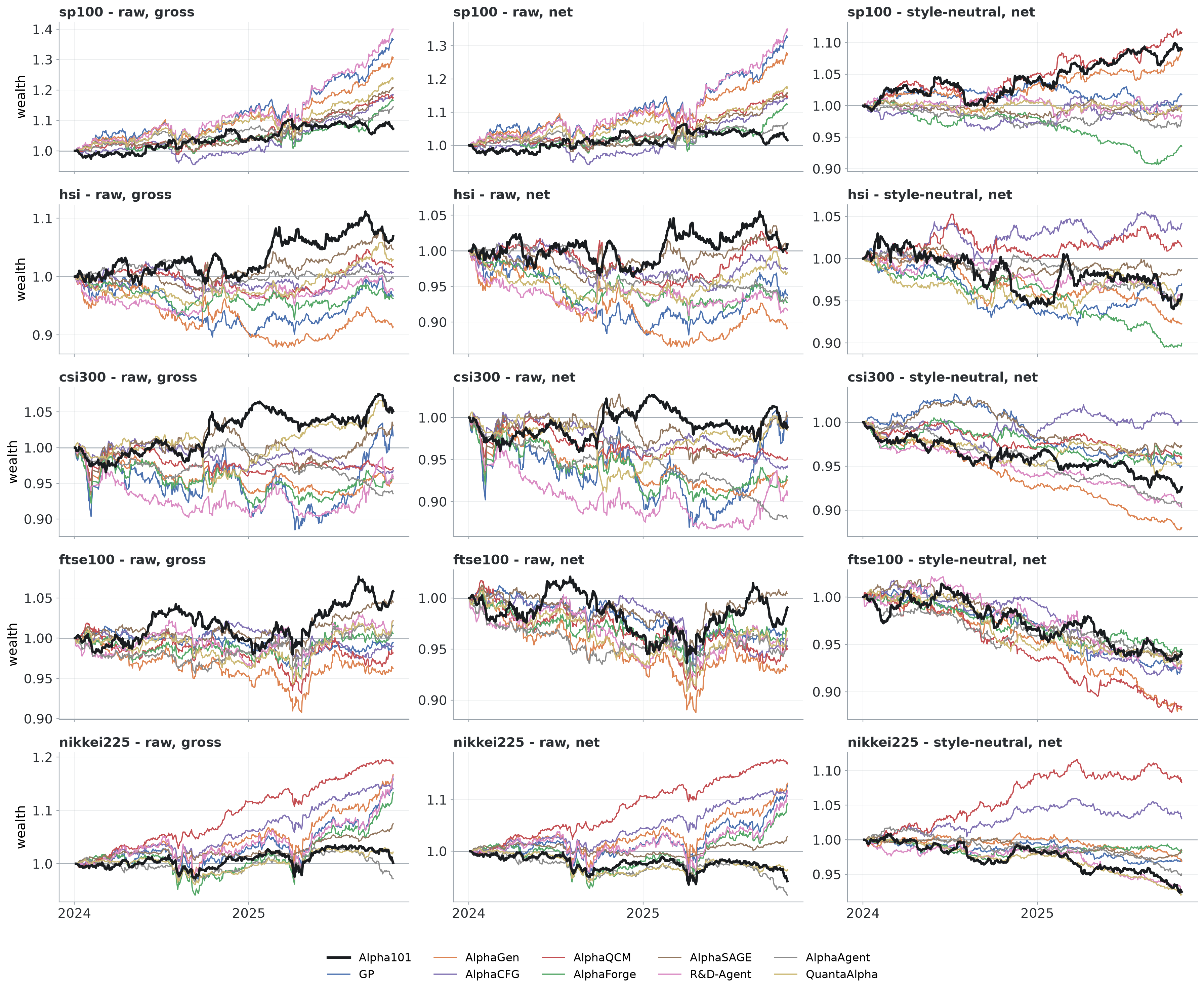}
    \caption{\textbf{Long--short test-period wealth under ridge combination.}
    Rows correspond to markets, while columns show raw gross, raw net, and
    style-neutralized net wealth. Unlike the long-only portfolios, the
    long--short portfolios reduce broad market-direction exposure and therefore
    provide a stricter test of the composite's cross-sectional ranking value.}
    \label{fig:long_short_wealth}
\end{figure*}
Figures~\ref{fig:long_only_wealth} and \ref{fig:long_short_wealth} trace the test-period wealth paths under ridge combination. The three columns distinguish raw gross performance, raw performance after costs, and performance obtained from style-neutralized composites after costs. This decomposition separates the existence of a gross return pattern from the losses introduced by trading and from the component associated with measured exposures.
The long-only paths are strongly influenced by the underlying market return, which explains why positive terminal wealth can persist even when composite predictiveness is weak. Long--short results are less uniform. Much of their positive aggregate performance is concentrated in the S\&P~100 and Nikkei~225, while several methods lose money in the remaining markets. The wealth paths therefore reinforce the cross-market heterogeneity visible in Table~\ref{tab:portfolio_market_sharpe}.

\begin{table*}[t]
    \centering
    \scriptsize
    \setlength{\tabcolsep}{3.2pt}
    \caption{\textbf{Portfolio performance before and after style
    neutralization.} Each cell reports \emph{raw/style-neutralized} performance under ridge combination, 20-day rebalancing, and a transaction cost of 10 basis points per dollar traded. Both variants are evaluated over the test period. CAGR is annualized and reported in percent; turnover is annualized and reported as multiples of portfolio value.}
    \label{tab:portfolio_neutralization}
    \resizebox{\textwidth}{!}{
    \begin{tabular}{lccc|ccc}
        \toprule
        & \multicolumn{3}{c}{Long-only}
        & \multicolumn{3}{c}{Long--short} \\
        \cmidrule(lr){2-4}\cmidrule(lr){5-7}
        Method
        & \shortstack{Net CAGR\\(\%)}
        & Net Sharpe
        & \shortstack{Turnover\\(\(\times\))}
        & \shortstack{Net CAGR\\(\%)}
        & Net Sharpe
        & \shortstack{Turnover\\(\(\times\))} \\
        \midrule
        Alpha101
        & \(26.5/25.3\) & \(1.241/1.276\) & \(8.80/9.02\)
        & \(2.6/3.0\) & \(0.319/0.540\) & \(8.73/8.93\) \\
        GP
        & \(30.4/20.9\) & \(1.173/1.029\) & \(6.32/7.98\)
        & \(3.7/-2.0\) & \(0.340/-0.462\) & \(6.37/8.03\) \\
        AlphaGen
        & \(28.0/19.1\) & \(1.127/0.977\) & \(5.52/7.79\)
        & \(2.3/-2.4\) & \(0.297/-0.569\) & \(5.57/8.03\) \\
        AlphaCFG
        & \(23.5/21.9\) & \(1.048/1.137\) & \(7.53/8.51\)
        & \(1.0/-1.4\) & \(0.169/-0.325\) & \(7.48/8.41\) \\
        AlphaQCM
        & \(29.6/26.0\) & \(1.313/1.323\) & \(4.74/5.93\)
        & \(3.7/1.3\) & \(0.668/0.283\) & \(4.58/6.02\) \\
        AlphaForge
        & \(28.2/21.9\) & \(1.208/1.124\) & \(7.68/8.70\)
        & \(2.9/-0.8\) & \(0.380/-0.092\) & \(7.62/8.64\) \\
        AlphaSAGE
        & \(29.2/22.1\) & \(1.244/1.114\) & \(7.45/8.23\)
        & \(2.9/-0.9\) & \(0.394/-0.167\) & \(7.54/8.25\) \\
        R\&D-Agent
        & \(31.2/20.6\) & \(1.193/1.057\) & \(7.89/9.85\)
        & \(6.5/-1.6\) & \(0.679/-0.449\) & \(7.84/9.81\) \\
        AlphaAgent
        & \(22.4/20.8\) & \(1.027/1.064\) & \(9.82/9.99\)
        & \(0.3/-2.2\) & \(-0.087/-0.521\) & \(9.81/10.05\) \\
        QuantaAlpha
        & \(25.3/23.7\) & \(1.106/1.190\) & \(8.87/9.56\)
        & \(2.7/-0.4\) & \(0.324/-0.096\) & \(8.85/9.52\) \\
        \bottomrule
    \end{tabular}}
\end{table*}
Table~\ref{tab:portfolio_neutralization} compares raw and style-neutralized performance on the same style-eligible test dates. These numbers differ slightly from Tables~\ref{tab:long_only_full_results} and \ref{tab:long_short_full_results}, which use the complete test window. The matched window is required because construction of market beta uses a 252-trading-day warm-up.
Style neutralization reduces long-only CAGR for every method, although its effect on long-only Sharpe is smaller and mixed because these portfolios remain fully invested and continue to participate in broad market returns. The long--short results provide a clearer view of residual predictive performance. AlphaQCM is the only automated method that remains profitable after neutralization: its net CAGR decreases from \(3.7\%\) to \(1.3\%\), its Sharpe ratio declines from \(0.668\) to \(0.283\), and its turnover rises from \(4.58\) to \(6.02\). It nevertheless maintains the lowest style-neutralized turnover among the automated methods. All other automated methods exhibit negative style-neutralized long--short CAGR and Sharpe. For example, R\&D-Agent's net CAGR falls from \(6.5\%\) to \(-1.6\%\), its Sharpe ratio falls from \(0.679\) to \(-0.449\), and its turnover increases from \(7.84\) to \(9.81\).

\begin{figure*}[t]
    \centering
    \begin{subfigure}[t]{0.325\textwidth}
        \centering
        \includegraphics[width=\linewidth]
            {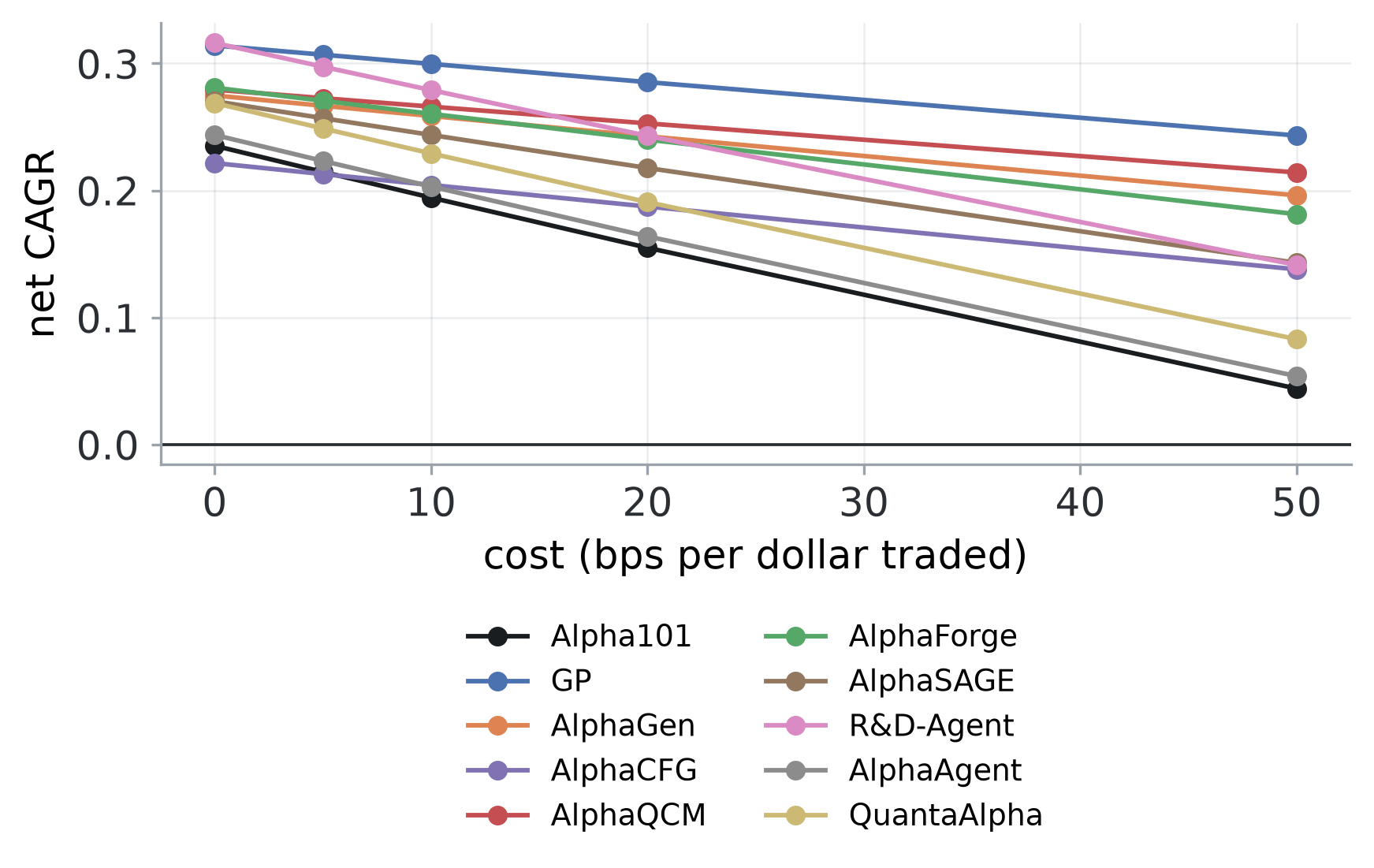}
        \caption{Transaction cost}
        \label{fig:lo_raw_cost}
    \end{subfigure}
    \hfill
    \begin{subfigure}[t]{0.325\textwidth}
        \centering
        \includegraphics[width=\linewidth]
            {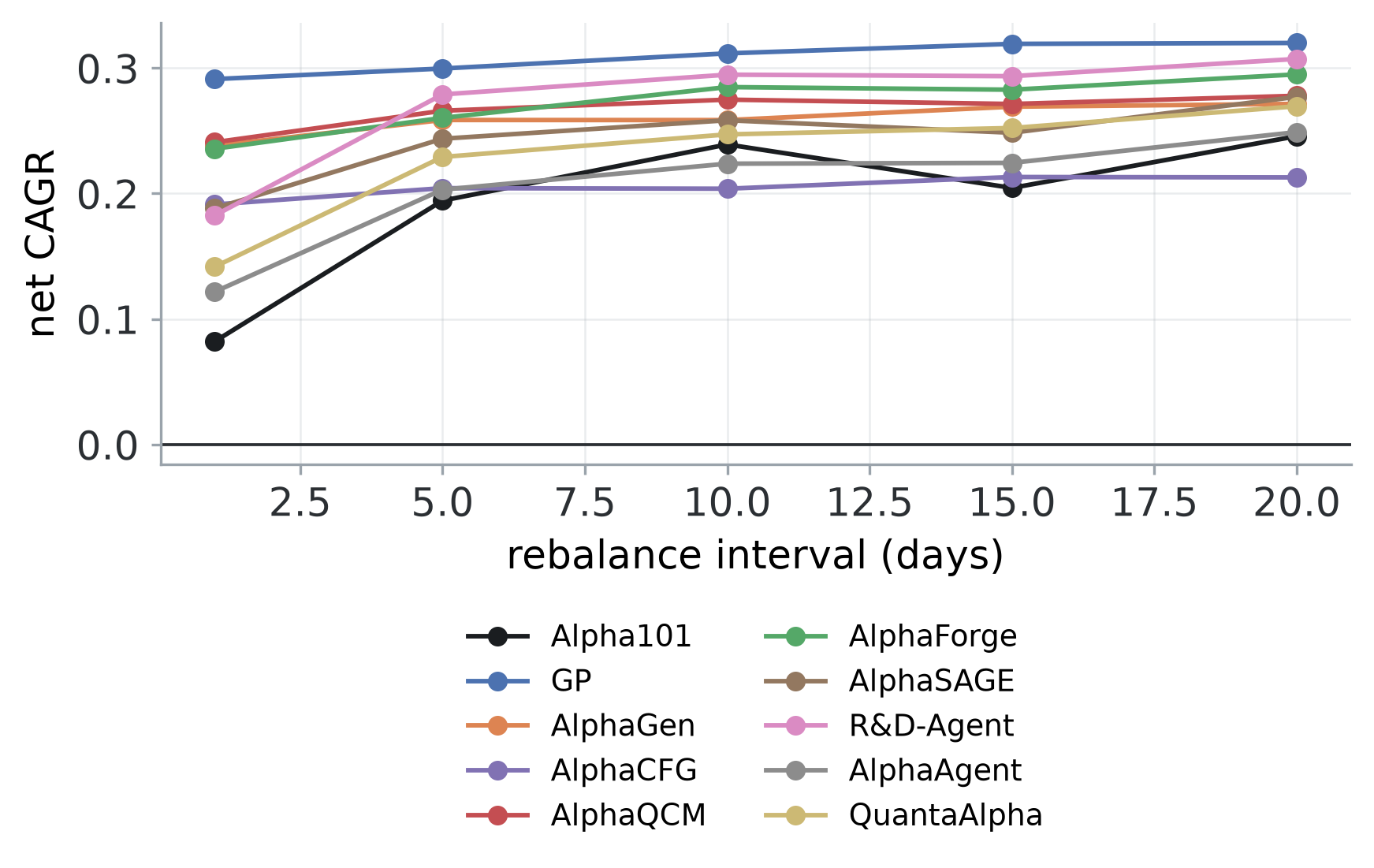}
        \caption{Rebalance interval}
        \label{fig:lo_raw_rebalance}
    \end{subfigure}
    \hfill
    \begin{subfigure}[t]{0.325\textwidth}
        \centering
        \includegraphics[width=\linewidth]
            {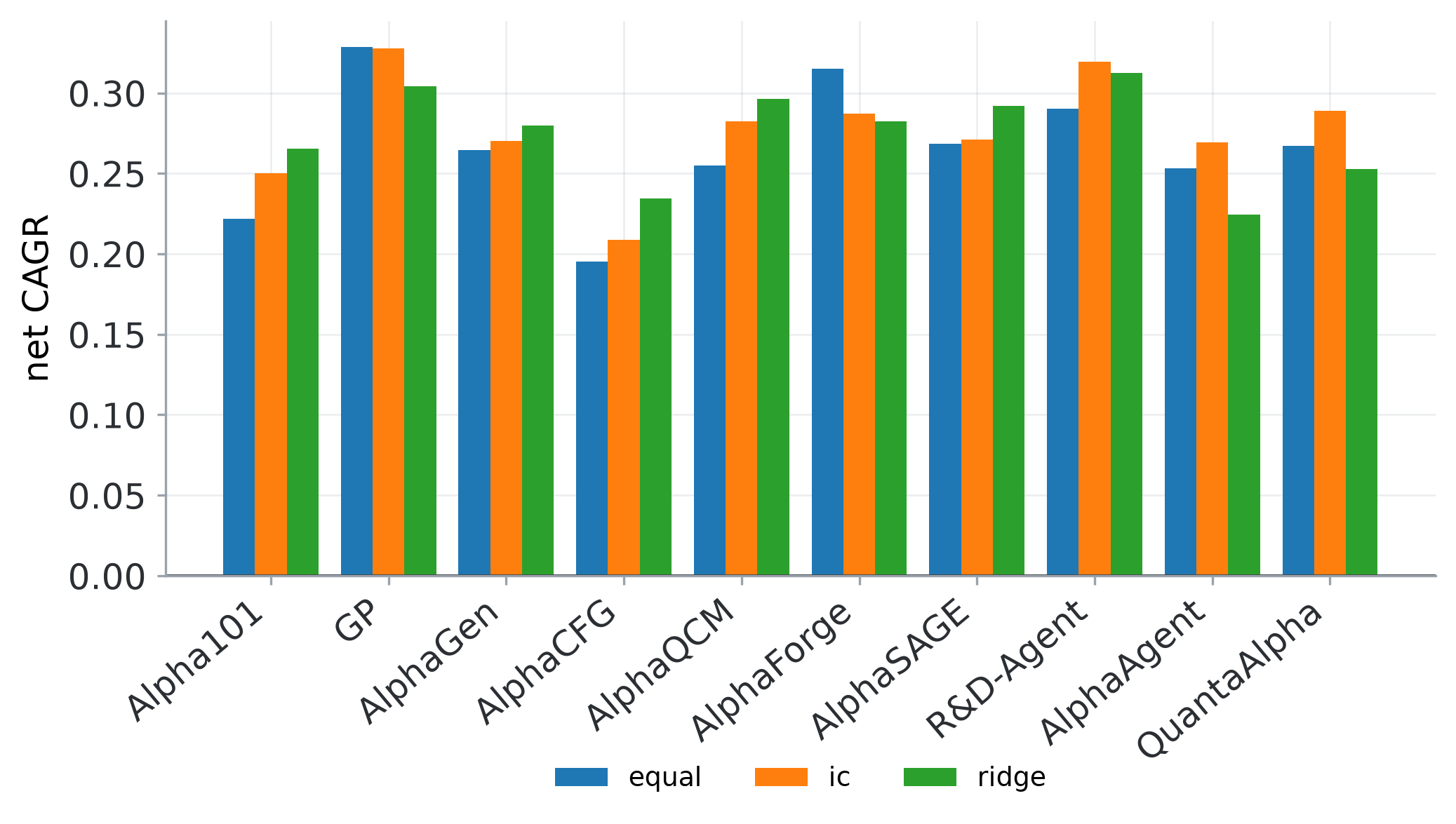}
        \caption{Combination rule}
        \label{fig:lo_raw_combo}
    \end{subfigure}
    \caption{\textbf{Sensitivity of long-only portfolios.}
    Panels report market-aggregated net CAGR as transaction costs,
    rebalancing frequency, or signal combination is varied individually.}
    \label{fig:long_only_raw_sensitivity}
\end{figure*}
\begin{figure*}[t]
    \centering
    \begin{subfigure}[t]{0.325\textwidth}
        \centering
        \includegraphics[width=\linewidth]
            {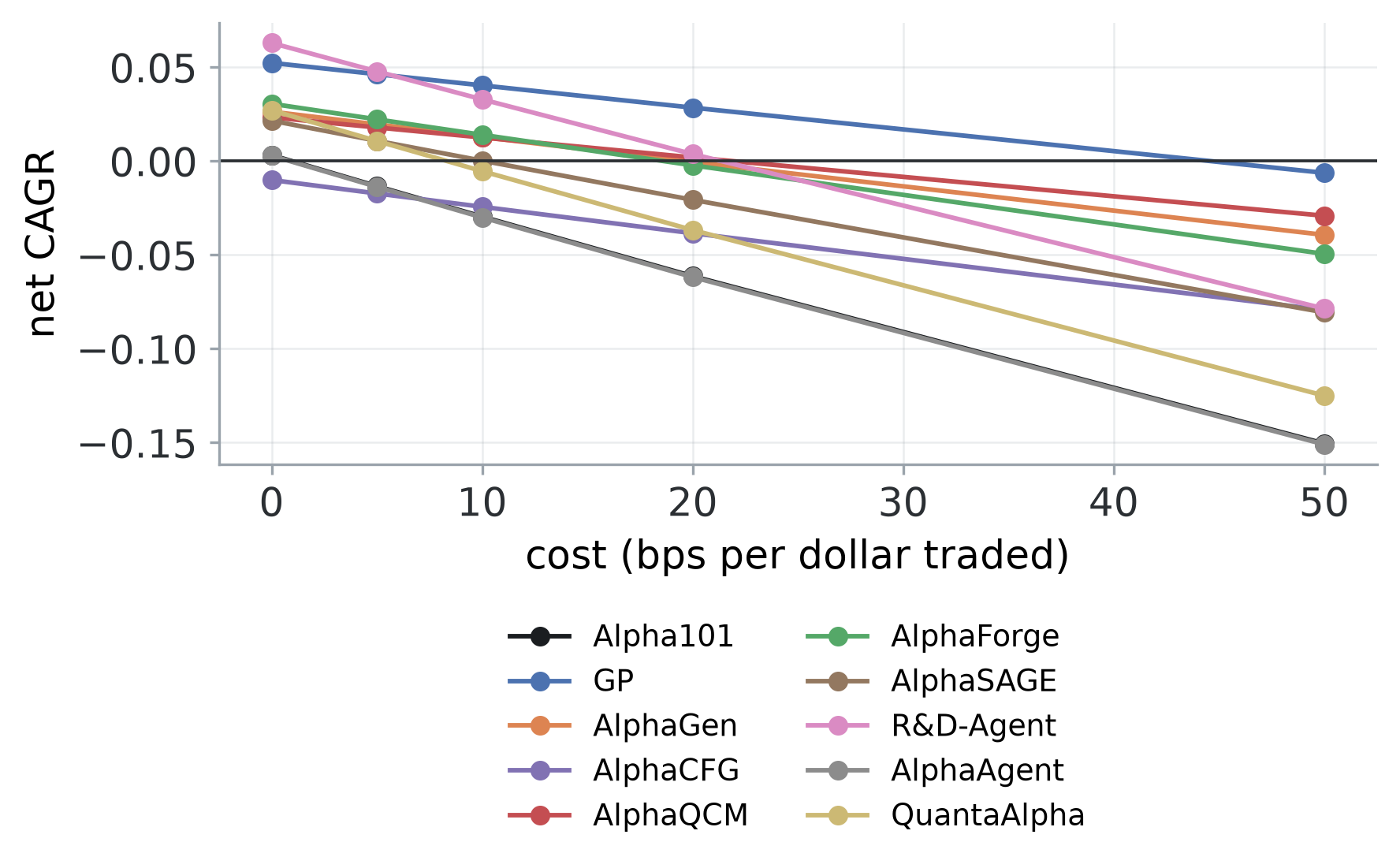}
        \caption{Transaction cost}
        \label{fig:ls_raw_cost}
    \end{subfigure}
    \hfill
    \begin{subfigure}[t]{0.325\textwidth}
        \centering
        \includegraphics[width=\linewidth]
            {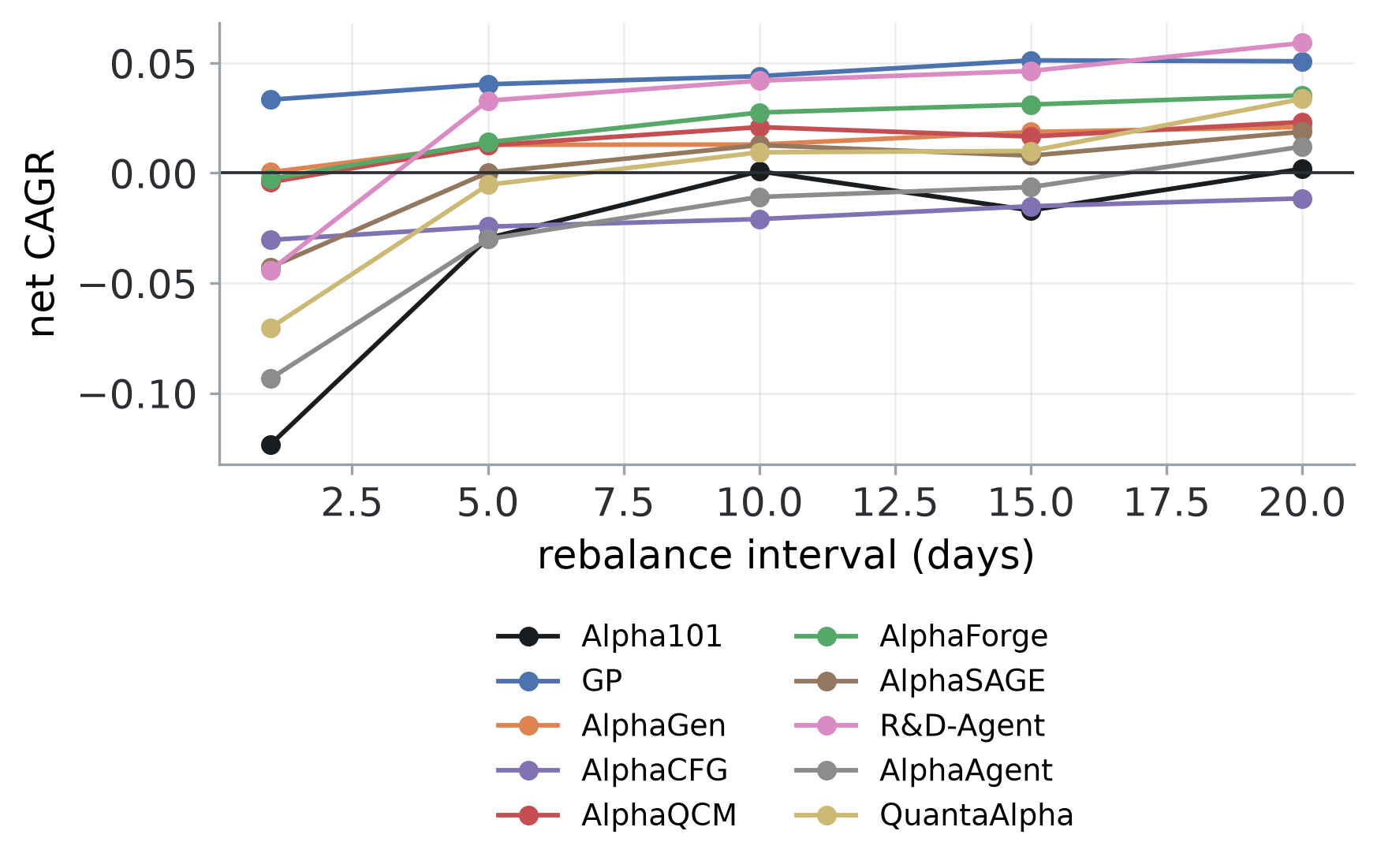}
        \caption{Rebalance interval}
        \label{fig:ls_raw_rebalance}
    \end{subfigure}
    \hfill
    \begin{subfigure}[t]{0.325\textwidth}
        \centering
        \includegraphics[width=\linewidth]
            {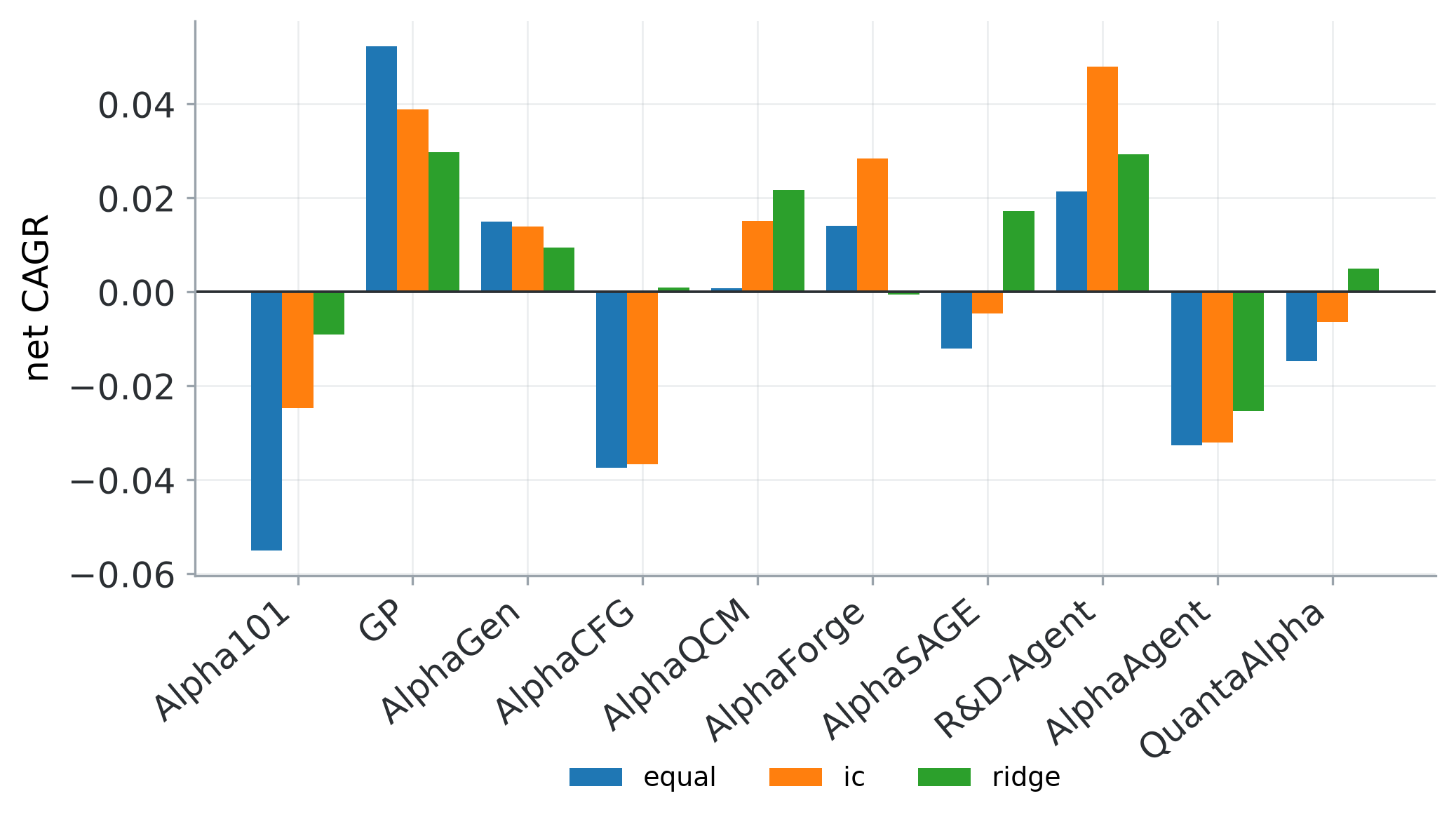}
        \caption{Combination rule}
        \label{fig:ls_raw_combo}
    \end{subfigure}
    \caption{\textbf{Sensitivity of long--short portfolios.}
    Long--short outcomes are substantially more sensitive to costs,
    rebalancing frequency, and signal combination than long-only outcomes.}
    \label{fig:long_short_raw_sensitivity}
\end{figure*}
\begin{figure*}[!t]
    \centering
    \begin{subfigure}[t]{0.325\textwidth}
        \centering
        \includegraphics[width=\linewidth]
            {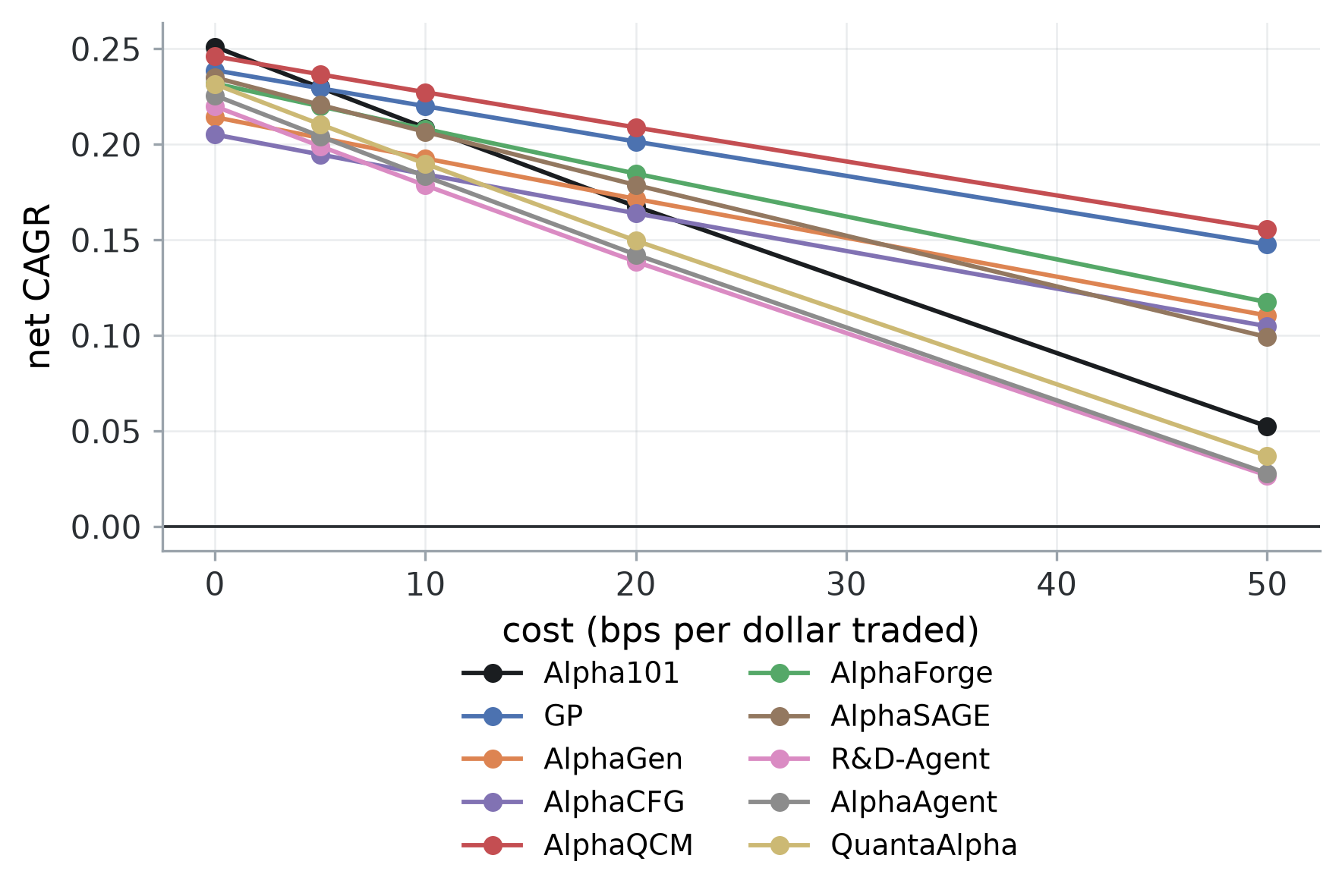}
        \caption{Transaction cost}
        \label{fig:lo_neutral_cost}
    \end{subfigure}
    \hfill
    \begin{subfigure}[t]{0.325\textwidth}
        \centering
        \includegraphics[width=\linewidth]
            {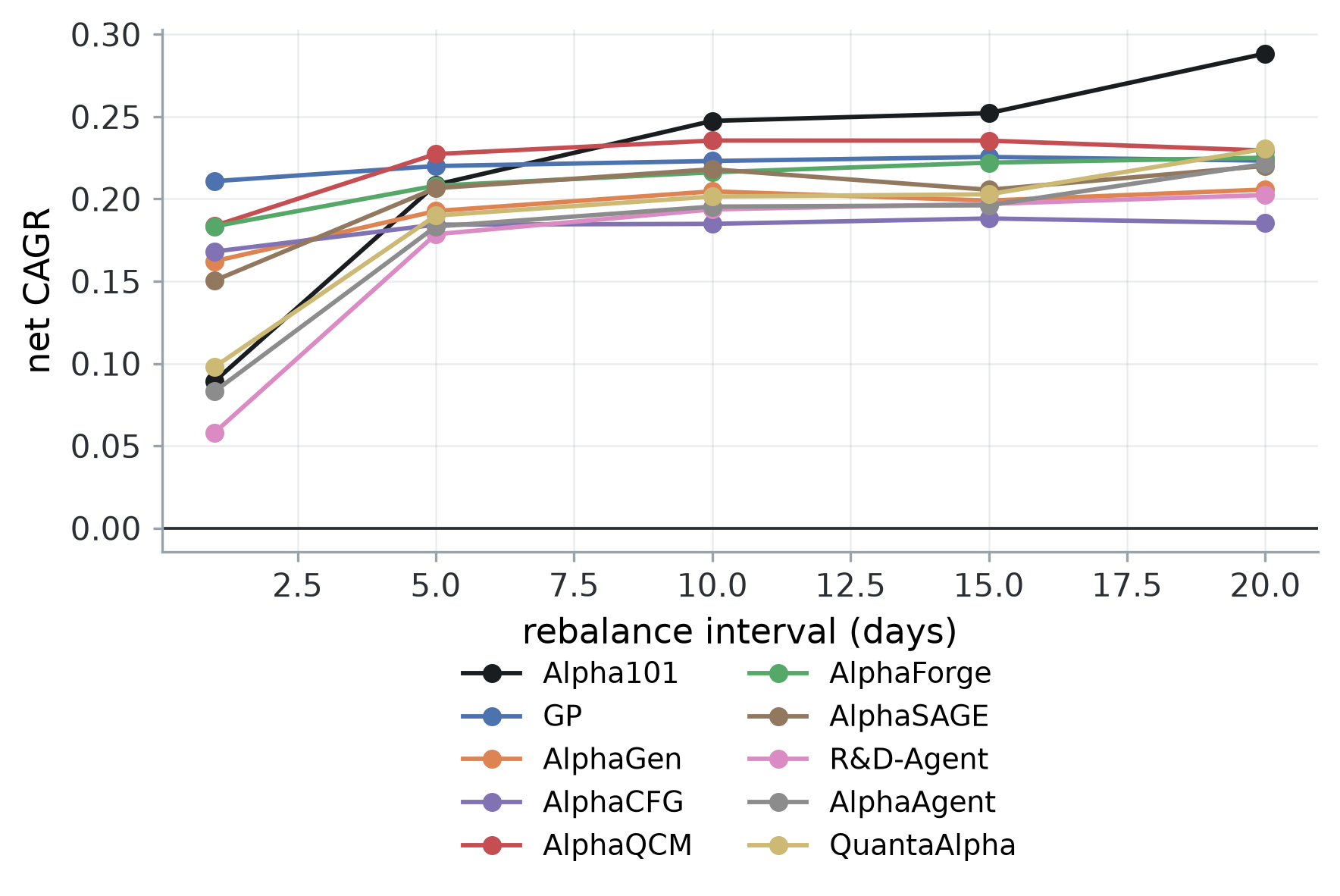}
        \caption{Rebalance interval}
        \label{fig:lo_neutral_rebalance}
    \end{subfigure}
    \hfill
    \begin{subfigure}[t]{0.325\textwidth}
        \centering
        \includegraphics[width=\linewidth]
            {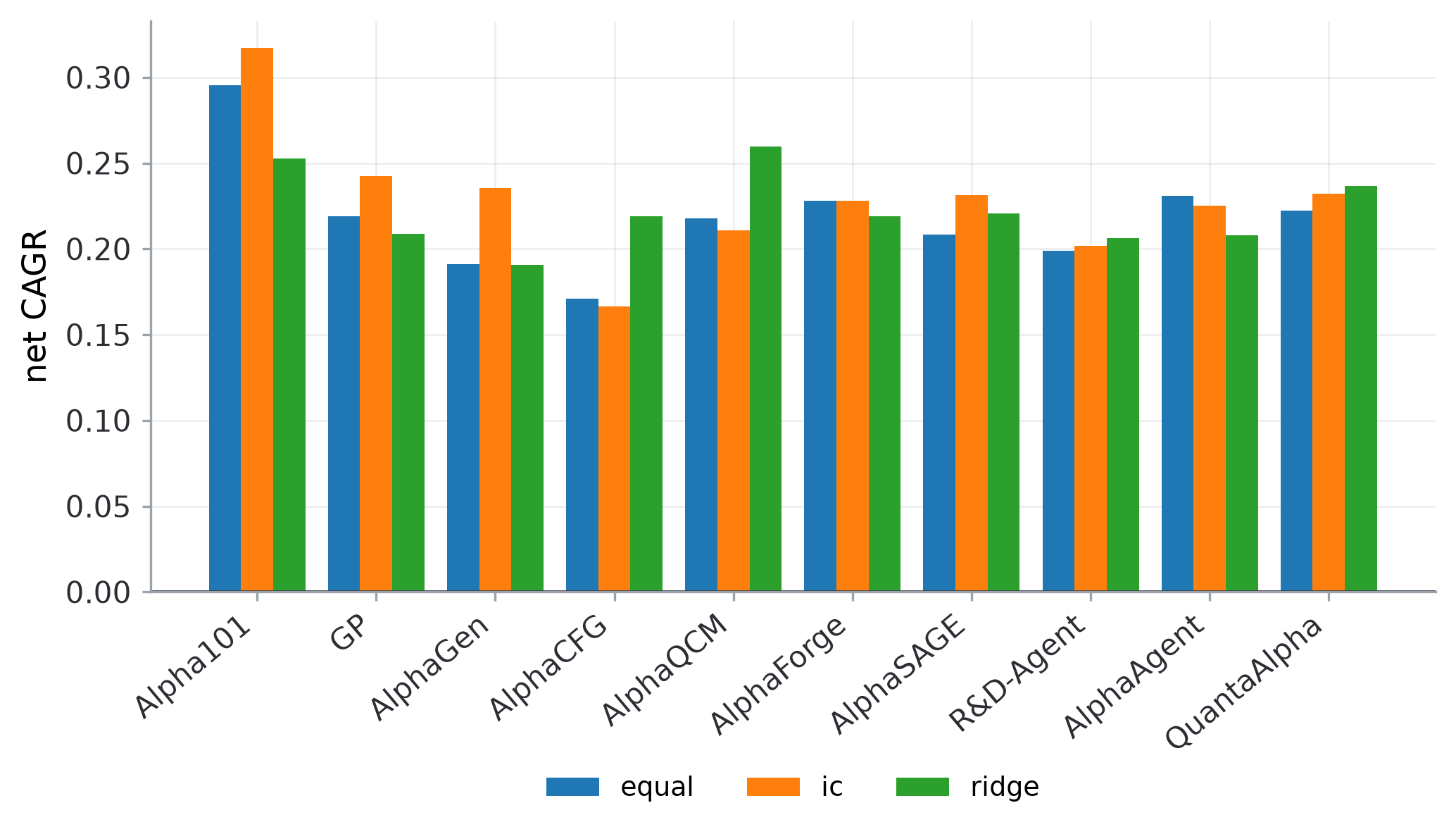}
        \caption{Combination rule}
        \label{fig:lo_neutral_combo}
    \end{subfigure}
    \caption{\textbf{Sensitivity of style-neutralized long-only portfolios.}
    Neutralization generally lowers CAGR and increases turnover sensitivity,
    while the fully invested construction retains substantial common market
    performance.}
    \label{fig:long_only_neutral_sensitivity}
\end{figure*}
\begin{figure*}[!t]
    \centering
    \begin{subfigure}[t]{0.325\textwidth}
        \centering
        \includegraphics[width=\linewidth]
            {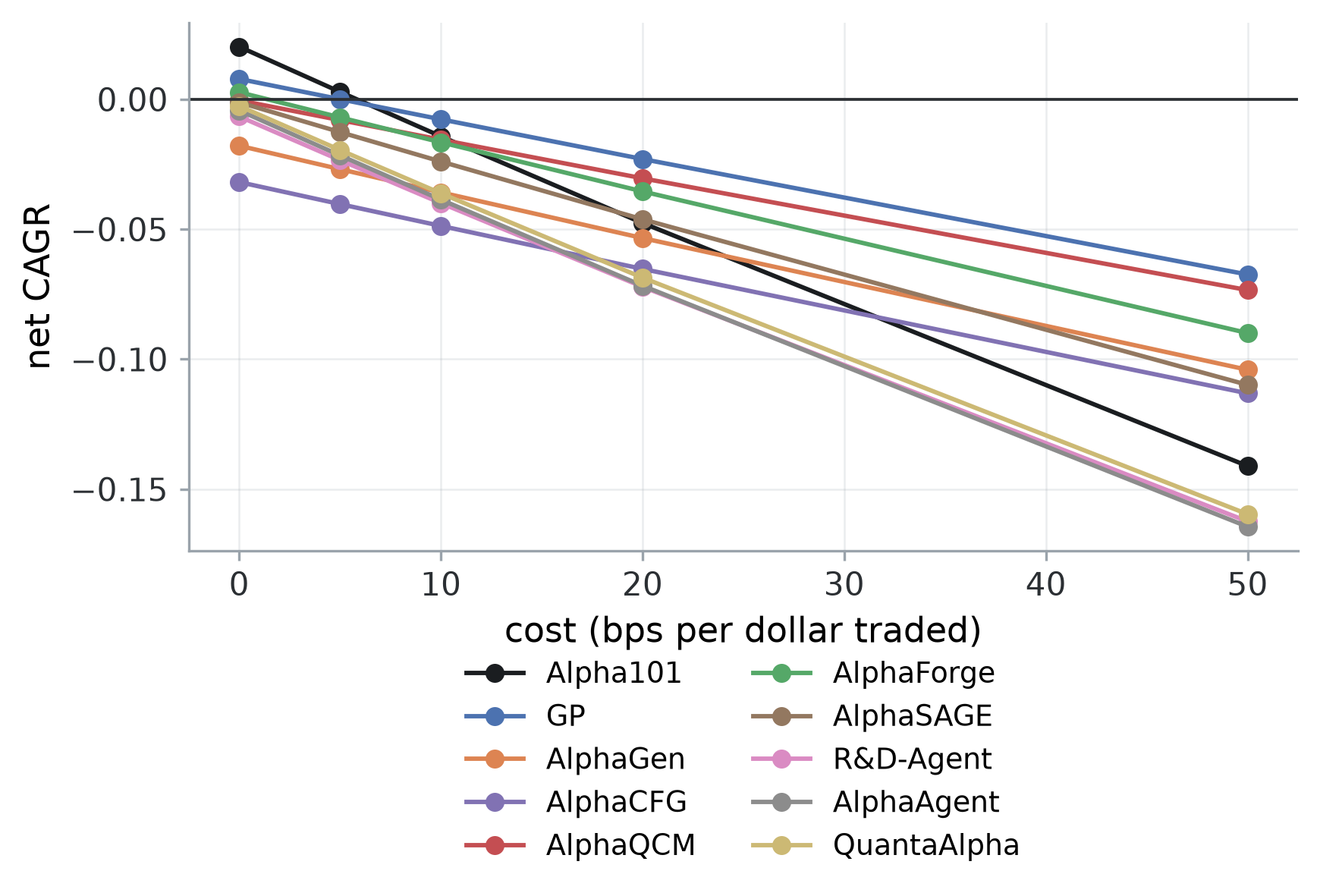}
        \caption{Transaction cost}
        \label{fig:ls_neutral_cost}
    \end{subfigure}
    \hfill
    \begin{subfigure}[t]{0.325\textwidth}
        \centering
        \includegraphics[width=\linewidth]
            {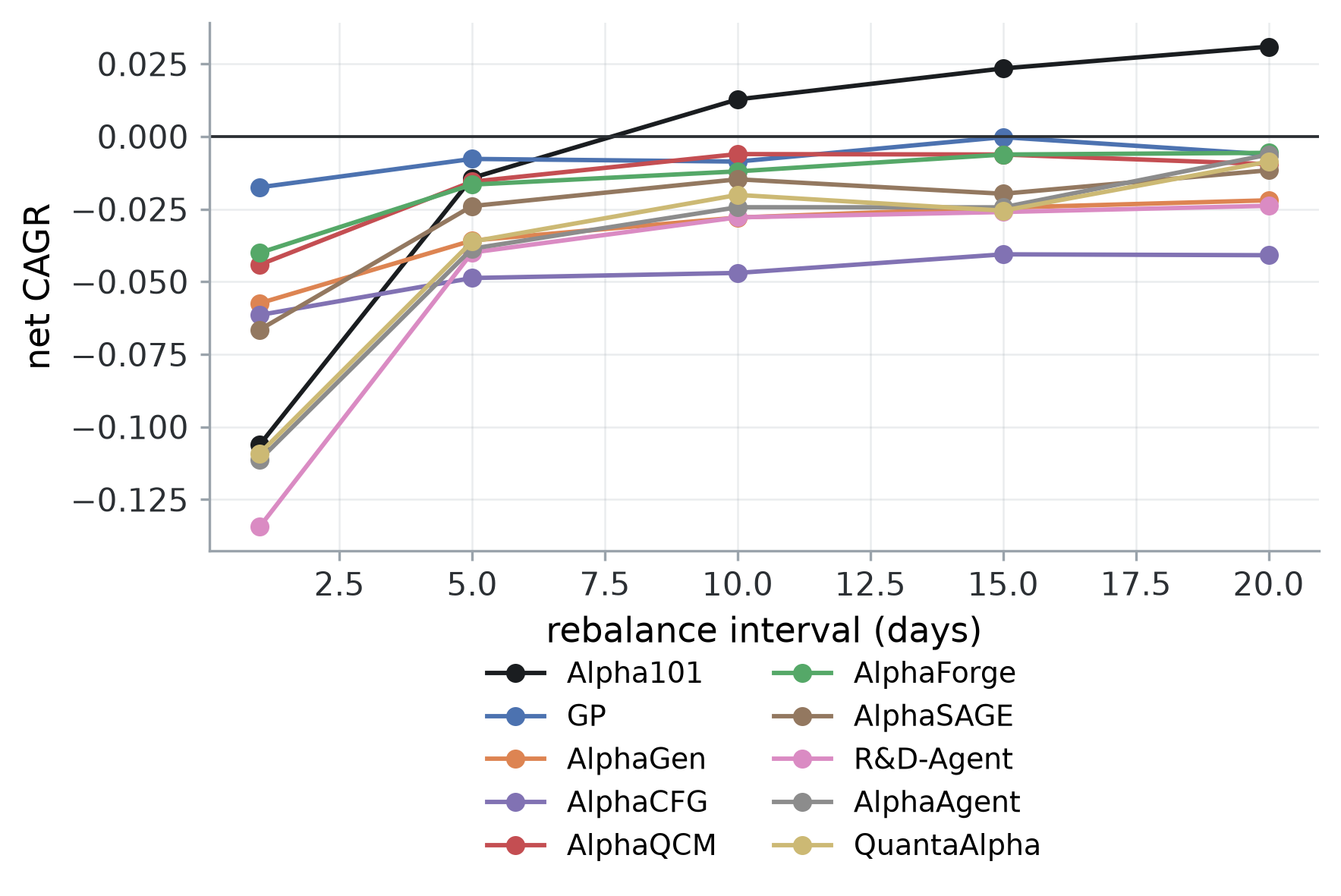}
        \caption{Rebalance interval}
        \label{fig:ls_neutral_rebalance}
    \end{subfigure}
    \hfill
    \begin{subfigure}[t]{0.325\textwidth}
        \centering
        \includegraphics[width=\linewidth]
            {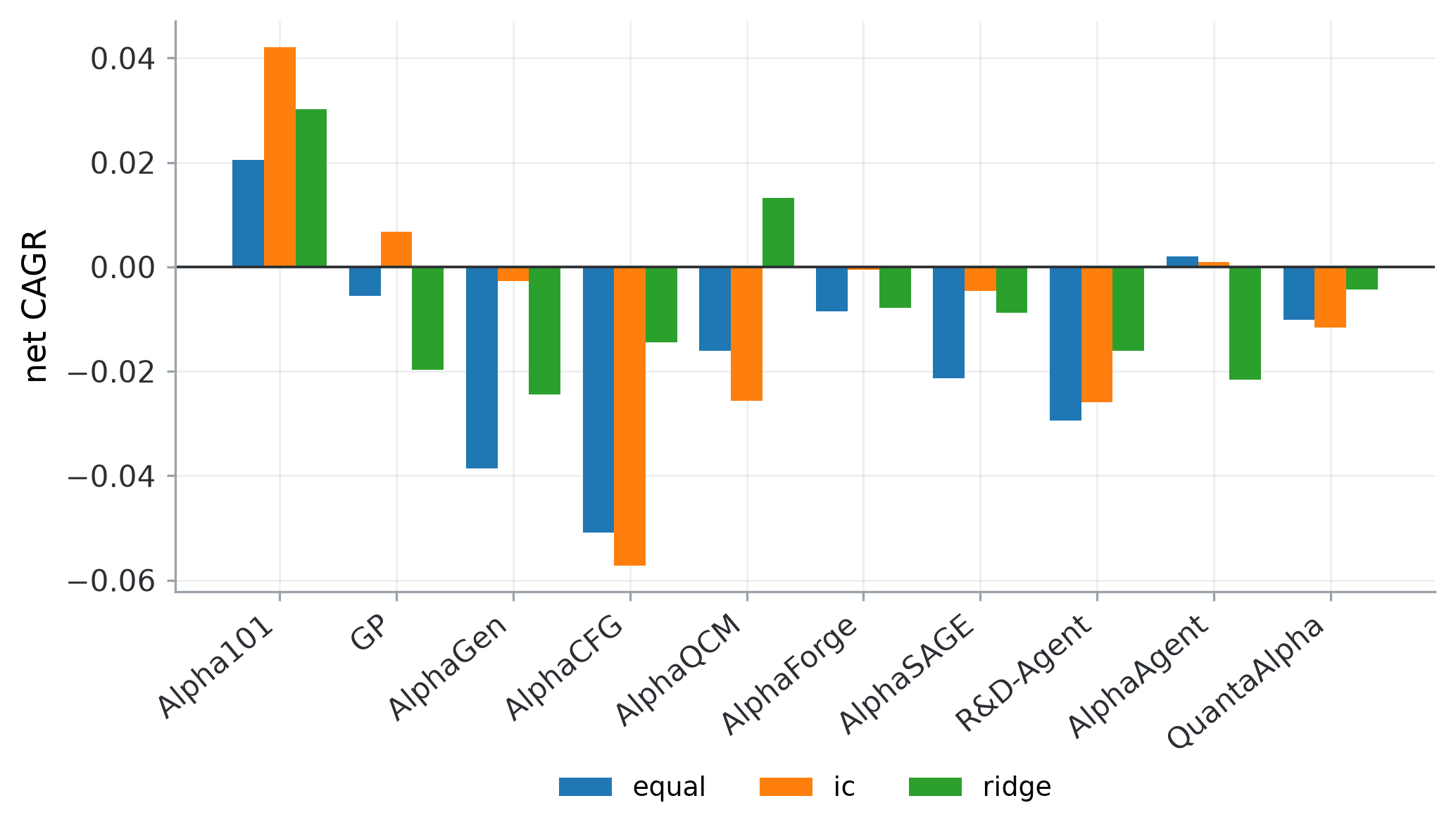}
        \caption{Combination rule}
        \label{fig:ls_neutral_combo}
    \end{subfigure}
    \caption{\textbf{Sensitivity of style-neutralized long--short portfolios.}
    After measured exposures are removed, most methods fail to sustain positive
    net CAGR across trading assumptions.}
    \label{fig:long_short_neutral_sensitivity}
\end{figure*}

\subsection{Sensitivity to Portfolio Construction}
\label{app:portfolio_sensitivity}

The following figures vary one design choice at a time while holding the remaining choices at their default values. Transaction costs vary from 0 to 50 basis points per dollar traded, the rebalance interval varies from 1 to 20 trading days, and signal combination varies among equal weighting, validation-IC weighting, and ridge estimation. These experiments are intended to attribute changes in performance to individual portfolio-design choices; they are not used to select the best test-period configuration.

Figures~\ref{fig:long_only_raw_sensitivity} and \ref{fig:long_short_raw_sensitivity} demonstrate that positive gross predictiveness does not determine the investable result. Long-only CAGR declines steadily with cost, although it still remains positive. Long--short profitability is considerably more fragile: most methods approach or fall below zero as costs increase. Longer rebalance intervals generally improve net performance by reducing turnover and aligning the holding period more closely with the 20-day prediction target.
The preferred combination rule also differs across methods and strategies. No single rule dominates for every method. Several method rankings change, and some long--short outcomes change sign, when the composite moves from equal weighting to IC weighting or ridge estimation. The measured value of a delivered pool therefore depends partly on how its constituent signals are combined.

The style-neutralized sensitivities in Figures~\ref{fig:long_only_neutral_sensitivity} and ~\ref{fig:long_short_neutral_sensitivity} strengthen this conclusion. Long-only portfolios continue to earn positive absolute returns because they remain fully invested, but their advantage relative to the raw composites narrows. For long--short portfolios, most methods produce negative net CAGR after neutralization under the default combination and cost assumptions. Slower rebalancing mitigates these losses by reducing turnover, but does not produce a common positive result across methods. Thus, the gap between raw and style-neutralized performance cannot be attributed to a single trading parameter.

\end{document}

%% file: math_commands.tex
\usepackage{amsmath,amsfonts,bm}

\def\eqref#1{equation~\ref{#1}}

\def\1{\bm{1}}

\DeclareMathAlphabet{\mathsfit}{\encodingdefault}{\sfdefault}{m}{sl}
\SetMathAlphabet{\mathsfit}{bold}{\encodingdefault}{\sfdefault}{bx}{n}



%% file: iclr2027_conference.bib
@article{kakushadze2016101,
  title={101 formulaic alphas},
  author={Kakushadze, Zura},
  journal={Wilmott},
  volume={2016},
  number={84},
  pages={72--81},
  year={2016},
  publisher={Wiley Online Library}
}

@inproceedings{yu2023generating,
  title={Generating synergistic formulaic alpha collections via reinforcement learning},
  author={Yu, Shuo and Xue, Hongyan and Ao, Xiang and Pan, Feiyang and He, Jia and Tu, Dandan and He, Qing},
  booktitle={Proceedings of the 29th ACM SIGKDD conference on knowledge discovery and data mining},
  pages={5476--5486},
  year={2023}
}

@article{yang2026alpha,
  title={Alpha Discovery via Grammar-Guided Learning and Search},
  author={Yang, Han and Hao, Dong and Wang, Zhuohan and Shi, Qi and Li, Xingtong},
  journal={arXiv preprint arXiv:2601.22119},
  year={2026}
}

@InProceedings{zhu2025alphaqcm,
  title={AlphaQCM: Alpha Discovery in Finance with Distributional Reinforcement Learning},
  author={Zhu, Zhoufan and Zhu, Ke},
  booktitle={Proceedings of the 42nd International Conference on Machine Learning},
  pages={80463--80479},
  year={2025}
}

@inproceedings{shi2025alphaforge,
  title={Alphaforge: A framework to mine and dynamically combine formulaic alpha factors},
  author={Shi, Hao and Song, Weili and Zhang, Xinting and Shi, Jiahe and Luo, Cuicui and Ao, Xiang and Arian, Hamid and Seco, Luis Angel},
  booktitle={Proceedings of the AAAI conference on artificial intelligence},
  volume={39},
  number={12},
  pages={12524--12532},
  year={2025}
}

@inproceedings{chen2026alphasage,
  title={Alphasage: Structure-aware alpha mining via gflownets for robust exploration},
  author={Chen, Binqi and Ding, Hongjun and Shen, Ning and Guo, Taian and Huang, Jinsheng and Liu, Luchen and Zhang, Ming},
  booktitle={International Conference on Learning Representations},
  volume={2026},
  pages={2955--2982},
  year={2026}
}

@article{li2026r,
  title={R\&D-Agent-Quant: a multi-agent framework for data-centric factors and model joint optimization},
  author={Li, Yuante and Yang, Xu and Yang, Xiao and Wang, Xisen and Liu, Weiqing and Bian, Jiang},
  journal={Advances in Neural Information Processing Systems},
  volume={38},
  year={2026}
}

@inproceedings{tang2025alphaagent,
  title={Alphaagent: Llm-driven alpha mining with regularized exploration to counteract alpha decay},
  author={Tang, Ziyi and Chen, Zechuan and Yang, Jiarui and Mai, Jiayao and Zheng, Yongsen and Wang, Keze and Chen, Jinrui and Lin, Liang},
  booktitle={Proceedings of the 31st ACM SIGKDD Conference on Knowledge Discovery and Data Mining V. 2},
  pages={2813--2822},
  year={2025}
}

@article{han2026quantaalpha,
  title={Quantaalpha: An evolutionary framework for llm-driven alpha mining},
  author={Han, Jun and Zhang, Shuo and Li, Wei and Dong, Yifan and Hu, Tu and Zhu, Yumo and Yu, Xiaomin and Guo, Xin and Liu, Zhaowei and Wang, Kunyi and others},
  journal={arXiv preprint arXiv:2602.07085},
  year={2026}
}

@inproceedings{shi2026navigating,
  title={Navigating the alpha jungle: An llm-powered mcts framework for formulaic alpha factor mining},
  author={Shi, Yu and Duan, Yitong and Li, Jian},
  booktitle={Proceedings of the AAAI Conference on Artificial Intelligence},
  volume={40},
  number={2},
  pages={997--1005},
  year={2026}
}

@inproceedings{liu2026cognitive,
  title={Cognitive alpha mining via llm-driven code-based evolution},
  author={Liu, Fengyuan and Huang, Yi and Luo, Sichun and Wang, Yuqi and Yang, Yazheng and Li, Xinye and Hu, Zefa and Feng, Junlan and Liu, Qi},
  booktitle={Proceedings of the 64th Annual Meeting of the Association for Computational Linguistics (Volume 1: Long Papers)},
  pages={11715--11749},
  year={2026}
}

@article{islam2023financebench,
  title={Financebench: A new benchmark for financial question answering},
  author={Islam, Pranab and Kannappan, Anand and Kiela, Douwe and Qian, Rebecca and Scherrer, Nino and Vidgen, Bertie},
  journal={arXiv preprint arXiv:2311.11944},
  year={2023}
}

@article{xie2024finben,
  title={Finben: A holistic financial benchmark for large language models},
  author={Xie, Qianqian and Han, Weiguang and Chen, Zhengyu and Xiang, Ruoyu and Zhang, Xiao and He, Yueru and Xiao, Mengxi and Li, Dong and Dai, Yongfu and Feng, Duanyu and others},
  journal={Advances in neural information processing systems},
  volume={37},
  pages={95716--95743},
  year={2024}
}

@article{yang2020qlib,
  title={Qlib: An ai-oriented quantitative investment platform},
  author={Yang, Xiao and Liu, Weiqing and Zhou, Dong and Bian, Jiang and Liu, Tie-Yan},
  journal={arXiv preprint arXiv:2009.11189},
  year={2020}
}

@article{liu2022finrl,
  title={FinRL-Meta: Market environments and benchmarks for data-driven financial reinforcement learning},
  author={Liu, Xiao-Yang and Xia, Ziyi and Rui, Jingyang and Gao, Jiechao and Yang, Hongyang and Zhu, Ming and Wang, Christina and Wang, Zhaoran and Guo, Jian},
  journal={Advances in Neural Information Processing Systems},
  volume={35},
  pages={1835--1849},
  year={2022}
}

@article{wang2025quantbench,
  title={Quantbench: Benchmarking ai methods for quantitative investment},
  author={Wang, Saizhuo and Kong, Hao and Guo, Jiadong and Hua, Fengrui and Qi, Yiyan and Zhou, Wanyun and Zheng, Jiahao and Wang, Xinyu and Ni, Lionel M and Guo, Jian},
  journal={arXiv preprint arXiv:2504.18600},
  year={2025}
}

@article{zhu2026knowing,
  title={From Knowing to Doing: A Memory-Controlled Benchmark for LLM Trading Agents on Stock Markets},
  author={Zhu, Taojie and Zhao, Wentao and Sun, Rui and Luan, Beidi and Lu, Jiacheng and Wang, Sinuo and Li, Jing and Jiang, Daxin and He, Yonghong and Bai, Zuo},
  journal={arXiv preprint arXiv:2605.28359},
  year={2026}
}

@inproceedings{luo2026alphabench,
  title={Alphabench: Benchmarking large language models in formulaic alpha factor mining},
  author={Luo, Haochen and Ko, HoTin and Chen, Jiandong and Sun, David and Zhang, Yuan and Liu, Chen},
  booktitle={International Conference on Learning Representations},
  volume={2026},
  pages={47768--47827},
  year={2026}
}

@inproceedings{zhang2026alphaforgebench,
  title={AlphaForgeBench: Benchmarking End-to-End Trading Strategy Design with Large Language Models},
  author={Zhang, Wentao and Zhao, Mingxuan and Gao, Jincheng and You, Jieshun and Jia, Huaiyu and Zhao, Yilei and An, Bo and Sun, Shuo},
  booktitle={Proceedings of the 32nd ACM SIGKDD Conference on Knowledge Discovery and Data Mining V. 2},
  pages={10268--10277},
  year={2026}
}

@inproceedings{ding2026alphaeval,
  title={Alphaeval: A comprehensive and efficient evaluation framework for formula alpha mining},
  author={Ding, Hongjun and Chen, Binqi and Huang, Jinsheng and Guo, Taian and Mao, Zhengyang and Shao, Guoyi and Zou, Lutong and Liu, Luchen and Zhang, Ming},
  booktitle={Proceedings of the 32nd ACM SIGKDD Conference on Knowledge Discovery and Data Mining V. 2},
  pages={897--908},
  year={2026}
}

@misc{qwen2026qwen38,
  title        = {{Qwen3.8-27B}},
  author       = {{Qwen Team}},
  year         = {2026},
  howpublished = {\url{https://huggingface.co/Qwen/Qwen3.8-27B}},
  note         = {Model card}
}

@article{sharpe1964capital,
  title={Capital asset prices: A theory of market equilibrium under conditions of risk},
  author={Sharpe, William F},
  journal={The journal of finance},
  volume={19},
  number={3},
  pages={425--442},
  year={1964},
  publisher={Wiley Online Library}
}

@article{ang2006cross,
  title={The cross-section of volatility and expected returns},
  author={Ang, Andrew and Hodrick, Robert J and Xing, Yuhang and Zhang, Xiaoyan},
  journal={The journal of finance},
  volume={61},
  number={1},
  pages={259--299},
  year={2006},
  publisher={Wiley Online Library}
}

@article{jegadeesh1993returns,
  title={Returns to buying winners and selling losers: Implications for stock market efficiency},
  author={Jegadeesh, Narasimhan and Titman, Sheridan},
  journal={The Journal of finance},
  volume={48},
  number={1},
  pages={65--91},
  year={1993},
  publisher={Wiley Online Library}
}

@article{jegadeesh1990evidence,
  title={Evidence of predictable behavior of security returns},
  author={Jegadeesh, Narasimhan},
  journal={The Journal of finance},
  volume={45},
  number={3},
  pages={881--898},
  year={1990},
  publisher={Wiley Online Library}
}

@article{chordia2001trading,
  title={Trading activity and expected stock returns},
  author={Chordia, Tarun and Subrahmanyam, Avanidhar and Anshuman, V Ravi},
  journal={Journal of financial Economics},
  volume={59},
  number={1},
  pages={3--32},
  year={2001},
  publisher={Elsevier}
}
